# Computation Of Microbial Ecosystems in Time and Space (COMETS): An open source collaborative platform for modeling ecosystems metabolism

Ilija Dukovski[1,2,*], Djordje Bajić[3,4,*], Jeremy M Chacón[5,6,*], Michael Quintin[1,2,*], Jean CC Vila[3,4], Snorre Sulheim[1,7,8], Alan R Pacheco[1,2], David B Bernstein[2,10], William J Riehl[9], Kirill S Korolev[1,2,11], Alvaro Sanchez[3,4], William R Harcombe[5,6], Daniel Segrè[1,2,10,11,12,#]

[1] Bioinformatics Program, Boston University, Boston, MA, USA
[2] Biological Design Center, Boston University, Boston, MA, USA
[3] Department of Ecology & Evolutionary Biology, Yale University, New Haven, CT, USA
[4] Microbial Sciences Institute, Yale University, West Haven, CT, USA
[5] Department of Ecology, Evolution and Behavior, University of Minnesota, St. Paul, MN, USA
[6] BioTechnology Institute, University of Minnesota, St. Paul, MN, USA
[7] Department of Biotechnology and Food Science, Norwegian University of Science and Technology, Trondheim, Norway
[8] Department of Biotechnology and Nanomedicine, SINTEF Industry, Trondheim, Norway
[9] Environmental Genomics and Systems Biology Division, Lawrence Berkeley National Laboratory, Berkeley, CA, USA
[10] Department of Biomedical Engineering, Boston University, Boston, MA, USA
[11] Department of Physics, Boston University, Boston, MA, USA
[12] Department of Biology, Boston University, Boston, MA, USA
* Equally contributing authors
# Corresponding author. Email: dsegre@bu.edu

## Abstract

Genome-scale stoichiometric modeling of metabolism has become a standard systems biology tool for modeling cellular physiology and growth. Extensions of this approach are also emerging as a valuable avenue for predicting, understanding and designing microbial communities. COMETS (Computation Of Microbial Ecosystems in Time and Space) was initially developed as an extension of dynamic flux balance analysis, which incorporates cellular and molecular diffusion, enabling simulations of multiple microbial species in spatially structured environments. Here we describe how to best use and apply the most recent version of this platform, COMETS 2, which incorporates a more accurate biophysical model of microbial biomass expansion upon growth, as well as several new biological simulation modules, including evolutionary dynamics and extracellular enzyme activity. COMETS 2 provides user-friendly Python and MATLAB interfaces compatible with the well-established COBRA models and methods, and comprehensive documentation and tutorials, facilitating the use of COMETS for researchers at all levels of expertise with metabolic simulations. This protocol provides a detailed guideline for installing, testing and applying COMETS 2 to different scenarios, with broad applicability to microbial communities across biomes and scales.

## Introduction

Microbial communities, from the simplest synthetically constructed[1–6] to the most complex naturally occurring ones[7–10] have significant impact on multiple aspects of human life, and have therefore become a key focus of interdisciplinary research in different fields, including microbial ecology and evolution[11,12], human health[10,13–15], biogeochemistry[7,9,16–18] and metabolic engineering[19,20]. These communities may involve extensive interactions of different microbial species with each other , and with the surrounding environment[11,21–24]. Often, short-term metabolic strategies employed by individual organisms can have long-term effects on environmental structure and composition, leading to complex processes and cycles that span multiple spatial and temporal scales. An emerging challenge in systems biology is the development of quantitative predictive framework that can help understand, control and design microbial communities across these different scales - a task with a myriad of practical implications[11,25–27]. The protocol introduced here describes Computation of Microbial Ecosystems In Time and Space (COMETS)[28], a multiscale, open access, collaborative platform for predicting the complex emergent properties that result from intracellular metabolism of individual species, and ensuing microbe-microbe and microbe-environment interactions (http://runcomets.org, Figure 1, Table 1).

In recent years, genome-scale stoichiometric models of metabolism (such as Flux Balance Analysis, FBA) have made it possible to produce testable predictions of all metabolic rates (or fluxes) in individual organisms, based on the knowledge of their genomes, and on simplifying assumptions (steady state and optimality) that do not require the knowledge of thousands of kinetic parameters necessary for kinetic models[29–32]. Manually curated and automatically constructed genome-scale stoichiometric models are now available for hundreds of prokaryotic and eukaryotic organisms, from a growing number of resources[33–37]. Genome-scale metabolic modeling and flux balance methods can be expanded from individual microbes to multi-species communities, through a variety of approaches that are still the subject of active research[38]. Some of these approaches assume specific community-level optimization or balanced growth across organisms to predict ecosystem-level fluxes at steady state[39–41]. Another class of approaches (Table 2), including the one at the core of COMETS, has instead taken advantage of an iterative dynamical variant of FBA known as Dynamic Flux Balance Analysis (dFBA)[42–45]. In dFBA, intracellular metabolism is still assumed to be at steady state, but the abundance of the different species and of environmental metabolites are treated as dynamical variables. Thus, dFBA has in principle the capacity to predict both population dynamics and ecological interactions as emergent properties that arise purely from the physiology of their constituent species[28,43,46–48]. The landscape of opportunities in this arena has been also largely catalyzed by the availability of an increasing number of models for hundreds of different species, based on manual reconstructions and automated computational pipelines[49,50].

In addition to being dynamical systems, naturally occurring communities typically occupy heterogeneous, structured environments, rather than well-mixed bioreactors. Spatial structure can have major consequences on interspecies interaction, and on community structure and function, both at the microscopic scale (e.g. the structure of a biofilm[51]), and at macroscopic ones (e.g. the distribution of colonies on a Petri dish[52]). COMETS was developed at its onset as a spatially-structured simulation engine[28], making an important step towards realistic modeling of microbial communities. In parallel, other studies have also implemented different versions of spatially explicit dFBA, often taking different approaches which are tailored for different applications (Table 2).

COMETS was first developed as a flexible tool for research on natural and synthetic microbial communities[28]. Nontrivial predictions about the taxonomic distribution and functional role of different species in a 3-species artificial consortium were successfully tested experimentally[28], paving the way for a number of follow-up studies. Since then several works have used COMETS to address disparate scientific questions on microbial metabolism and microbiomes such as identifying ecological interactions among microbes[21,53,54], studying the evolution of mutualism[55,56], eco-evolutionary dynamics[57], microbial community engineering[58], gut microbiome function[59], and spatial distribution of colonies on a surface[52]. COMETS is being further developed as a free open access software platform for predictive modeling of microbes, microbial communities and complex cellular populations. COMETS could become a practical computational tool for research on microbial communities across a large range of disciplines, including microbiology, synthetic biology, metabolic engineering and biophysics. We envisage that it will continue to grow as a community effort, for which different research teams will be able to add their own modules, and benefit from the collective effort of everyone else. In this paper we (i) summarize the theoretical background that underlies COMETS simulations, (ii) describe in detail how to install and get started with simulations, and (iii) present a number of representative test cases, from the simplest to the more sophisticated. For each case study we provide a full step-by step description of how to implement simulations and interpret the results. The case studies included in the manuscript, which exemplify possible applications of COMETS are summarized in the following list:

- Creation and analyses of COMETS simulations using the MATLAB and Python toolboxes
- Simulation of bacterial growth in well mixed conditions
- Simulations of growth of multiple microbial strains in well-mixed conditions
- Modelling the diurnal cycle in photosynthesis
- Simulating competition assays and competitive exclusion
- Simulating a chemostat and crossfeeding
- Simulating evolutionary processes in microbial populations
- Simulating the sequence of mutations involved in an evolutionary innovation
- Simulations including extracellular reactions
- Modeling of spatial growth and propagation of colonies on flat surfaces
- Simulations of a virtual Petri dish
- Simulations involving demographic noise and cooperative biomass propagation
- Soil-air interface simulation

With the inclusion of a user-friendly graphical interface and versatile scripting toolboxes (e.g. Python and MATLAB, Fig. 1), our aim is to make COMETS accessible to a wide range of computational and experimental researchers as well as educators and students. For instance, students can use COMETS to simulate "experiments" and learn-by-doing core concepts in biology such as those related to microbial growth, competition for resources, metabolic exchange, and evolution (Table 1). By integrating the scripting toolboxes with COBRA, we have enabled a seamless workflow from building COBRA metabolic networks to running COMETS simulations, which greatly simplifies the process of testing predictive metabolic reconstructions. An online, hands-on tutorial is also available (https://comets-manual.readthedocs.io/en/latest/), and is regularly updated as new functionalities are added to COMETS.  Here we present several protocols for case studies utilizing COMETS 2.

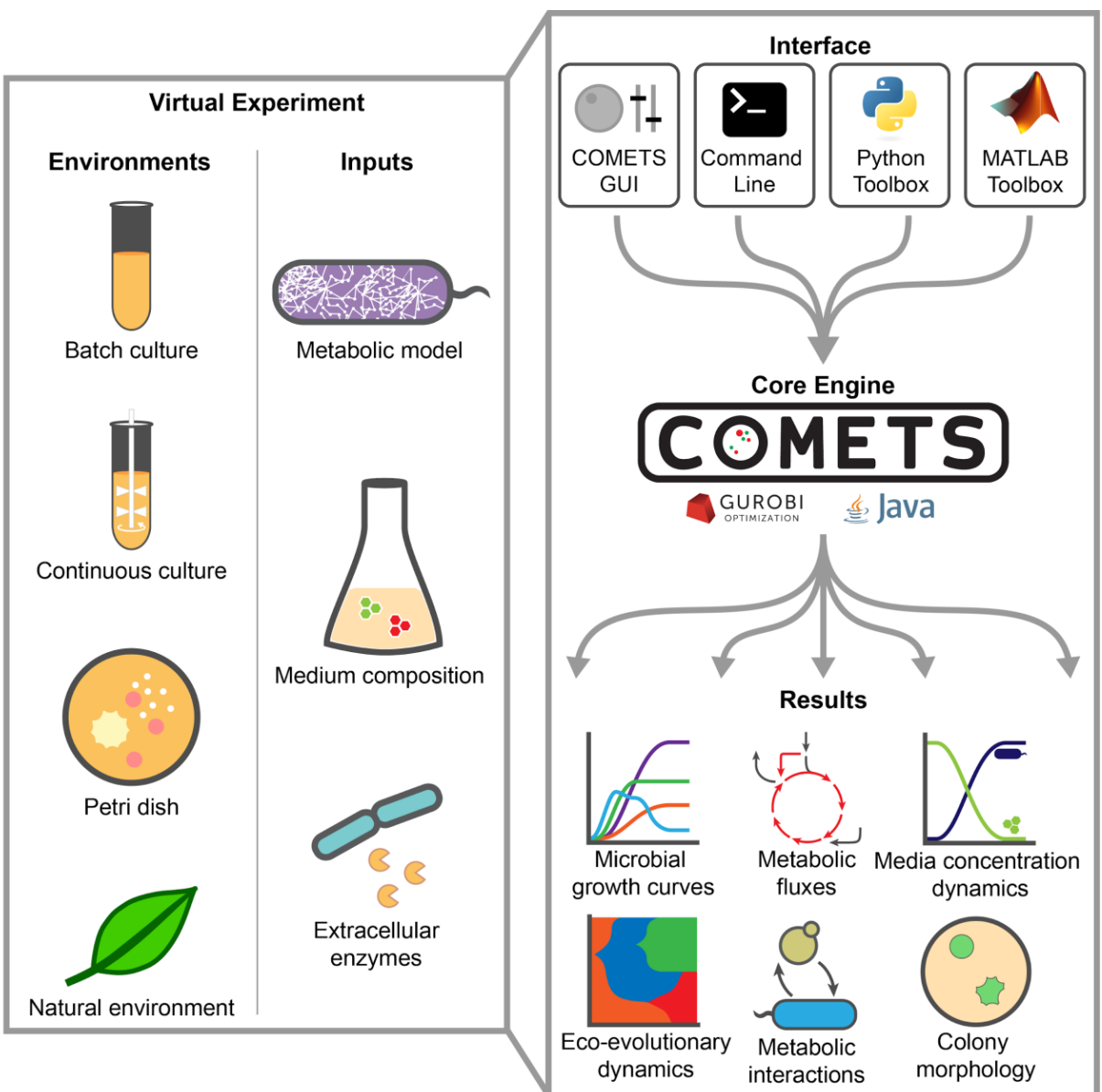

**Figure 1. Overview of the COMETS platform.** Virtual experiments in COMETS combine a variety of environments and biochemical inputs. These combinations can be quickly generated using one of the provided interfaces, which feed into the COMETS core engine. The engine simulates the spatio-temporal dynamics of the ecosystem and outputs microbial biomass information, metabolic fluxes, and media concentration over time. Downstream analysis, either within the toolboxes or with the user's software of choice, can then be applied to further visualize and characterize the results.

| General category | Specific category | Capability | Example rationale |
|---|---|---|---|
| Spatial capabilities | Biomass motion | Linear diffusion | Run-and-tumble motion |
| | | Non-linear diffusion dependent on local properties, deterministic or stochastic | Metabolite lubrication, surfactant secretion in the environment, cooperative motion |
| | | Convective (pushing) motion, deterministic or stochastic | Colony growth via pushing forces, non-motile motion |
| | | Impenetrable barriers | Barriers such as rocks or beads |
| | | Model mixing or enforced non-overlap | Cells can swim into the same general space, or create layered biofilms which prevents penetration by other cell types |
| | Metabolite motion | Linear diffusion | Normal property of chemicals |
| | | Impenetrable barriers | Barriers such as rocks or beads |
| Biological capabilities | Cell growth and death | Growth rate via flux-balance analysis with or without pFBA secondary optimization | Optimal metabolic growth with or without minimized sum of absolute values of all fluxes |
| | | Standard FBA, Monod, or Hill uptake rates | Linear, saturating, or cooperative uptake mechanisms |
| | | Lag phases via activation rate | Variable time to exit from stationary state |
| | | Fixed, species-specific death rates | Cell death, proportional to population size |

| | | Gene cost | Size of the genome represents an energy cost that is applied to lower the biomass growth rate. |
|---|---|---|---|
| | Stochastic changes | Evolution by generation of related models with altered flux bounds | Mutations arising during ecosystem lifetime |
| | | Stochastic fluctuations (gaussian or demographic) | Random fluctuations in growth rate due to environmental or demographic fluctuations |
| Environmental capabilities | Metabolite sources and sinks | Fixed local concentration of a given metabolite | Buffered source of a metabolite, such as oxygen at an air/liquid interface |
| | | Fixed local environmental metabolite replenishment rate | Spatially structured: interaction of community with nutrient-producing host cells. Spatially unstructured: Chemostat |
| | | Constant dilution rates (for all biomass and environmental metabolites) | Simulation of bioreactor in chemostat mode |
| | | Time-dependent variation on the abundance of a given extracellular metabolite, according to a predefined function | Periodic availability of light for day-night cycles |
| | Bottlenecks | Abrupt dilution events of biomass followed by replenishment of nutrients | Batch transfer experiments or seasonality |
| Other | | Extracellular enzyme costly secretion and activity | Secretion of cellulase and cellulolytic activity by diffusing enzymes in the environment |
| | | Capacity to handle many (>>100) stoichiometric models | Simulations of complex communities/microbiomes |

Table 1: List of key COMETS capabilities and application examples.

| | COMETS[28] | COMETS 2 (this work) | BacArena[60] | Matnet[61] | 3DdFBA[62] | Spatial dFBA lab[42] | IndiMesh[63] |
|---|---|---|---|---|---|---|---|
| | | | | | | | |
| Multiple genotypes | ✓ | ✓ | ✓ | ✓ | ✗ | ✓ | ✓ |
| Diffusive bacterial / metabolite spread | ✓ | ✓ | ✓ | ✓ | ✓ | ✓ | ✓ |
| Convective and/or collective bacterial spread | ✗ | ✓ | ✓ | ✓ | ✓ | ✓ | ✓ |
| Chemotaxis | ✗ | ✗ | ✓ | ✓ | ✗ | ✓ | ✓ |
| Spatially-varying diffusion | ✗ | ✓ | ✓ | ✗ | ✗ | ✗ | ✓ |
| Evolution | ✗ | ✓ | ✗ | ✗ | ✗ | ✗ | ✗ |
| Extracellular enzymes | ✗ | ✓ | ✗ | ✗ | ✗ | ✓ | ✗ |
| Approachable toolboxes | ✗ | ✓ | ✓ | ✓ | ✗ | ✗ | ✗ |
| GUI | ✓ | ✓ | ✗ | ✗ | ✗ | ✗ | ✗ |
| Population-based or agent-based | population | population | agent | agent | population | population | agent |

Table 2. Comparison of COMETS capabilities with previous version and other FBA based software packages. The black checkmark labels a fully functional capability. The blue checkmark labels a limited capability, one that requires additional programming/script writing.

## Development of the protocol

### Flux Balance Analysis: A brief overview

Flux Balance Analysis (FBA) is a constraint-based computational method used to predict the function or phenotype of an organism by simulating its metabolism. Although it has been described extensively elsewhere[31,32], here we give a brief overview of the basic principles of FBA.

The network of metabolic chemical reactions is represented by the stoichiometric matrix S. In this matrix, rows represent metabolites and columns represent reactions; $S_{ij}$ represents the moles of metabolite $i$ consumed ($S_{ij}<0$) or produced ($S_{ij}>0$) by reaction $j$. FBA, like many other stoichiometry-based models of metabolism, relies on the assumption that cellular metabolism is at steady state. This assumption should be thought of as pertaining to a population of cells over a certain period of time, such that, on average, the concentrations of metabolites inside cellular biomass do not change in time. This steady state assumption imposes the following linear constraints on the fluxes through the metabolic reactions:

$$S\nu = 0$$

where v is the vector of reaction fluxes, whose *i*-th component $v_i$ is the flux through reaction *i* (typically in units of mmol/grDW*h). Additionally, a lower (*lb*) and upper (*ub*) bounds can be set to constrain each flux between a minimal and a maximal value:

$$lb_j^\alpha \leq \nu_j^\alpha \leq ub_j^\alpha$$

These bounds may be used to define a reaction as irreversible by setting:

$$lb_j^\alpha = 0$$

In the case of "exchange reactions" (reactions representing the availability of nutrients from the environment), we use these bounds as tuning knobs to define the maximal uptake rate of the corresponding nutrients. The bulk of the metabolic fluxes are left virtually unbounded. Thus, in practice, the main constraint to internal metabolic fluxes arises from the requirement of mass balance, defined by the stoichiometric matrix and ultimately by the structure of the metabolic network, and by the boundary conditions of nutrient availability and thermodynamic infeasibility. Note that, as described later, the nutrient availability flux bounds will be dealt with in a substantially different way in dynamic FBA and in COMETS.

Mathematically, these constraints define a convex polytope, i.e. a volume of permitted fluxes in high dimensional space, with the number of dimensions defined by the number of reactions, i.e. the number of columns of the stoichiometric matrix *S*. Note that the reconstruction of a metabolic network from an organism genome (described in detail elsewhere[64]) involves substantially more complicated steps, including a detailed mapping between genes and reactions. These steps are not described here, but are an important component for the usage of FBA methods towards making accurate predictions.

In order to predict a specific set of fluxes for a given metabolic network, FBA requires an additional step , in which the feasible space is searched for a point (or set of points) that maximizes (or minimizes) a given objective function, represented in the form of a linear combination of the flux variables. Usually, this objective function is the production of a set of molecules (building blocks, energy and redox currency) that metabolism needs to provide in precise proportions in order for other cellular processes (synthesis of macromolecules, membranes, DNA replication, transcription, etc) to generate of new biomass[65]. The use of linear objective functions makes it possible to solve this mathematical problem through well-established efficient linear programming algorithms, available through a number of libraries. A typical FBA optimization for a genome-scale model, on a standard laptop computer, takes on the order of a few milliseconds. Biologically, the search for a set of fluxes that optimizes a given objective implies the hypothesis that an organism has evolved to be able to regulate its metabolic fluxes to approach that optimum under a set of environmental conditions. In other words, the model assumes an "optimal regulation". This assumption is partly justified by evolution, but it does not necessarily hold in all conditions[66–69]. COMETS can accommodate arbitrary objective functions, in addition to maximization of biomass production. Moreover, it supports multiple objectives optimized iteratively, including the minimization of the sum of the absolute values of fluxes (also known as parsimonious FBA)[69].

## Dynamic Flux Balance Analysis

Dynamic Flux Balance Analysis[43] is an iterative extension of FBA that explicitly includes the dynamics of the organisms as they grow, and the effects of this growth in the environment. dFBA produces piecewise-linear approximations of the microbial growth curve (i.e., biomass as a function of time), and of the environmental abundance of metabolites, that can change due to external factors, or through uptake/secretion fluxes. Notably, in dFBA, while extracellular metabolites can dynamically change, intracellular ones are still assumed to be at steady state (through fast equilibration). In COMETS, for each microbial species $\alpha$, we implement dFBA by numerically solving its biomass equation:

$$\frac{\partial B^\alpha}{\partial t} = \nu^\alpha_{growth} B^\alpha$$

where $B^\alpha$ is the biomass of species $\alpha$ and $\nu^\alpha_{growth}$ is the growth rate, as computed through FBA. Effectively, upon fixing a finite $\Delta t$, a change of biomass for each species in the next time step is computed as $\Delta B^\alpha = \nu^\alpha_{growth} B^\alpha \Delta t$. The dynamics of each external metabolite is governed by the equation:

$$\frac{\partial Q^i}{\partial t} = \sum_\alpha \nu^\alpha_i B^\alpha$$

$Q^i$ where is the abundance of external metabolite and $\nu^\alpha_i$ is the exchange flux of metabolite $i$ in species $\alpha$. Similar to the biomass equation, the changes in extracellular metabolites are computed based on FBA-inferred fluxes and the finite time interval.

At the beginning of the simulation, the starting molecular composition of the environment is initialized, based on boundary conditions set by the user. At each iteration, the program estimates each model's uptake bounds based on the external concentration of nutrients, and solves each model's FBA, obtaining an estimate of the growth rate and all other fluxes, including uptake and secretion. These fluxes are used as inputs in the above equations to compute the changes in biomass and extracellular metabolites. One of the important outcomes of this process is the fact that different organisms may compete for common resources and/or exchange metabolites as an outcome of their own objective function. Microbe-environment and microbe-microbe interactions are emergent properties of the physiology of each species[28].

As mentioned above, a key aspect of dFBA is that it requires a mapping between the external nutrient concentration and the maximal uptake rate for each metabolite in each organism. COMETS includes three possible choices for such mapping functions. Bounds can be either a linear function of the concentration, a Monod (or Michaelis-Menten) function, or a pseudo-monod uptake type (i.e. linear until a given threshold, then constant). The type of uptake can be specified in the parameter exchangeStyle.

## Spatial structure and dynamics

The classical implementation of dFBA described above (which can be implemented in COMETS) corresponds to a well-mixed system, in which all microbes and metabolites are uniformly distributed and have access to each other in proportion to their concentration. In addition to this dynamics in time, COMETS is able to take into account the spatial structure of microbial colonies and communities, simulating arbitrary two-dimensional spatial structures (a 3D version is in principle available, but has not been thoroughly tested yet). Spatial structure in COMETS is implemented as a 2D grid of cubic "boxes" with a given dimension and volume. Inside each of these "boxes", a well-mixed scenario is assumed. The biomass of different species and the environmental metabolites can propagate from a given box to neighboring boxes based on physics laws of convection-diffusion, as described in detail below.

### Biomass propagation

The core of the COMETS method is the simulation of the propagation of the biomass present in the system. The simulations are performed by numerically solving the partial-differential equations that govern the dynamics of the system. The dynamical variable of biomass (formally biomass density) is spatially continuous. Although the natural unit of biomass is a single cell of an organism, we implemented the biomass dynamics as one of a locally averaged continuous quantity. The reason for this choice is to be able to simulate macroscopic systems on the order of centimeters and larger. A methodology based on individual cells would significantly hinder the extent in both size and time of the simulations.

The partial differential equation for biomass propagation written in the general form is:
$$\frac{\partial B^\alpha}{\partial t} = \vec{\nabla}(D^\alpha \vec{\nabla} B^\alpha) - \vec{\nabla}(B^\alpha \vec{u}^\alpha) + f^\alpha(B^\alpha, Q^m)$$

Here $B^\alpha = B^\alpha(\vec{r}, t)$ is the biomass of species $\alpha$ at spatial position $\vec{r}$ and at time $t$. The operator $\vec{\nabla}$ is the vector differential operator, $D^\alpha = D^\alpha(\vec{r}, t)$ is the diffusivity of species α, and it can vary in space and time explicitly, or as a function of the local biomass. $Q^m = Q^m(\vec{r}, t)$ is the local nutrient/metabolite content (density). $\vec{u}^\alpha = \vec{u}^\alpha(B^\alpha; \vec{r}, t)$ is the local velocity of the bulk biomass of the corresponding species. The biomass velocity can be a function of the biomass (as a mechanistic model) or explicitly a function of the time and spatial position. Finally, $f^\alpha(B^\alpha, Q^m)$ is the biomass growth/death term. This term has the same form as the corresponding one for dFBA.

The temporal dynamics of the biomass at a spatial point is governed by the three terms on the right-hand side of the equation. The first term is a diffusive one, and it models the free movement of the individual bacterial cells. The diffusivity may be an explicit function of time and/or spatial position. In this case the local diffusivity depends on the external conditions, such as material in the region where the biomass is propagating, etc. The diffusivity may also be a function of the biomass, modeling the cooperativity in the propagation of the bacterial colony. The second term on the right-hand side of the equation is the advective one and models the bulk motion of the biomass with a local velocity $u^\alpha$. The local velocity may explicitly depend on the spatial point and time. This would be a model of biomass motion in external flow.

The biomass velocity however may be a function of the biomass itself given via a mechanistic model, such as a model of propagation by mutual mechanical pushing of the cells.

In COMETS we have implemented the mechanistic model of biomass propagation by cellular pushing [70]. As individual cells grow and divide, the local density of the biomass is increased. At the point when the density reaches the value of densely packing, the cells are in mechanical contact, and a field of pressure develops due to the mechanical interaction, i.e. pushing of neighboring cells. The local velocity of biomass $B^\alpha$ is given by the gradient of the local pressure developed due to cells pushing on each other:

$$\vec{u}^\alpha = \vec{\nabla} P / \mu^\alpha$$

where $\mu^\alpha$ is a friction constant and $P$ is the pressure field given by:

$$P = E^\alpha (1 - \rho_0/\rho)^{3/2}$$

where $\rho$ is the local density of the total biomass, $E^\alpha$ is the elastic constant for species $\alpha$ and $\rho_0$ is the density of the biomass at closed packing, i.e. when the bacterial cells are touching, but not pushing each other. If the density $\rho < \rho_0$, the pressure field is equal to zero.

Another model for propagation of bacteria biomass that we implemented in COMETS simulates the cooperative behavior in a dense bacterial colony. Based on the fact that certain bacterial species secrete a lubricant [71,72] which changes the local mobility of the bacterial cells. This secretion is typically dependent on the local density of cells. In COMETS we simulate this phenomenon by modeling the biomass diffusivity of species $\alpha$ as:

$$D^\alpha = D_0^\alpha + D_k^\alpha \rho^k$$

where $D_0^\alpha$ is a general linear term, and $D_k^\alpha \rho^k$ depends on the local biomass density to the power of $k$.

In addition to the dependence of $D^\alpha$ on the biomass density, the diffusivity of the biomass may optionally be restricted to the parts of the biomass field that are actively growing. We implement this feature by multiplying $D^\alpha$ with the Hill function:

$$H(\Delta B^\alpha) = (\Delta B^\alpha)^n / ((\Delta B^\alpha)^n + K^n)$$

where $\Delta B^\alpha$ is the local biomass change due to growth in a single discrete time step, and $n$ and $K$ are the Hill function parameters.

The last term in the equation for biomass propagation models the local growth (and/or death) of the biomass. This term typically is proportional to the biomass, and a pre-factor given by the growth rate of the biomass. In COMETS, the growth rate of the local biomass is determined by the metabolic activity in the model, and is a function of the local quantity (formally concentration) of the external nutrients/metabolites. In COMETS, we calculate the growth rate utilizing the Flux Balance Analysis methodology (as described above). The local growth rate can also be augmented by a death rate that removes a fraction of the biomass at each timestep.

### Nutrient propagation

In addition to the biomass propagation, COMETS simulates the spatio-temporal dynamics of the metabolites/nutrients that are uptaken and/or secreted by the organisms. The dynamics of the external metabolite $m$ is determined by their uptake and secretion by the organisms, as well as their convection over the spatial layout:

$$\frac{\partial Q^m}{\partial t} = \vec{\nabla}(D^m \vec{\nabla} Q^m) - \vec{\nabla}(Q^m \vec{u}^m) + \sum_\alpha B^\alpha \nu^{\alpha,m}$$

Here the diffusivity of the metabolites may also be locally defined, or/and can depend on the local biomass content.

### Demographic and growth noise

Two types of stochastic noise are implemented in COMETS.

#### Growth rate noise

This is a simple broadening of the growth rate with a gaussian noise term. Instead of implementing the growth rate as calculated by the FBA algorithm, it is sampled from a gaussian distribution centered at the FBA obtained growth rate.

#### Demographic (shot) noise

Demographic noise is given as a stochastic term in:

$$\frac{\partial B^\nu}{\partial t} = ... + \mu B^\nu + \sigma \sqrt{B^\nu} \eta$$

where *B$^\nu$* is the biomass of species *v*, *η* is white noise and *σ* is a parameter that determines the magnitude of the noise. The demographic noise is implemented in COMETS according to the method described in [73].

The change of the biomass from the growth term is first calculated, then it is resampled in two steps according to the procedure in[73]. If , first we sample the shape parameter of the Gamma distribution from the Poisson distribution:

$$P(\alpha; \lambda) = e^{-\lambda} \frac{\lambda^\alpha}{\alpha!}$$

Where:

$$\lambda = \frac{2B^\nu}{\Delta t (\mu\sigma)^2}.$$

Then, with the sampled *α*, we sample from the Gamma distribution:

$$\Gamma(x; \alpha, \beta) = \beta^\alpha \frac{x^{\alpha-1} e^{-\beta x}}{\Gamma(\alpha)}$$

where the scale parameter *β=1* and *Γ(α)* is the Gamma function.
The new biomass is given by:

$$B^\nu = \frac{1}{2} \Delta t (\mu\sigma)^2 x$$
.

### Extracellular reactions

COMETS includes the capability to use kinetic rate laws to simulate two types of reactions involving extracellular media components. The first are elementary reactions of arbitrary order with any number of reactants or products, of the form

$$x_1 M_1 + x_2 M_2 + \to y_1 P_1 + y_2 P_2 + ...$$

for reactants *M* and products *P*, with respective stoichiometries *x* and *y*, and with a reaction rate

$$r = k \Pi [M_i]^{x_i}$$

given the rate constant *k*. The second type are enzyme-catalyzed reactions with a single substrate and any number of products, of the form

$$S \to y_1 P_1 + y_2 P_2 + ...$$

where the stoichiometry of the substrate *S* is always assumed to be 1. The reaction rate is determined by the Michaelis-Menten equation

$$r = k_{cat}[E] * [S]/(K_M + [S])$$

that accounts for the concentrations of the enzyme *E* and the substrate *S* the turnover rate $k_{cat}$, and the half-saturation concentration $K_M$.

Changes in metabolite concentrations over the course of a single simulation timestep are calculated by converting the set of all extracellular reactions into a system of ordinary differential equations, then approximating the solution with the classical Runge-Kutta integrator from the Apache Commons Math library[74]. The process of updating metabolite concentrations by applying the effects of extracellular reactions happens once during each simulation timestep, after metabolites have been updated by the dFBA process and before diffusion occurs.

### Evolution

In addition to ecological dynamics, COMETS also has the capability of mutating species during the simulation, which results in the capability of simulating evolutionary dynamics. Mutations occur during growth: at each iteration and for each species α, COMETS first computes the number of new individual cells arising in the previous time interval *Δt* as

$$N_G = [B_\alpha(t) - B_\alpha(t - \Delta t)] C_S$$

where $C_S$ is the size of a single cell (in grs. dry weight, specified by the parameter cellSize). Given the total population growth $N_G$ and mutation rate *μ*, COMETS stochastically samples a number from a Poisson distribution with mean $N_G μ$ (or a binomial if populations contain less than 10 cell divisions). The resulting mutants - new stoichiometric models with modified stoichiometry based on a set of rules (see below) - are then placed randomly in cells containing biomass of the ancestor, with a probability per cell of the simulation grid proportional to the fraction of $N_G$ in that cell.

Two types of mutations are implemented in COMETS, reaction knock out and reaction knock-in. The knock-out rate $μ_{KO}$ is set using the parameter mutRate, and represents the knock-out rate per generation and per reaction. Thus, $μ=Rμ_{KO}$, where *R* is the number of reactions of a given model. In contrast to knock outs, the knock in rate is computed per generation and is set up using the addRate parameter. In order to simulate knock-in mutations (i.e. reaction additions) models must be previously prepared by adding all the reactions that we want to be potential additions to the model or models, with both upper and lower bounds equal to 0. These reactions will initially be unavailable to the optimizer, and become available only once "added", i.e. once their upper bound is set to 1000 by COMETS during the simulation.

## Numerical integration of spatio-temporal equations

The method used for numerical integration of the partial differential equations in COMETS depends on the type of equation, i.e. the type of model of spatio-temporal propagation, that is being solved. The three different models for propagation of biomass, the simple diffusion, propagation by pushing and non-linear cooperative diffusion, cannot be optimally solved by a single method.

For the simple (linear) diffusion model of biomass propagation the user can choose between two implemented numerical methods for its solution. One is using an alternating direction implicit (ADI) scheme with a central difference formulation[28] and the other is an 8-point integration scheme. The other two models of biomass propagation, the model of convection (pushing) and the non-linear diffusion, due to the presence of the nonlinear terms, are solved by implementing the predictor-corrector Adams-Bashford-Moulton scheme[75]. The diffusion of the media is solved by the standard implicit method, the same as for the linear diffusion of the biomass.

Detailed description of the numerical methods implemented in COMETS can be found in the COMETS user manual (https://comets-manual.readthedocs.io).

## Software architecture of COMETS

### Core architecture

The core of COMETS is written in Java and performs its main functionalities: dynamic FBA, propagation of biomass and media in time and space, extracellular reactions or evolution.

The code is organized in several Java packages, each containing several classes. The core of the code is in the following packages:

comets: This package contains the top-most super-classes and interfaces. This organization of the superclasses was done with a future development in mind, including the possibility of modules that compute growth with algorithms other than FBA.

fba: This package contains the core of the program, including most of the data structure as well as the run methods that perform the core procedures of the simulation. Here are most of the FBA specific subclasses of the superclasses found in the comets package.

ui: This package contains the classes related to the graphical user interface.

util: The util package contains general mathematical utilities, independent of the FBA methodology, such as several PDE solvers.

The Java core of the FBA methodology in COMETS is structured in four main classes, organized hierarchically:

Model: Instances of this class are metabolic models which are optimized using FBA in the simulations. This class also has a method to mutate models, i.e. add or remove reactions, used for evolutionary simulations.

Cell: The spatial structure in COMETS is structured as a grid (either 2 or 3-dimensional). The cell class represents what we refer to earlier as a "box", i.e. a single location on this grid, with defined dimensions. Note that this cell should not be confused with a biological cell. A cell class contains models and media, whose biomasses and concentration are updated in each iteration of the dynamic FBA simulation. This class also performs the extracellular reactions, if present.

World: The world contains all the cells (i.e. boxes, see cell class above) with their models and media. After each iteration, this class performs the computations necessary to propagate biomass and media between cells. In evolutionary simulations, it also decides stochastically which models mutate and in which cells.

Parameters: This class contains all the necessary parameters for running a simulation. These parameters, their units, and default (or alternative) values are listed in the Table in Appendix 2.

Comets: This is the main class of COMETS. It integrates all of the above, runs a simulation and produces the output.

The software contains some additional "helper" classes that deal with file loading or optimizers used by the simulation. The central class that does the FBA optimization is the abstract class FBAOptimizer, which has two subclasses, FBAOptimizerGurobi and FBAOptimizerGLPK.

## Optimizers

A central computational part of the FBA method as implemented in COMETS is the optimization of the objective function in each genome scale metabolic network. COMETS currently supports two optimizers that perform this task. The optimizer is defined at the model level and the user can choose one of them for each model. The implemented optimizers are:

Gurobi: This commercial software package is available at http://www.gurobi.com. Users with academic affiliation can obtain a free academic license. Gurobi is set as the default optimizer in COMETS.

glpk: This free open source package is available at https://www.gnu.org/software/glpk/.

### Parallelized dFBA

The simulation algorithm implemented in COMETS at each time step performs a single FBA optimization for each model present at the spatial grid point. We have implemented an option to run these optimizations in parallel on a multi-core CPU. This provides significant improvement in performance when computations are run on high-performance computers. The parallelization is implemented by a multithreading algorithm.

### Key inputs to a COMETS simulation

At the basic level, the simulation parameters in COMETS are loaded from input text files. The parameters files can also be loaded from the Graphical User Interface (GUI). Some of the key simulation parameters can be set directly in the GUI. If a parameter is not listed in the input file it is assigned the default value.

There are three types of input: models, parameters and layout. All three types of input are stored in text files files with a specific COMETS format.

Models: Contain the stoichiometric metabolic reconstructions for organisms to be used, and some model-specific COMETS parameters such as model propagation constants and type of optimizer.

Layout: Contains information about the environment, including the media and spatial structure at each location in space, and the possible presence of extracellular enzymatic reactions.

Parameters: A list of simulation parameters. A comprehensive set of parameters that can be specified can be found in Appendix 2 and in the online documentation.

### Output observables

A COMETS simulation can produce several types of quantitative outputs at all or at selected time steps. Output files can be written in a MATLAB .mat binary format, as MATLAB .m scripts containing variable definitions, or as tab-separated files. In the python toolbox, these files will be automatically read by the toolbox and results data will be stored within python, to facilitate downstream analyses.

Total biomass
This output consists of a tab-delimited text table containing the iteration number in the first column, and the total biomass of each model present in the simulation in the successive columns.

Biomass: This contains detailed information about the spatial distribution of biomass for each model and each simulation grid cell.

Media: This output consists of the detailed information of the amount of extracellular metabolites/nutrients on the spatial grid. Alternatively, one can select specific metabolites to track using the specificMedia parameter set, which returns a tab-separated file.

Fluxes: The collection of all fluxes, including the exchange fluxes, are recorded for each model at each time point, and at each grid point of the spatial layout.

### User Interfaces

While the actual computation is performed by the core COMETS Java code, efficient use relies on using either the MATLAB or Python COMETS-toolbox to generate layout and model files. These toolboxes are not required to run COMETS itself if suitable models and layouts are available, but it allows the user to create advanced environments and settings in an efficient and intuitive manner. The layout and model file created by these toolboxes are then subsequently provided as input files to the core COMETS software as illustrated in Fig. 1. The COMETS Graphical User Interface (GUI) provides some basic visualization of the simulation results, but it is highly recommended to analyze the result files created by COMETS in MATLAB, Python or any other data visualization tool.

## Limitations

Some fundamental limitations of COMETS are inherited from basic limitations of the stoichiometric modeling and FBA methodology. These include the fact that availability of genome scale metabolic models for the organisms of interest is not guaranteed. The process of creating a new genome scale metabolic model from the sequenced genome of an organism has its own protocols[49], and is not discussed here in further details. A number of approaches for the automated reconstruction and gap filling of stoichiometric models are now available, though one should be careful in using such gap-filled models without further testing or manual curation. Another limitation inherent to FBA, which is propagated to COMETS, is the lack of explicit gene regulation. Some approaches previously developed to cope with this limitation (such as rFBA, caFBA, Gimme) can be in principle added to COMETS, but are not part of the current releaseHowever, classical FBA perturbations, such as deletion of specific genes or blockage of specific reactions can be easily implemented in COMETS, by setting appropriate constraints.

Other limitations of COMETS are specifically arising from the complexity of numerical integration of the convection-diffusion equations. Internally, the COMETS engine uses a fixed time step integration rather than a variable time-step integrator such as in DFBAlab[53]. This requires users to choose a time step that is small enough to avoid numerical errors and instabilities that can propagate in growth and spatial solutions. This sometimes requires the choice of very small time-steps (less than a second in the branching colony protocol), which can result in long simulation times for complex layouts. COMETS will let the user know if the time step is too big to guarantee correct numerical integration of the biomass propagation equation (through a message: Warning: Negative biomass at X,Y, reduce the time step.), or the media diffusion equation (through a message: Unstable diffusion constants. Try setting the time step below <Δt> seconds) where the estimate of $\Delta t$ is given by the stability criteria *Δt<Δx$^2$/D*, where Δx is the size of the spatial grid spacing and *D* is the diffusion constant.

Another important limitation that users should be aware of is that, by design, COMETS takes a population-level approach, rather than an individual-based approach. In other words, FBA, as computed at each coordinate in space, is meant to estimate the average metabolic behavior of the population of cells present in a box at those coordinates. This approximation, which is a natural extension of the steady state nature of FBA, implies that COMETS is currently not ideal for studying phenotypic cell-to-cell variability in a population. Notably, however, by using appropriate statistical calculations, COMETS can be used to model some important aspects of microbial dynamics that depend on the discrete nature of cellular populations, including demographic noise (see section Demographic noise and cooperative biomass propagation) and evolutionary dynamics stemming from genetic variations (see section Simulating evolution).

Finally, while we have incorporated many biological processes into the COMETS engine, there is a long list of important biological processes that can impact microbial growth, many of which can be incorporated in future versions or extensions of COMETS. These include chemotaxis, toxin release and sensitivity, quorum sensing, and other kinds of signaling processes.

## Materials: Hardware requirements and Installation

### Recommended Hardware

- A computer with a 64-bit processor and either Linux, Windows or Mac OS system. Memory recommendation is 8Gb or higher. A multicore processor is recommended to run multithreaded simulation of a large spatial layout grid, but not strictly necessary.
- Java, including Java Development Kit. Version 8 or higher is recommended.
- Gurobi Optimizer (recommended; free academic license available). Alternatively, the GNU Linear Programming Kit (GLPK) can be used as an alternative, open source, optimizer.
- Python (recommended > 3.4) is required to use the Python toolkit. Additionally, COBRApy[76] is recommended to facilitate model creation.
- MATLAB and Cobra Toolbox[64] are required to use the MATLAB toolkit.

COMETS has been tested on:
- Microsoft Windows 10 Home, version 10.0.18362
- Linux Ubuntu 20.04.1 LTS
- MacOS X Catalina 10.15.5 and MacOS X El Capitan Version 10.11.6

## COMETS Installation

### Java install ● TIMING 10-20 min

COMETS is written in the Java programming language. Installed Java 64-bit platform is a prerequisite for running COMETS. The minimum required version of 64-bit Java is 1.8. Java can be downloaded and installed from: https://www.java.com/. The Java Development Kit can be downloaded and installed from: https://www.oracle.com/technetwork/java/javase/downloads/index.html

### Gurobi install ● TIMING 15-30 min

The primary and default optimization software in COMETS is Gurobi. The package can be downloaded and installed from: http://www.gurobi.com/ The installation of Gurobi requires obtaining a license from: http://www.gurobi.com/downloads/licenses/license-center. Academic users that will use COMETS on an individual basis may obtain a free academic license. When the installation of Gurobi is finished, it is very important to have the environment variable GUROBI_HOME set to the directory where Gurobi was installed. In Windows this variable is set automatically during the installation process. In Linux and MacOS however, depending on the system, sometimes this is not the case. It is therefore important to make sure that a line such as the following example:

export GUROBI_HOME=/usr/gurobi/gurobi902/linux64/

is included in the user's .bashrc file in Linux, or the corresponding file for an alternative shell. The version name and number in gurobi902 should be set to the one installed.
In MacOS the line:

export GUROBI_HOME=/Library/gurobi902/mac64/

should be included in the .bash_profile file in older versions of MacOS, or .zshrc in the latest version of MacOS. The version name and number in gurobi902 should be set to the one installed. It is important to source these files before attempting to run COMETS. The easiest way to do that is to close the terminal and open another one.

Also, the COMETS installer will add lines:

export COMETS_HOME=/home/username/comets
export PATH=$PATH:$COMETS_HOME
export GUROBI_COMETS_HOME=$(echo $(ls -d /usr/gurobi/gurobi*/linux64)|awk {'print $NF'})
export LD_LIBRARY_PATH=$LD_LIBRARY_PATH:$GUROBI_COMETS_HOME/lib/

in the .bashrc file in Linux, and

export COMETS_HOME=/home/username/comets

```
export PATH=$PATH:$COMETS_HOME
export GUROBI_COMETS_HOME=$(echo $(ls -d /Library/gurobi*/mac64)|awk {'print $NF'})
export LD_LIBRARY_PATH=$LD_LIBRARY_PATH:$GUROBI_COMETS_HOME/lib/
```

in the .bash_profile file in older versions of MacOS, and .zshrc in the latest version of MacOS, where username is replaced with the specific one for the user. This should make Gurobi libraries available to COMETS. It is our experience however, that in some instances Gurobi cannot be started if GUROBI_HOME is not set explicitly in the .bashrc, .bash_profile or the .zshrc file.
If you custom installed Gurobi in a directory other than the standard one as described above, please change the above lines to point to it. If you have more than one Gurobi package installed, please change the above lines to point to the installation for which you have a Gurobi license.

In Windows, the Gurobi installer adds the GUROBI_HOME environment variable automatically. However, if the Gurobi cannot be started by COMETS, the most likely reason is that this variable was not set properly, or the user did not obtain a license. In Windows, this variable can be set in the Control Panel.

More information can be found in the manual: https://comets-manual.readthedocs.io.

### GLPK install ● TIMING 15-30 min

The installer for the alternative optimizer used by COMETS can be found here: https://www.gnu.org/software/glpk/ This is an open source package. The details on how to install it can be found in the COMETS manual: https://comets-manual.readthedocs.io.

### COMETS installation ● TIMING 2-10 min

There are two ways to install COMETS:
1. Using the COMETS installer.
2. Unpacking the comets_2.10.0.tar.gz file.

The easiest way is to use the installer, especially recommended for individual use on a laptop or desktop. The installer can be downloaded from: https://www.runcomets.org. The users are required to register, after which they can obtain the installer appropriate for their system. The installer guides the user through a standard GUI installation procedure that includes accepting the license agreement, choosing the directory where COMETS will be installed (the default directory is recommended), the option to create a desktop shortcut, etc. The installer is available for the Windows (comets_windows-x64_2_10_0.exe), MacOS (comets_macos_2_10_0.dmg) and Linux (comets_unix_2_10_0.sh) systems. The installer is invoked either by a double-click on their icon (Windows, MacOS), or running them from a command line (Windows, Linux). If the installer cannot be started in Linux, the following command should be executed:

```
chmod a+x comets_unix_2_10_0.sh
```

before running it.

In MacOS, if the system does not allow the application to be installed, holding the command button and clicking on the installer, or going to the apple menu → System Preferences → Security & Privacy → General → Open Anyway, will solve this problem. If another window with security warning opens up, clicking on Open Anyway will allow the installation.

The installer will install the GUI version and a shortcut to it on the desktop. Also, the installer will add the script comets_scr to the user's PATH variable, so COMETS can be started in all three systems from a command line by:

comets_scr comets_script

where comets_script contains description of the input files:

```
load_comets_parameters      global_params.txt
load_package_parameters     package_params.txt
load_layout                 layout.txt
```

As mentioned above, the COMETS installer will add the lines such as:

```
export COMETS_HOME=/home/username/comets
export PATH=$PATH:$COMETS_HOME
export GUROBI_COMETS_HOME=$(echo $(ls -d /usr/gurobi/gurobi*/linux64)|awk {'print $NF'})
export LD_LIBRARY_PATH=$LD_LIBRARY_PATH:$GUROBI_COMETS_HOME/lib/
```

in the .bashrc file in Linux, and the lines:

```
export COMETS_HOME=/home/username/comets
export PATH=$PATH:$COMETS_HOME
export GUROBI_COMETS_HOME=$(echo $(ls -d /Library/gurobi*/mac64)|awk {'print $NF'})
export LD_LIBRARY_PATH=$LD_LIBRARY_PATH:$GUROBI_COMETS_HOME/lib/
```

in the .bash_profile file in older versions of MacOS, and .zshrc in latest versions of MacOS, where username is replaced with the specific one for the user. In order to source the above lines, the user in Linux or MacOS should open a new terminal, or source the .bashrc, .bash_profile or .zshrc files by the dot command:

. ~/.bashrc

If the user chooses to install COMETS or Gurobi in directories other than the default ones, then the lines above should be edited accordingly.

In Windows the user may want to check if COMETS_HOME and GUROBI_HOME are properly set, and set them if needed.

In addition to the GUI installer, we provide the comets_2.10.0.tar.gz file for custom installation, typically on a Linux system. The file should be unpacked in the directory were COMETS will be installed with:

```
tar -xzvf comets_2.10.0.tar.gz   ./
```

This will create the comets installation directory. In this case, the user should add the above lines to the .bashrc file manually.

COMETS can easily be uninstalled by running the uninstaller that can be found in the directory where COMETS was installed. In Windows it can also be uninstalled from the Control Panel.

For more information, the user can consult the manual: https://comets-manual.readthedocs.io.

COMETS MATLAB Toolbox install ● TIMING 10-30 min

The prerequisite for this toolbox is to have MATLAB installed from https://www.mathworks.com.

The COMETS toolbox for MATLAB can be downloaded from https://github.com/segrelab/comets-toolbox. The user may download the toolbox as an archive from the GitHub repository, or execute the following command from the command line (the folder ./comets-toolbox will be created in the working directory):

git clone https://github.com/segrelab/comets-toolbox.git comets-toolbox

A prerequisite to install the toolbox this way is to have installed git which can be found here: https://git-scm.com/. Once git is installed, the user can download the toolbox either using the command line above, or by using the GitHub Desktop application that can be obtained here: https://desktop.github.com/.

Once this folder has been created, run the following commands in MATLAB to add the toolbox and its subfolders to the MATLAB path:

addpath(genpath("comets-toolbox"),"-end");
savepath();

where comets-toolbox is the full path to the directory where the toolbox was installed. On a Windows system, for example, this path may be:

C:\Users\username\comets-toolbox

where username is replaced with the specific one for the user.

Alternatively, the code can be downloaded as a .zip file by clicking on the green "Code" button at the repository. Then it needs to be extracted by right click→Extract all or using the unzip command:

unzip comets-toolbox-master.zip

and run the addpath and savepath commands as above, with the path to the directory comets-toolbox-master.

In addition, this toolbox requires the installation of the COBRA toolbox, available at https://opencobra.github.io/ [64]. Many functions of the COMETS toolbox will not work before loading the COBRA toolbox using the *initCobraToolbox()* command. The detailed instructions for installing the COBRA toolbox can be found here: https://opencobra.github.io/cobratoolbox/stable/installation.html. A prerequisite to install the COBRA toolbox is to have installed git which can be found here: https://git-scm.com/. Once git is installed, the toolbox can be installed by running:

git clone --depth=1 https://github.com/opencobra/cobratoolbox.git cobratoolbox

As mentioned above, the package can be fetched by using GiHub Desktop: https://desktop.github.com/. Once this folder has been created, run the following commands in MATLAB to add the toolbox and its subfolders to the MATLAB path:

addpath(genpath("cobratoolbox"),"-end");
savepath();

where cobratoolbox is the full path to the directory where the toolbox was installed. On a Windows system, for example, this path may be:

C:\Users\username\cobratoolbox

where username is replaced with the one specific for the user.

COMETS Python toolbox (cometspy) install ● TIMING 2-10 min

We recommend installing Python (version >= 3.6) using the Anaconda distribution which can be installed from https://www.anaconda.com/products/individual. Anaconda will conveniently install the jupyter notebook, which is how we have provided our protocols. Alternatively Python can be installed from https://www.python.org/. Note that If the downloaded installer does not run in Linux or MacOS, execute the following command line:

chmod a+x  Anaconda3-X.sh

where X is replaced with the specific version name and number.

The COMETS Python toolbox (cometspy) is available from the package manager PyPI cometspy using the pip command. In Linux systems, the pip command is usually installed through available repositories (e.g. sudo apt-get install python3-pip in Debian-based distributions). If python has been installed through Anaconda pip will have been installed during this installation.

Once pip is installed, cometspy can be installed by running the command line:

pip3 install cometspy

If this command does not run, close the terminal and open another one, then run it again.

In Windows this is best done by going to the start menu, and running "Anaconda Powershell Prompt". The above command can be run from the Anaconda Powershell.

# Expertise needed to implement the protocol: COMETS User Experience

## Levels of expertise

COMETS can be utilized in several different modes, requiring different levels of expertise. In its most basic form, the expertise required is a general familiarity with the simulated biological processes and basic computational skills. At this level, the user can easily utilize the Graphical User Interface (GUI) or manually write text input files to run simple COMETS simulations. This approach may be appropriate for introducing COMETS to undergraduate students in classes on biology and related discipline.

A slightly more advanced level is required to use the MATLAB or the Python toolbox to prepare and run the models and layouts, given available stoichiometric models in other standard formats such as SBML and/or .mat (also used in COBRA). This level would require a basic knowledge of MATLAB or Python, depending on the toolbox of choice. Using these toolboxes however greatly expands the possibilities of interaction with COMETS. For example, one can easily run simulations in a loop that changes environmental conditions in each loop, and plot and analyze the results in a fully custom way. These modes of running COMETS, in addition to being relevant for a number of research applications, could be used in educational settings, e.g. for teaching systems biology in graduate student classes.

Users comfortable with command line applications and scripting can run COMETS using a command-line approach, which can facilitate use in a high-performance computing facility.

Finally, the most advanced level of expertise is the development of custom COMETS features, which requires knowledge of the Java programming language, and version control systems Git and GitHub. In this case, a desirable skill is familiarity with one of the integrated development environments such as Eclipse. This level of expertise is required if the user would like to develop novel capabilities, which are beyond the scope of this protocol.

In the following sections we provide a detailed overview of the different COMETS interfaces the user may choose. In the Procedures section, examples are given for running simulations using each of these interfaces.

## GUI

The basic level of expertise required to run COMETS with the GUI is the ability to download the installer from the COMETS website, download and install Java and Gurobi as well as obtain and install the Gurobi license. The installation instructions for obtaining Gurobi license require the ability of opening a command line terminal and running a single command line. The COMETS installer provides an option for creating a desktop shortcut. This in practice means that the user is required to only be able to navigate the COMETS GUI. Downloading a layout and model files, editing the run parameters, running the simulations etc. are all based on a standard windows GUI and do not require any specialized computational skills. The GUI however currently supports the very basic COMETS capabilities, with limited access to only a subset of the modeling parameters and it can only save the layout images. Any visualization of the results, such as the plot of biomass vs. time, requires knowledge of some of the standard plotting and data manipulation software such as MATLAB or R. This level of expertise would be typical for a high-school or undergraduate biology student.

The GUI has limited capabilities and should be understood only as an educational tool, introduction to COMETS to novice users. The full simulation capabilities of COMETS are accessed through the MATLAB and Python toolboxes, and/or direct manipulation of the input files.

## COMETS Python Toolbox (cometspy)

The COMETS Python toolbox is a full environment for COMETS simulation and analysis. It integrates with existing Python scientific computing packages (scipy, matplotlib, etc) and offers a full compatibility with established methods in the field such as COBRAPy[64]. We recommend the usage of the COMETS Python toolbox in the context of Jupyter (https://jupyter.org/), and we provide Jupyter notebooks for all the protocols presented here in the supplement, although of course other platforms can also be used (e.g. IPython).

## COMETS MATLAB Toolbox

The COMETS MATLAB Toolbox is a collection of classes and functions intended to facilitate the processes involved in creating layouts for simulations, and includes utilities to execute COMETS within scripts from the command line and to parse output files. The MATLAB Toolbox uses metabolic models in the format of the COBRA Toolbox for MATLAB[64]. The user of this toolbox needs to be familiar with the basics of using MATLAB. A knowledge of the COBRA Toolbox is desired but not necessary.

## Command line user

This level of expertise will be required from users that plan to perform a large "production" level of sets of simulations most likely on a computational cluster. This level of expertise would require knowledge of Linux operating system, remote connection software to a computational cluster, such as ssh, Putty, MobaXterm is required. Skills in submitting and deleting jobs on a queue scheduler such as qsub are required at this level of expertise.

### Open source development

COMETS (https://www.runcomets.org) is an open source code and it is available at https://github.com/segrelab/comets. The code is distributed under the GNU General Public Licence Version 3. The Matlab toolbox is available at https://github.com/segrelab/comets-toolbox, distributed under the GNU General Public License Version 3. The COMETS Python toolbox is available at https://github.com/segrelab/cometspy, distributed under the GNU General Public License Version 3. Interested developers can contribute to the growing COMETS community and collaboration is facilitated by a public forum at https://gitter.im/segrelab/comets. Contributors, as well as users, should follow the license requirements, the contributing guidelines and the code of conduct found in the GitHub repository. For further information about the code, both users and contributors can contact the development team at comets@bu.edu.

### Online Documentation, Tutorials and Data Availability

The documentation for COMETS and for its Python and Matlab toolboxes is available at https://comets-manual.readthedocs.io. The documentation is structured as a tutorial and contains all the examples shown in this protocol. Raw documentation files are stored at https://github.com/segrelab/comets-manual. The COMETS Protocols GitHub repository (https://github.com/segrelab/COMETS_Protocols) also contains all input files and Jupyter notebooks for all these examples/case studies.

# Procedure: running COMETS simulations, with illustration of specific case studies

The following subsections describe detailed instructions on how to operate the different toolboxes and interfaces, and different case studies of COMETS simulations, from very basic and simple ones to more advanced. The individual case studies do not need to be implemented sequentially. Moreover, users may choose to follow examples from specific case studies that fit their expertise and needs. To facilitate navigation of these case studies, we use a letter labelling code that describes what user interface the application is based on ([M]= Matlab, [P]=Python, [GUI]=Graphical User Interface, [CL]=command line).

*The basics of COMETS using the MATLAB toolbox* [M]

The COMETS MATLAB Toolbox is a collection of classes and functions intended to facilitate the processes involved in creating layouts for simulations, and includes utilities to execute COMETS within scripts from the command line and to parse output files. Similarly to the Python Toolbox's use of COBRAPy, the MATLAB Toolbox uses metabolic models in the format of the COBRA Toolbox for MATLAB[64].

A brief overview of the most important components of the COMETS MATLAB Toolbox follows. For more in-depth documentation, see http://segrelab.github.io/comets-toolbox/ and https://comets-manual.readthedocs.io.

Primary Tasks

Manipulating metabolic models: The MATLAB Toolbox uses stoichiometric metabolic models in the format of the popular COBRA Toolbox for MATLAB[64], allowing users familiar with COBRA to quickly get up to speed, and allowing us to begin from model source files in an already wide-spread format. Because COBRA is not intended to support dynamic Flux Balance Analysis, we have added fields to the model structure that capture temporal behaviors: the setBiomassRxn() function can be used to specify a reaction that determines the growth rate, and the setKm() and setVmax() functions allow individual uptake reactions to be given a rate that is dependent on metabolite concentrations in a Michaelis-Menten-like manner.

Creating layouts: A "layout" is the structure which represents the simulated world in a COMETS analysis, containing sets of metabolic models with individual biomasses, as well as metabolites, across the space of a 2-dimensional grid. A layout object contains a CometsParams object which can be saved and loaded in order to conveniently ensure that all simulations are performed with a user's preferred set of simulation parameters and default values. The layout also contains instructions for the addition or removal of media over the course of the simulation, the diffusion properties for each organism and metabolite, the kinetic parameters of extracellular reactions (such as decay or enzymatic degradation), and the locations of "barrier" spaces which block diffusion. Utility functions exist to automate the placement of some layout elements, for example placing bacterial colonies at equidistant points or applying barrier spaces at the edges of the grid in order to create a circular plate.

Creating COMETS inputs and executing simulations: The formats of the text files required by COMETS for models and layouts are not amenable to editing by hand, as doing so requires the user to track the indexes of multiple elements in several lists and refer to the documentation for the details of each field. The COMETS MATLAB Toolbox provides scripts to generate these files so that users no longer have to concern themselves with the contents of these files, and can save input files sets using the createCometsFiles() command or directly execute a simulation using a layout object with the runComets() command.

Handling COMETS outputs: The various log files generated by COMETS for biomass, metabolite concentrations, and fluxes can be loaded into MATLAB tables for easier filtering and analysis through the functions parseBiomassLog(), parseMediaLog(), and parseFluxLog(). Utility scripts are provided which make it simple to generate plots of these measurements over time with the functions plotBiomassTimecourse() and plotMediaTimecourse().

Classes and Data Structures

CometsLayout: The main class which encapsulates all information involved in a single COMETS simulation by containing media contents, the list of COBRA models, spatial information, and a single CometsParams object. Contents of the layout should be manipulated by using the methods of the CometsLayout class instead of being directly modified when possible, for example by editing initial media through setMedia() or adding metabolic models through addModel().

CometsParams: A class to contain the parameters for global simulation parameters as well as model-level default values.

File I/O

createCometsFiles(cometsLayout, [directory, layoutFileName, separatePamsFiles]): Creates the COMETS script, layout, and model files. If separateParmasFiles is true, it creates separate files to contain the global and package parameters. Otherwise, all parameters are included in the body of the layout file.

parseBiomassLog(fileName): Processes a MATLAB-format biomass log from a COMETS simulation. Returns a table with the following columns:
- t: Timestep.
- X: X coordinate.
- Y: Y coordinate.
- Z: Z coordinate. Excluded in 2D simulations.
- model: ID number of the model. Arranged as in the layout file, starting with 0.
- biomass: biomass value in grams.

parseFluxLog(fileName): Processes a MATLAB-format reaction flux log from a COMETS simulation. Returns a table with the following columns:
- t: Timestep.
- X: X coordinate.
- Y: Y coordinate.
- Z: Z coordinate. Excluded in 2D simulations.
- model: ID number of the model. Arranged as in the layout file, starting with 0.
- rxn: ID number of the reaction. Arranged as in the metabolic model, starting with 0.
- flux: Flux through the reaction.

parseMediaLog(fileName, [metNames]): Processes a MATLAB-format media log from a COMETS simulation. If a cell array is provided as metNames, only records for the corresponding metabolites will be loaded. Returns a table with the following columns:
- t: Timestep.
- X: X coordinate.
- Y: Y coordinate.
- Z: Z coordinate. Excluded in 2D simulations.
- met: ID number of the metabolite. Arranged as in the layout file, starting with 0.
- amt: Concentration of the metabolite in millimoles.
- metname: Name of the metabolite.

Standard Workflow

createLayout({cobraModels}): Initialize a COMETS layout with default properties as stored in the CometsParams class, and add any metabolic models provided as arguments as though invoking CometsLayout.addModel().

addModel(*cometsLayout*, *cobraModel*): Attaches the given model to the layout, and adds any of the model's exchange metabolites to the layout's list of media components.

setMedia(*cometsLayout*,*{metaboliteNames}*,*concentrations*): Set the initial media concentrations for every space in the simulation. To alter the initial media in an individual grid cell, use the method setInitialMediaInCell() instead.

setInitialPop*(cometsLayout, [format, amount, resize])*: Sets the initial population for each metabolic model in the given *cometsLayout*, arranged according to the *format* parameter, to the value or values provided in the amount parameter in grams. If *resize* is not false, the dimensions of the *cometsLayout* will be adjusted as well. *Format* may be one of two options:
- 'Colonies' (default): Up to four evenly spaced colonies will be created, one for each model in the cometsLayout. Default dimensions 100 by 100 grid cells.
- '1x1': Biomass for all models will be placed in the center grid cell. Default dimensions 1 by 1 grid cell.

RunComets(cometsLayout, [directory]): Creates layout and model files in the given directory (defaulting to the current working directory) and executes a simulation by invoking COMETS through the command line.

*Case Study #1: Growth of bacteria in well mixed conditions using the MATLAB toolbox and COMETS GUI* [M][GUI] ● TIMING setup: 30 min - 1hr, simulation: 1-5 min

One of the first successful simulations of the time dynamics of bacterial metabolism was the classical study of *E. coli* batch culture by Varma and Palsson[43,77]. Here we reproduce one of the results in the study, the anaerobic fermentation in minimal media with glucose as the only carbon source using the core model of *E. coli*[78]. This model consists only of a small but key subset of the reactions present in the metabolic network of *E. coli*. However, it is sufficiently complex to reproduce some of the fundamental metabolic activities in a bacterial cell, such as glycolysis, TCA cycle, pentose phosphate shunt.

The spatial layout in this elementary case consists of a single grid point of 1 cubic cm of volume, thus modeling well mixed, i.e. non spatially structured conditions. We seeded the batch culture with $5 \cdot 10^{-6}$ grams of *E. coli* biomass. The initial composition of the substrate was 11 mM of glucose and infinite amounts of ammonia and phosphate. The nutrient uptake was modeled with the standard Michaelis-Menten kinetics, using the typical Monod parameter for anaerobic uptake of glucose by *E. coli*.

Figure 2a shows the growth of the *E. coli* biomass, while Fig. 2b shows the concentrations of glucose and the three products of glucose fermentation. The small value of the $K_M$ parameter results in the characteristic exponential curves with abrupt stop at the point of depletion of glucose.

The prerequisite for this protocol is the installation of COMETS, MATLAB and Cobra, as outlined above. The procedure consists of three main groups of steps:
1. Creation of COMETS format input files using the MATLAB toolbox.
2. Running COMETS
3. Visualization and analysis of the output.

The first of the three is of a general nature, and can be used as an initial procedure for a large number of COMETS simulation setups. We will follow the procedure with the command lines executed in LInux and MacOS. Most of the command lines however are simple change of directory and copying of files, so they can easily be adopted to Windows.

Download the protocol files

1. First we will download the protocol files from https://github.com/segrelab/COMETS_Protocols/. The files can be downloaded either as a single zip file, or using the GitHub Desktop application. If downloaded as a zip file it needs to be extracted in Windows by right clicking on the file, clicking on "Extract all". This will create a directory COMETS_Protocols-master. In Linux and Mac OS it can be unzipped by the command line:

   unzip COMETS_Protocols-master.zip

2. The next step is to navigate to the working directory.

   cd COMETS_Protocols-master/COMETS_Protocols-master/COMETS_Protocols/
   COMETS_example_Ecoli_CoreModel_WellMixed/Matlab/Create_layout_MATLAB_toolbox/

Create the COMETS format input files

3. The E. coli core model used in this simulation can be downloaded from the BIGG Models database: http://bigg.ucsd.edu/models/e_coli_core. In this example we will download the uncompressed e_coli_core.xml SBML file, and move it to the working directory.
▲ **CRITICAL STEP** The availability of a stoichiometric model is a prerequisite for any COMETS simulation.

4. To create COMETS format input files, we follow the MATLAB script: COMETS_example_Ecoli_CoreModel_WellMixed/Create_layout_MATLAB_toolbox/create_comets_model_and_layout.m. The file can be opened and run in MATLAB, or each step can be entered at the MATLAB command line, as follows.

5. Assuming the COBRA toolbox has been initialized with:

```
>>initCobraToolbox()
```

we first use the Cobra function readCbModel to read the SBML model file:

```
>> ecoli = readCbModel('e_coli_core.xml');
```

This creates a MATLAB structure ecoli that contains the *E. coli* core model:

```
S: [72×95 double]
mets: {72×1 cell}
b: [72×1 double]
csense: [72×1 char]
rxns: {95×1 cell}
lb: [95×1 double]
ub: [95×1 double]
c: [95×1 double]
osenseStr: 'max'
genes: {137×1 cell}
rules: {95×1 cell}
geneNames: {137×1 cell}
compNames: {2×1 cell}
comps: {2×1 cell}
proteins: {137×1 cell}
metFormulas: {72×1 cell}
metNames: {72×1 cell}
metHMDBID: {72×1 cell }
metKEGGID: {72×1 cell }
metChEBIID: {72×1 cell}
metMetaNetXID: {72×1 cell}
rxnNames: {95×1 cell}
```

rxnECNumbers: {95×1 cell}
rxnKEGGID: {95×1 cell}
rxnMetaNetXID: {95×1 cell}
rxnSBOTerms: {95×1 cell}
subSystems: {95×1 cell}
description: 'e_coli_core'
modelVersion: [1×1 struct]
modelName: 'Escherichia coli str. K-12 substr. MG1655'
modelID: 'e_coli_core'
...

Optionally, we can change the name/description of the model file. This will be the name of the COMETS format input model file.

\>\> ecoli.description='e_coli_core';
? TROUBLESHOOTING

6. Next, we will use the COMETS MATLAB toolbox commands to create an empty COMETS layout and add the model to it:

\>\> world = CometsLayout();
\>\> world = world.addModel(ecoli);

This creates a default COMETS simulation spatial layout structure with one model in it:

models: {[1×1 struct]}
xdim: 1
ydim: 1
mets: {20×1 cell}
media_amt: [20×1 double]
params: [1×1 CometsParams]
diffusion_constants: [20×2 double]
global_media_refresh: [20×1 double]
media_refresh: [20×1 double]
global_static_media: [20×2 double]
static_media: [20×1 double]
initial_media: 0
barrier: 0
initial_pop: 1.0000e-06
external_rxns: [0×0 table]
external_rxn_mets: [0×0 table]

Here the layout is populated with the metabolites world.mets that are exchanged by the model.

```
>> world.mets

{'ac[e]'   }
{'acald[e]' }
{'akg[e]'  }
{'co2[e]'  }
{'etoh[e]' }
{'for[e]'  }
{'fru[e]'  }
{'fum[e]'  }
{'glc__D[e]'}
{'gln__L[e]'}
{'glu__L[e]'}
{'h2o[e]'  }
{'h[e]'    }
{'lac__D[e]'}
{'mal__L[e]'}
{'nh4[e]'  }
{'o2[e]'   }
{'pi[e]'   }
{'pyr[e]'  }
{'succ[e]' }
```

7. Next we set the dimensions of the simulation grid to 1x1, i.e. a trivial grid consisting of a single point. This corresponds to a spatially homogeneous, i.e. well-mixed conditions.

```
>> x = 1;
>> y = 1;
>> world = world.setDims(x,y);
```

8. In this step we will set the amounts of the metabolites in the media to 0.011 mmol for glucose, and 1000 mmol for ammonia, phosphate, water and hydrogen. The amount for oxygen is set to zero to simulate anaerobic conditions.

```
>> world = world.setMedia('glc__D[e]',0.011);
>> world = world.setMedia('nh4[e]',1000);
>> world = world.setMedia('pi[e]',1000);
>> world = world.setMedia('h2o[e]',1000);
>> world = world.setMedia('h[e]',1000);
>> world = world.setMedia('o2[e]',0);
```
▲ **CRITICAL STEP** Setting the correct amount of nutrients for the organism will have a strong impact on the final result.

9. Add the initial population to $5 \cdot 10^{-6}$ grams.

```
>> world = setInitialPop(world, '1x1', 5e-6);
```
▲ **CRITICAL STEP** The initial population must be set.

10. And finally write the layout and model files.

>> writeCometsLayout(world,'./','Ecoli_batch_layout.txt',0);

The outcome of the procedure are the layout file Ecoli_batch_layout.txt and the model file e_coli_core_txt.

The COMETS format layout file Ecoli_batch_layout.txt consists of the following fields relevant for this simulation:

```
model_file e_coli_core.txt
  model_world
      grid_size 1 1
        world_media
          ac[e] 0
          acald[e] 0
          akg[e] 0
          co2[e] 0
          etoh[e] 0
          for[e] 0
          fru[e] 0
          fum[e] 0
          glc__D[e] 0.011
          gln__L[e] 0
          glu__L[e] 0
          h2o[e] 1000
          h[e] 1000
          lac__D[e] 0
          mal__L[e] 0
          nh4[e] 1000
          o2[e] 1000
          pi[e] 1000
          pyr[e] 0
          succ[e] 0
     ...
       //
     //
     initial_pop
     0 0 5.000000e-06
  //
```

The COMETS format model file e_coli_core.txt consists of the following fields:

```
SMATRIX  72  95
       1   49   1.000000
       ...
       //
BOUNDS  -1000  1000
       1   -1000.000000   1000.000000
       ...
       //
OBJECTIVE
       13
//
METABOLITE_NAMES
       13dpg[c]
       ...
       //
REACTION_NAMES
       ACALD
       ...
       //
EXCHANGE_REACTIONS
 20 21 22 23 24 25 26 27 28 29 30 31 32 33 34 35 36 37 38 39
//
OBJECTIVE_STYLE
   MAX_OBJECTIVE_MIN_TOTAL
//
```

11. In this step we will copy the layout and model files to the working directory where we will run COMETS. This is done with the command line:

```
cp e_coli_core.txt ../COMETS_simulation_Ecoli_core
cp Ecoli_batch_layout.txt ../COMETS_simulation_Ecoli_core
```

12. In addition to the required layout and model files, in this protocol we used the optional parameters files. These files are loaded to COMETS after the layout. The parameters are separated as global, defining the total number of simulation steps, output files, and package ones that define the dFBA specific simulation parameters such as the size of the spatial grid box and the time step.

    The global parameters file global_params.txt:

```
maxCycles = 1000
pauseonstep = false
writeFluxLog = false
FluxLogName = ./flux.m
FluxLogRate = 10
writeMediaLog = true
MediaLogName = ./media.m
MediaLogRate = 10
writeBiomassLog = false
BiomassLogName = ./biomass.txt
BiomassLogRate = 1
writeTotalBiomassLog = true
totalBiomassLogRate = 1
TotalbiomassLogName = ./total_biomass.txt
useLogNameTimeStamp = false
```

The FBA package parameters file package_params.txt contains:

```
numDiffPerStep = 10
numRunThreads = 1
exchangestyle = Monod Style
defaultVmax = 18.5
defaultKm = 0.000015
defaultHill = 1
timeStep = 0.01
deathRate = 0.0
spaceWidth = 1.0
maxSpaceBiomass = 10
minSpaceBiomass = 1e-11
showCycleTime = true
showCycleCount = true
```

Run the COMETS simulation

13. In this protocol we will run COMETS from the graphical user interface (GUI). We start the program by running COMETS from the installed shortcut icon in Windows, Mac OS or Linux. Once COMETS has started, the user will see the window of the COMETS application such as the one shown in Supplementary figure 1.
    ? TROUBLESHOOTING

14. We load the layout file by clicking on the 'Load layout file' button.
    ? TROUBLESHOOTING

15. When the layout is loaded, we can load both of the the parameters files: global_params.txt and package_params.txt, from the File tab:
    File->LoadParametersFile.

16. The simulation is started from the Simulation tab. To start the simulation, click on the Run/Pause Simulation button. To run the simulation continuously, uncheck the Pause after cycle field.

17. The COMETS run above produces the total_biomass.txt and media.m output files. We can visualize the time dynamics of the bacterial biomass production with the MATLAB script PlotTotalBiomass.m that produces Fig. 2**a**:

```
%Load the biomass output
load 'total_biomass.txt'

%The timestep (in hours) is defined in the package_params.txt file
timeStep=0.01;

%Spatial grid point volume in litres
volume=1e-3;

%Plot the biomass density (in g/l) as a function of time (in hours)
biomass_plot = plot(timeStep*total_biomass(:,1),total_biomass(:,2)/volume)

set(gca,'FontName','Helvetica');
set(gca,'FontSize',15);
set(biomass_plot,'LineWidth',2);
xlabel 'Time [h]'
ylabel 'Biomass density [g/l]'
print('Ecoli_core_batch_biomass','-dpng')
```

The expected results, shown in Fig. 2b were produced by the MATLAB script for visualization of the relevant metabolites, PlotMedia.m:

```matlab
%Load the media output
%It contains the metabolite names in the string array media_names
media

%Choose the metabolites to be plotted
%Here we choose glucose, acetate, formate and ethanol
metabolites_to_plot = [9 1 6 5];
metabolites_names = ["glucose" "acetate" "formate" "ethanol"];

%The time step (in hours) is defined in the package_params.txt file
timeStep=0.01;

%Spatial grid point volume in litres
volume=1e-3;

%Media write step
mediaLogRate = 10;

%Number of simulation steps
maxCycles = 1000;

%Number of media write steps
total_write_steps = maxCycles/mediaLogRate;

%Line styles for the plot
lines_vec = {'-', '--', ':', '-.'};

for j=1:length(metabolites_to_plot)
        for i=0:total_write_steps
        varname=genvarname(['media_',num2str(i*mediaLogRate)]);
        var=eval(varname);
        metabolite_variable=var(metabolites_to_plot(j));
        metabolite=full(metabolite_variable{1});

        metabolite_density(i+1,metabolites_to_plot(j))=metabolite/volume;
        time(i+1)=timeStep*mediaLogRate*i;
        end
```

```
metabolites_plot=plot(time,metabolite_density(:,metabolites_to_plot(j)),'LineStyle',lines_vec{j})
        set(metabolites_plot,'LineWidth',2);
        hold on
end

set(gca,'FontName','Helvetica');
set(gca,'FontSize',15);

xlabel 'Time (h)'
ylabel 'Concentration (mM)'
legend(metabolites_names,'Location','northwest');
print('Ecoli_core_batch_metabolites','-dpng')
```

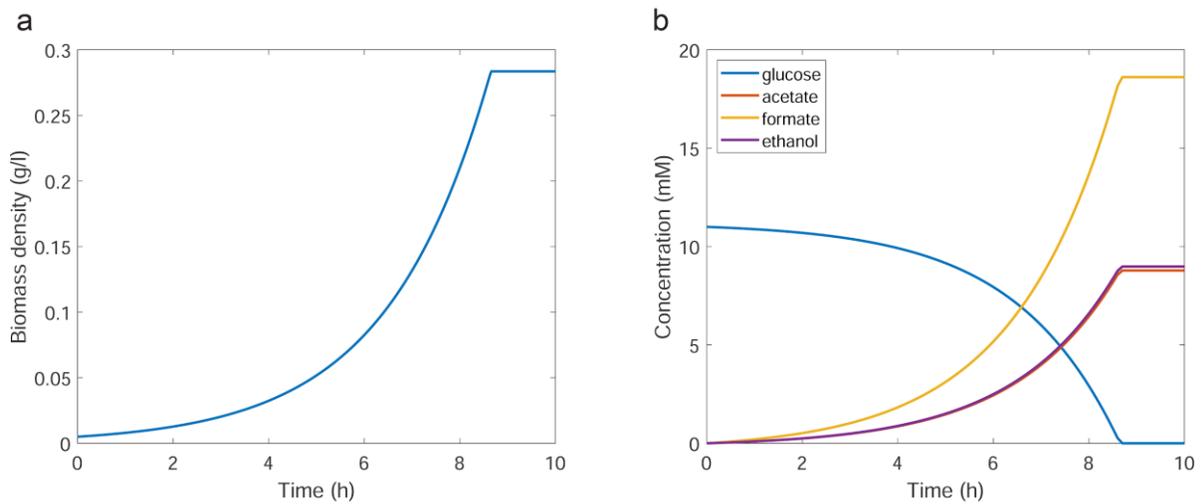

**Figure 2. Growth of *E. coli* (core model) batch culture in minimal medium, with glucose as the only carbon source. a**) Plot of biomass vs. time. **b**) Plot of the key metabolites vs. time. The biomass growth stops when the glucose is completely depleted. The production of the typical products of fermentation also coincides with the growth of the biomass.

*Case Study #2: Modeling bacterial colonies: growth and propagation on flat surfaces* [CL] ● TIMING setup: 30min-1hr, simulation: 1-3 hr

With this protocol we will illustrate the growth and propagation of a colony of non-motile bacteria. As explained in the methods section, a growing colony of non-motile bacteria propagates by mechanical pushing among the growing cells. One of the characteristics of the implemented model is that it undergoes a transition in the colony morphology when the dense packing parameter is varied. In this protocol we will illustrate this transition, by performing two simulations for the dense packing parameter value of $0.5 \cdot 10^4$ and $1.5 \cdot 10^4$. The two produced morphologies are shown on Fig. 3a and 3b and in Supplementary videos 1a and 1b. We will use a simple metabolic model with a single nutrient and single reaction of biomass growth from the uptake of the nutrient. The units of the nutrient and biomass in this case are arbitrary, since this is a theoretical toy model, rather than a realistic stoichiometric one.

A technical aspect illustrated in this protocol is the utilization of a computational cluster to perform the COMETS simulations. We will run the simulations in the computational environment of a standard queueing system on a Linux cluster.

The input files for this protocol are: the model file, the layout file and the two parameters files, the package and global parameters files. They are found in the two directories: COMETS_example_CircularAndBranchingColonies/Circular_colony and COMETS_example_CircularAndBranchingColonies/Branching_colony.

Create the COMETS format input files

1. The model file is tiny_model.txt. This file, in addition to the required metabolic model fields, contains the fields of the propagation model, as explained in the methods section. In this protocol we will perform two runs, as two alternative options with values of the packedDensity parameter set to (Optional) $0.5 \cdot 10^4$ and $1.5 \cdot 10^4$:
    (i) packedDensity 0.5e4
    (ii) packedDensity 1.5e4

    Also, we will add growth noise by setting the parameter noiseVariance to 20.

```
OPTIMIZER GUROBI
SMATRIX 1 2
    1    1       -1.0000
    1    2       -1.0000
//
BOUNDS -1000 1000
        2   0   1000
//
OBJECTIVE
   2
//
METABOLITE_NAMES
   carbon
//
REACTION_NAMES
   Carbon_exch
   Biomass
//
EXCHANGE_REACTIONS
   1
//
packedDensity 0.5e4
//
elasticModulus 1.0e-4
//
frictionConstant 1.0
//
convDiffConstant 0.0
//
noiseVariance 20.0
//
```

▲ **CRITICAL STEP** Creating a correct model is critical for the success of the simulation.

2. The layout file consists of a 100x100 grid, uniformly populated with the nutrient (carbon) and inoculated with initial biomass at its center. The units in this case are arbitrary.

```
model_file    tiny_model.mod
model_world
    grid_size 100 100
    world_media
            carbon    1.0
    //
    static_media 0 0
    //

    barrier
    //
//
    initial_pop
            50 50 1.0
    //
//
```

▲ **CRITICAL STEP** The initial biomass must be set.

3. The global_params.txt file contains the rate and name for saving the snapshot images of the colony during the simulation. In this example the images are saved in png format in a subdirectory (of the working directory) named images.

```
maxCycles = 2700
pauseonstep = false
pixelScale = 5
saveslideshow = true
slideshowExt = png
slideshowColorRelative = true
slideshowRate = 100
slideshowLayer = 1
slideshowName = ./images/image
writeFluxLog = false
FluxLogName = ./flux.txt
FluxLogRate = 1
writeMediaLog = false
MediaLogName = ./media.txt
MediaLogRate = 1
writeBiomassLog = false
BiomassLogName = ./biomass.txt
BiomassLogRate = 1
writeTotalBiomassLog = true
totalBiomassLogRate = 1
TotalbiomassLogName = ./total_biomass.txt
useLogNameTimeStamp = false
```

▲ **CRITICAL STEP** The user must make sure that the directory ./images exists in the working directory.

The package parameters file contains the simulation parameters, most notably the propagation mode of the colony set to biomassmotionstyle = Convection 2D.

```
biomassmotionstyle = Convection 2D
numDiffPerStep = 10
numRunThreads = 1
exchangestyle = Monod Style
defaultKm = 0.01
defaultHill = 1
defaultVmax = 100
timeStep = 0.00005
deathRate = 0.0
spaceWidth = 0.01
maxSpaceBiomass = 10
minSpaceBiomass = 0.25e-10
allowCellOverlap = true
toroidalWorld = false
showCycleTime = true
showCycleCount = true
```

▲ **CRITICAL STEP** Creating a correct value for biomassmotionstyle = Convection 2D is critical for the success of the simulation.

Running the COMETS simulation

4. We will run COMETS with the above input utilizing the command line (non-GUI or script) option. This script can be run on a linux command line either by running:

comets_scr comets_script

in Linux, MacOS and Windows, where the comets_script file contains the names of the input files:

```
load_comets_parameters     global_params.txt
load_package_parameters    package_params.txt
load_layout                layout.txt
```

Optionally, to submit it and run it to a cluster queuing system, one should use the queueing command provided on the computational cluster. For example:

qsub -pe omp 1 -l h_rt=24:00:00 qscript

Where qscript is the bash script to be submitted:
```
#!/bin/bash -l
module load gurobi/9.0.0
comets_scr comets_script
```

Here the second line loads the gurobi optimizer module. The specific name and version of the optimized should be set to the one installed on the user's system.
? TROUBLESHOOTING

Create a simulation video

5. The output of the run consists of a collection of figures representing the 2D morphology of the bacterial colony. In this example the figures are saved in the subdirectory images. Figures 3a and 3b show the final simulation outcome for the two cases. Optionally, the user can create a video of the growth of the colonies from the individual images using the command:

convert image_*.png movie.gif

The produced videos can be seen in the Supplementary videos 1a and 1b.
? TROUBLESHOOTING

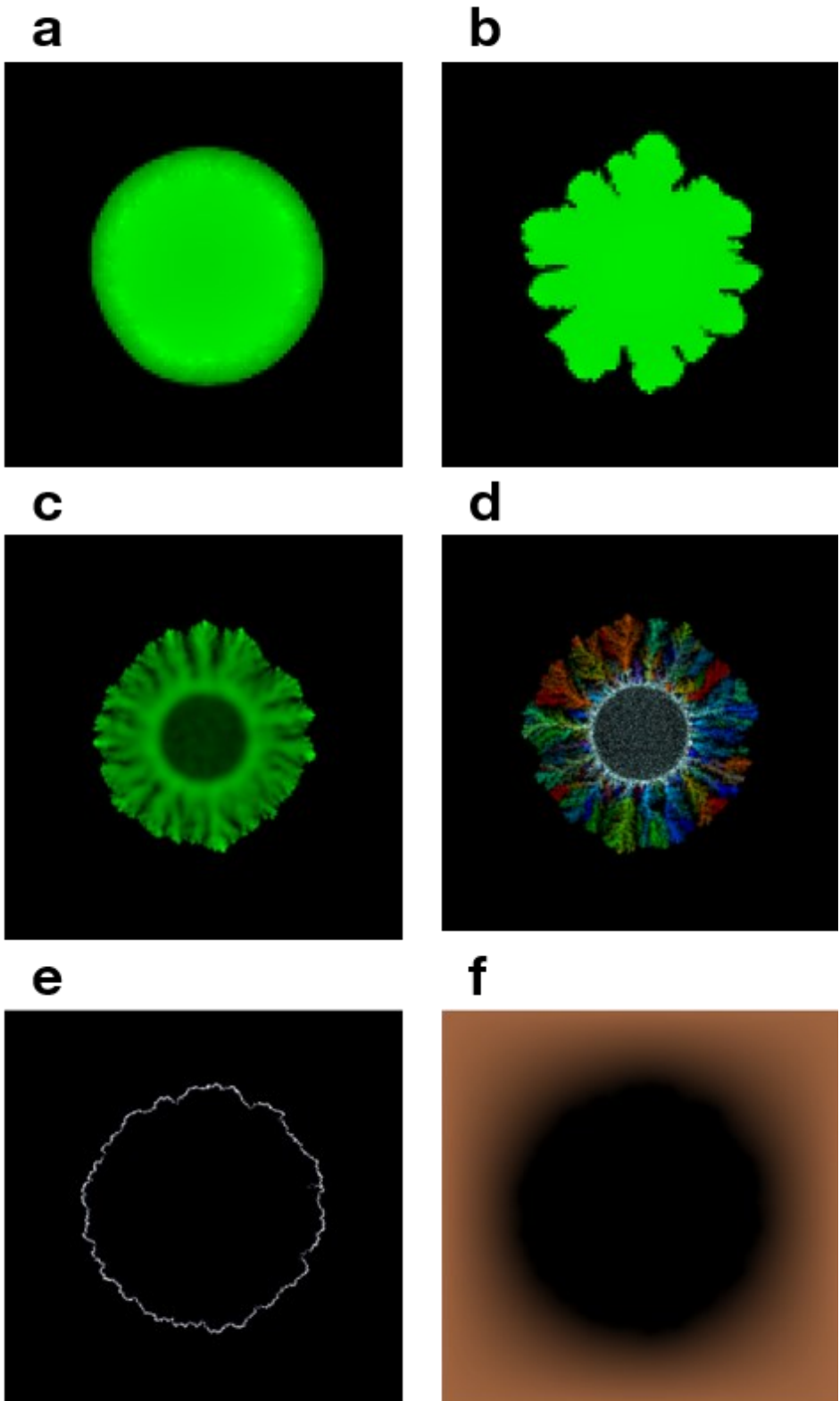

**Figure 3. A variety of morphologies simulated by COMETS.** Types of bacterial colony morphologies, simulated using the Convection 2D biomass propagation model: **a)** Circular colony with stable front

propagation simulated with the Convection 2D biomass propagation model. The value of the packedDensity parameter was set to a value below the critical for emergence of unstable growth front. **b**) Dendritic colony with unstable front propagation. In this case the value of packedDensity was greater than the critical. Panels **c**) and **d**) show two morphologies obtained by running the ConvNonLinDiff 2D biomass propagation model with demographic noise: **c**) Single strain colony, **d**) Five-strains colony. In the five strain colony case we see the segregation of the strains due to the presence of the demographic noise.

*Case Study #3: Virtual petri dish* [CL] ● TIMING setup: 30 min - 1hr, simulation: 4-5 hr

This protocol illustrates the visualization capabilities of COMETS as well as the capabilities of defining a heterogeneous spatial substrate/environment. The spatial layout is defined as a simulated Petri dish. The dish is split in two halves with different physical properties. The nutrient diffusivity in one half of the dish is ten times smaller than the other half. This simulation layout corresponds to an experimental layout with agar or natural substrate with two different densities, or other physical properties that make the diffusivity of the nutrient in the two parts to be different. The simulation of this layout results with colonies of different size in the two spatial parts.

In this protocol we will also provide detailed description of the spatial output of biomass density, nutrient concentration as well as biomass growth rate, visualized in MATLAB. The simulation produces the three output files that we will use to visualize the layout: biomass.m, media.m, fluxes.m. We will visualize the spatial distribution of the biomass, the growth rate of the biomass, as well as two key metabolites: glucose and acetate. The anticipated result is shown in Fig. 4. The top layer shows the biomass distribution after simulated 40 hours. The layout is split in the two regions with different friction constants, and consequently difference in the size of the colonies. Below, the growth rate visualization shows that the colonies are growing predominantly at the leading edges. Two two lower slices show the depletion of glucose and buildup of acetate in the spatial layout.

The input files for this protocol are: the model file, the layout file and the two parameters files, the package and global parameters files.

Create the COMETS input files

1. The basic model file in this simulation is the same as in protocol Case Study #1: e_coli_core.txt. This protocol however is a spatial simulation, therefore, in addition to the metabolic model, the file e_coli_core.txt contains the parameters of the propagation model as well:

packedDensity 0.022
//
elasticModulus 1.0e-10
//
convDiffConstant 0.0
//
noiseVariance 100.0
//

▲ **CRITICAL STEP** The biomass propagation parameters must be set correctly to obtain the desired outcome.

2. The layout file split_layout.txt consists of a 100x100 grid uniformly populated with the nutrients and inoculated with initial biomass at its center.

```
model_file    e_coli_core.txt
model_world
    grid_size 200 200
    world_media
        ac[e] 0
        acald[e] 0
        akg[e] 0
        co2[e] 0
        etoh[e] 0
        for[e] 0
        fru[e] 0
        fum[e] 0
        glc__D[e] 5.0e-5
        gln__L[e] 0
        glu__L[e] 0
        h2o[e] 1000
        h[e] 1000
        lac__D[e] 0
        mal__L[e] 0
        nh4[e] 1000
        o2[e] 0.0
        pi[e] 1000
        pyr[e] 0
        succ[e] 0
        //
    substrate_diffusivity
        5e-6   5e-6   5e-6   5e-6   5e-6   5e-6   5e-6   5e-6   5e-6   5e-6   5e-6   5e-6
        5e-6   5e-6   5e-6   5e-6   5e-6   5e-6   5e-6   5e-6
         5e-7   5e-7   5e-7   5e-7   5e-7   5e-7   5e-7   5e-7   5e-7   5e-7   5e-7   5e-7
        5e-7   5e-7   5e-7   5e-7   5e-7   5e-7   5e-7   5e-7
        //
        substrate_friction
        1.0
        1.0
        //
    substrate_layout
1   1   1   1   1           …
….
        2   2   2   2   2   2   2   2   2   2   2   2
//
//
    initial_pop
        140 50 1.0e-6
        58 162 1.0e-6
        28 55 1.0e-6
        132 78 1.0e-6
        85 152 1.0e-6
        73 123 1.0e-6
        34 76 1.0e-6
```

```
                150 45 1.0e-6
                120 73 1.0e-6
                34 138 1.0e-6
                44 180 1.0e-6
                180 150 1.0e-6
                165 135 1.0e-6
                24 38   1.0e-6
                44 45   1.0e-6
                99 108 1.0e-6
                95 23 1.0e-6
                35 101 1.0e-6
                25 102 1.0e-6
    //
//
```

3. The global_params.txt file contains the rate and name of saving the snapshot images of the colony during the simulation.

```
maxCycles = 20000
pixelScale = 5
saveslideshow = true
slideshowExt = png
slideshowColorRelative = true
slideshowColorValue = .05
slideshowRate = 1000
slideshowLayer = 20
slideshowName = images/image
writeFluxLog = true
FluxLogName = flux.txt
FluxLogRate = 1000
writeMediaLog = true
MediaLogName = media.txt
MediaLogRate = 1000
writeBiomassLog = true
BiomassLogName = biomass.txt
BiomassLogRate = 1000
writeTotalBiomassLog = true
totalBiomassLogRate = 1
TotalbiomassLogName = totalbiomass.txt
useLogNameTimeStamp = false
```

4. The package parameters file contains the simulation parameters, most notably the propagation mode of the colony set to biomassmotionstyle = Convection 2D

```
numDiffPerStep = 10
numRunThreads = 10
growthDiffRate = 0
flowDiffRate = 3e-10
exchangestyle = Monod Style
defaultKm = 0.01
defaultHill = 1
timeStep = 0.005
deathRate = 0.0
spaceWidth = 0.02
maxSpaceBiomass = 10.0
minSpaceBiomass = 1e-10
allowCellOverlap = true
toroidalWorld = false
showCycleTime = true
showCycleCount = true
defaultVmax = 10
biomassmotionstyle = Convection 2D
```

▲ **CRITICAL STEP** The biomass propagation style must be set correctly to obtain the desired outcome.

Run the COMETS simulation

5. To submit it and run it to a cluster queuing system, as option 5B, one should use the qsub command provided on the computational cluster. For example:

```
qsub -pe omp 10 -l h_rt=24:00:00 qscript
```

Where qscript is the bash script to be submitted:

```
#!/bin/bash -l
module load gurobi/9.0.0
comets_scr comets_script
```

Here the second line loads the gurobi optimizer module. The specific name and version of the optimized should be set to the one installed on the user's system.
? TROUBLESHOOTING

6. The visualization is done using the MATLAB script FourSlicesImage.m found in the subdirectory FourLayersImage.

```
clear all
azymuth=20;
elevation=80;
width=1.0;
hight=0.25;
xpos=0.25;
ypos=0;

fig=figure;
set(gcf, 'Position', [100, 100, 300, 600])

subplot('Position',[ypos xpos*3 width hight])

biomass_atStep8000;

var=full(biomass_8000_0);
s=pcolor(1:200,1:200,var)
hold on
divide(1:200)=100;
plot(1:200,divide,'w--');
s.EdgeColor = 'none';
ax=gca;
zlim([0 3]);
ax.Box = 'off';
ax.XGrid = 'off';
ax.YGrid = 'off';
ax.ZGrid = 'off';
ax.XTick = [];
ax.YTick = [];
ax.ZTick = [];
ax.ZColor = 'none';
ax.YColor = 'none';
ax.XColor = 'none';
colormap(ax,copper);
view(azymuth,elevation);
clear var

subplot('Position',[ypos xpos*2 width hight])

for i=1:200
   for j=1:200
        fluxes{8000}{i}{j}{1}=[];
        fluxes{8000}{i}{j}{1}(1:95)=0.0;
   end
end
zz(1:200,1:200)=0;
flux_atStep8000;
```

```matlab
for i=1:200
   for j=1:200
        zz(j,i)=fluxes{8000}{i}{j}{1}(13);
   end
end

s=pcolor(1:200,1:200,zz);
hold on
divide(1:200)=100;
plot(1:200,divide,'w--');
s.EdgeColor = 'none';
ax=gca;
zlim([0 20.0]);
ax.Box = 'off';
ax.XGrid = 'off';
ax.YGrid = 'off';
ax.ZGrid = 'off';
ax.XTick = [];
ax.YTick = [];
ax.ZTick = [];
ax.ZColor = 'none';
ax.YColor = 'none';
ax.XColor = 'none';
colormap(ax,gray);
view(azymuth,elevation);

clear fluxes;

subplot('Position',[ypos xpos*1 width hight])
zz(1:200,1:200)=0;
media_atStep8000;

for i=1:200
       for j=1:200
         zz(j,i)=media_8000{1}(i,j);
        end
end

s=pcolor(1:200,1:200,zz)
hold on
divide(1:200)=100;
plot(1:200,divide,'w--');
s.EdgeColor = 'none';
ax=gca;
zlim([0 1000]);
ax.Box = 'off';
ax.XGrid = 'off';
ax.YGrid = 'off';
ax.ZGrid = 'off';
```

```matlab
ax.XTick = [];
ax.YTick = [];
ax.ZTick = [];
ax.ZColor = 'none';
ax.YColor = 'none';
ax.XColor = 'none';
colormap(ax,jet);
view(azymuth,elevation);

subplot('Position',[ypos xpos*0 width hight])

zz(1:200,1:200)=0;

for i=1:200
        for j=1:200
        zz(j,i)=media_8000{9}(i,j);
         end
end

s=pcolor(1:200,1:200,zz)
hold on
divide(1:200)=100;
plot(1:200,divide,'w--');
s.EdgeColor = 'none';
ax=gca;
zlim([0 1000]);
ax.Box = 'off';
ax.XGrid = 'off';
ax.YGrid = 'off';
ax.ZGrid = 'off';
ax.XTick = [];
ax.YTick = [];
ax.ZTick = [];
ax.ZColor = 'none';
ax.YColor = 'none';
ax.XColor = 'none';
colormap(ax,jet);
view(azymuth,elevation);
```

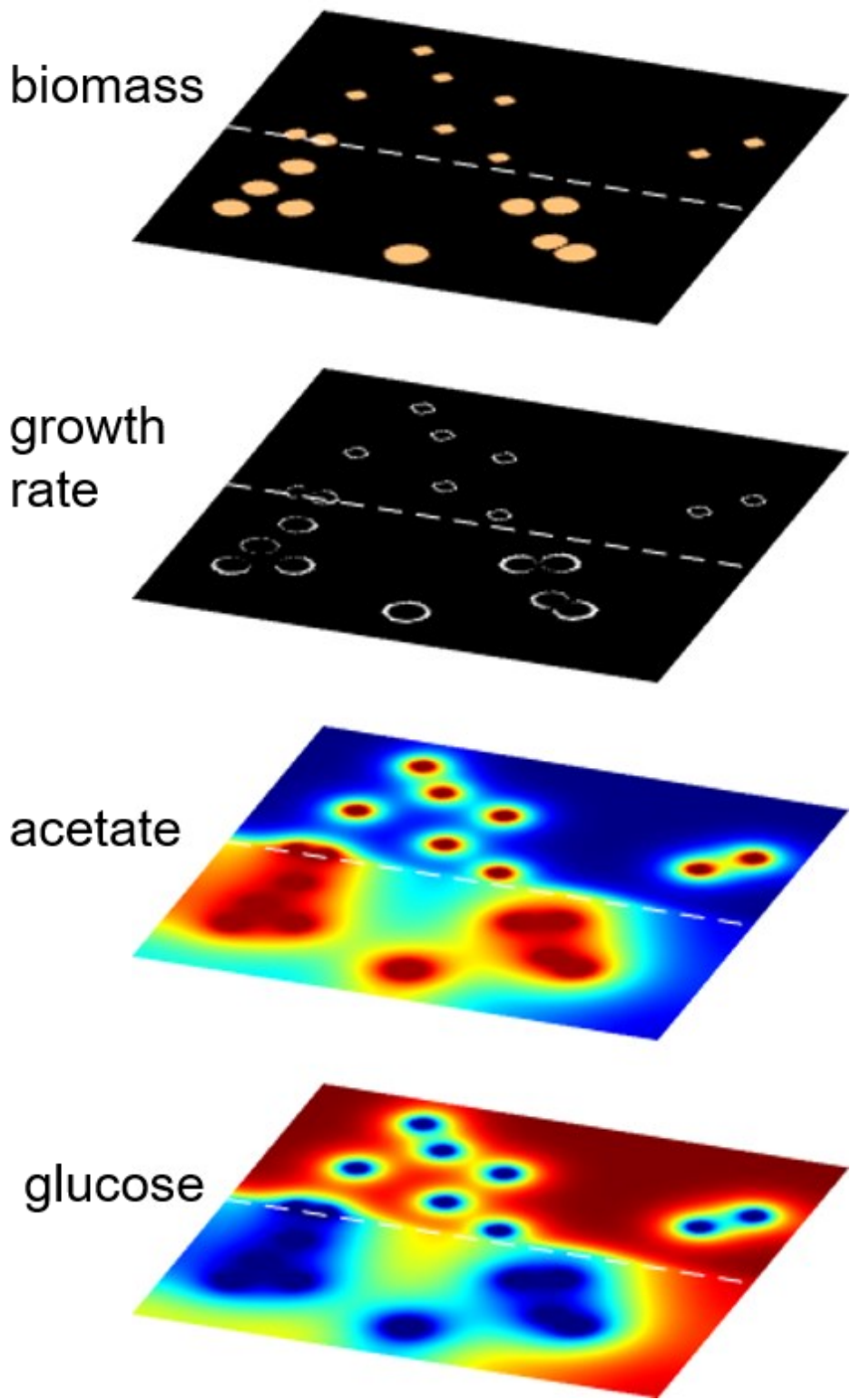

**Figure 4. Virtual Petri dish.** The four layers show the spatial distribution of the biomass, biomass growth rate, and concentrations of acetate and glucose, respectively from top to bottom. The blue color in the bottom two layers represents a depleted, and the red color represents an enriched metabolite region.

*Case Study #4: Demographic noise and cooperative biomass propagation: single strain* [CL] ● TIMING setup: 30min-1hr, simulation: 3-4 days

This protocol illustrates two biological features modeled in COMETS: demographic noise and cooperative biomass propagation. We illustrate the role of demographic noise on the formation of the colony morphology. Also, in this case we use the model of cooperative propagation of bacteria, as explained in the section dedicated to the development of the protocol, to simulate the spread of the bacterial biomass. This simulation provides an example of the visualization of the biomass, biomass growth rate and the glucose spatial profiles shown in Fig. 3c, 3e and 3f.

Create the COMETS input files

1. The stoichiometric model for this protocol is the same as the model in the procedure section dedicated to the growth of *E. coli* in well-mixed condition, the e_core_model.txt. The preparation of the model file in the COMETS format is the same as the procedure in the above section on growth of bacteria in well mixed condition. The difference in this case, is that this file, in addition to the required metabolic model fields, contains the fields for the cooperative propagation model, and the demographic noise:

```
convNonlinDiffZero  0.0
//
convNonlinDiffN  1.0e-5
//
convNonlinDiffExponent 1.0
//
convNonlinDiffHillN 2.0
//
convNonlinDiffHillK 0.01
//
noiseVariance 0.0
//
neutralDrift true
//
neutralDriftSigma 0.1
//
```

2. The layout file circular.txt, consists of a 400x400 grid initially uniformly populated with the nutrients and inoculated with initial biomass of E. coli in the center, simulating an initial drop.

3. The global_params.txt file, as in the previous examples, contains the rate and name of saving the snapshot images of the colony during the simulation.

4. The package parameters file contains the simulation parameters, most notably the propagation mode of the colony set to biomassmotionstyle = ConvNonlin Diffusion 2D

```
biomassmotionstyle = ConvNonlin Diffusion 2D
defaultdiffconst = 6.0E-6
exchangestyle = Monod Style
defaultVmax = 18.5
defaultKm = 0.000015
defaultHill = 1
```
▲ **CRITICAL STEP**  Setting the correct values of the parameters is critical for obtaining the correct outcome.

Run the COMETS simulation

5. To submit it and run it to a cluster queuing system, one should use the Grid Engine User command provided on the computational cluster. For example:

```
qsub -pe omp 10 -l h_rt=24:00:00 qscript
```

Where qscript is the bash script to be submitted:

```
#!/bin/bash -l
module load gurobi/9.0.0
comets_scr comets_script
```

Here the second line loads the gurobi optimizer module. The specific name and version of the optimized should be set to the one installed on the user's system.

The output of this run consists of a collection of figures representing the 2D morphology of the single species bacterial colony with one example shown in Fig. 3C.
? TROUBLESHOOTING

Create a simulation video and layout images

6. With the collection of image in the directory images/ one can create a gif format video of the growth of the colonies, with the additional step:

```
convert image_*.png movie.gif
```

or any protocol of choice for creation of movies from collection of png images.
The video is shown in Supplementary video 2.
? TROUBLESHOOTING

7. The data in the biomass.m, media.m and fluxes.m files can be further analysed and visualised. Here we provide an example of visualization of the biomass, biomass growth rate and the glucose spatial profiles in Figs. 3c, 3e and 3f. The figures were obtained by running the MATLAB script RateandNutrientImage.m. In order to manage the memory needed to run the script, we prepared the flux and metabolite output file in a way to contain only the results at the desired time step 12000 using the command line:

cat fluxes.m |grep {12000} >fluxes_atTime12000.m

and

cat media.m |grep _12000{9} >media_glucose_atTime12000.m

The MATLAB script RateandNutrientImage.m:

```
clear all

azymuth=0;
elevation=90;
width=0.3;
hight=1.0;
xpos=0.0;
ypos=0.33;
edgepos=0.02;

for i=1:400
        for j=1:400
        fluxes{12000}{i}{j}{1}=zeros(13);
        end
end

flux_atTime12000;
full(fluxes);

media_glucose_atTime12000;

time=12000

fig1=figure()
%set(gcf, 'Position', [100, 100, 1000, 500])
zz(1:400,1:400)=0;
for i=1:400
        for j=1:400
        a=fluxes{time}{i}{j}{1}(13);
        zz(j,i)=a;
        end
end
s1=pcolor(zz);
axis square
s1.EdgeColor = 'none';
ax=gca;
ax.Box = 'off';
ax.XGrid = 'off';
ax.YGrid = 'off';
ax.ZGrid = 'off';
ax.XTick = [];
ax.YTick = [];
ax.ZTick = [];
ax.ZColor = 'none';
ax.YColor = 'none';
ax.XColor = 'none';
colormap(ax,bone);
view(azymuth,elevation);
```

```
fig2=figure()

time_string=num2str(time);
zz(1:400,1:400)=0;
varname=genvarname(['media_' time_string]);
variable=eval(varname);
for i=1:400
        for j=1:400
        zz(j,i)=variable{9}(i,j);
        end
end
s=pcolor(zz);
axis square
s.EdgeColor = 'none';
ax=gca;
ax.Box = 'off';
ax.XGrid = 'off';
ax.YGrid = 'off';
ax.ZGrid = 'off';
ax.XTick = [];
ax.YTick = [];
ax.ZTick = [];
ax.ZColor = 'none';
ax.YColor = 'none';
ax.XColor = 'none';
colormap(ax,copper);
view(azymuth,elevation);
```

*Case Study #5: Demographic noise and cooperative biomass propagation: five strains* [CL] ● TIMING setup: 30min-1hr, simulation: 3-5 days

This protocol, as the previous one, shows the role demographic noise and cooperative biomass propagation in colony morphology formation. In this case, however, we illustrate the role of demographic noise by showing the formation of single strain sectors in a population consisting of mixed five metabolically identical strains. In this protocol we start with the initial population of the colony consisting of five metabolically identical strains. The presence of demographic noise, in this case, leads to formation of sectors populated by a lower number of strains. Eventually, as time progresses, the growing edge of the colony shows sectors with a single strain. This decay of the heterozygosity has been observed in experiments and studied theoretically[79,80]. The result is shown in Supplementary video 3. A sample image is shown in Fig. 3d.

The stoichiometric model used in this protocol is identical as the one in the previous section, the e_coli_core model, except that it differs in the value of a single parameters:

convNonlinDiffHillK 1.0e-3
//

The layout file, however, in this case contains five repetitions of the same model:

model_file e_coli_core.txt e_coli_core.txt e_coli_core.txt e_coli_core.txt e_coli_core.txt

Also, the initial_pop block in the layout file contains five identical columns, corresponding to each strain, with identical initial populations.

```
     Initial_pop
151  191 0.619774e-04 0.619774e-04 0.619774e-04    0.619774e-04 0.619774e-04
151  192 0.619650e-04 0.619650e-04 0.619650e-04    0.619650e-04 0.619650e-04
...
```

Create the COMETS input files

1. Create the input parameters, model and layout file, following the same procedure as in the previous protocol. In this case, write the layout file with five repetitions of the same model, as described above.
2. Change the convNonlinDiffHillK parameter value to 1.0e-3 in the e_coli_core.txt file.
   ▲ **CRITICAL STEP** Setting the correct value of the parameter is critical for obtaining the correct outcome.

Run the COMETS simulation

3. Run comets on a cluster by executing the commands:

qsub -pe omp 10 -l h_rt=24:00:00 qscript

   The output of this run consists of a collection of figures representing the 2D morphology of the single species bacterial colony with clear spatial segregation of the strains shown in Fig. 3d.
   ? TROUBLESHOOTING

Create a simulation video

4. With the collection of images in the directory images/, we can create a gif format video of the growth of the colonies, with the additional step:

convert image_*.png Movie_fiveStrains.gif

   or any protocol of choice for creation of movies from a collection of png images.
   ? TROUBLESHOOTING

*Case Study #6: Simulations including extracellular reactions* [M] ● TIMING setup: 30 min, simulation 1-2 min

This protocol demonstrates the capacity of COMETS to simulate reactions involving extracellular metabolites using the MATLAB toolbox. In the first case, a simple binding reaction is defined with the form A+B→C. In the second case, an enzyme, E, catalyzes the conversion of F→G. We will use the Scerevisiae_iMM904.mat[81] stoichiometric model.

Extracellular reaction information is included in the COMETS layout file, in blocks with the following format:

REACTANTS
        rxnIdx metIdx order k   //elemental rxn
        rxnIdx metIdx km      //enzyme catalyzed
ENZYMES
        rxnIdx metIdx kcat
PRODUCTS
        rxnIdx metIdx stoich
//

Each reactant or product in a given reaction should appear on its own line, using the same integer rxnIdx per extracellular reaction. The metIdx field corresponds to a metabolite's position in the world_media block. The system determines whether a reaction is elemental or enzymatic based on the presence of an entry in the ENZYMES block. In the case of elemental reactions, only the first entry for the rate constant k is used.

Create the COMETS input files

1. The first step is to load a model and create a layout. For the purposes of this demo we will not be concerned with the metabolites used by the metabolic model, but this step is included because attempting to execute COMETS without any biomass will return a warning instead of completing the simulation.

    ```
    load('Scerevisiae_iMM904.mat'); %as 'model'
    layout = createLayout(model);
    layout = setInitialPop(layout,'1x1');
    ```

2. Add a simple extracellular reaction of the form *A+B*, with a reaction rate $k=0.5s^{-1}$.

    ```
    layout = addExternalReaction(layout,'reaction1',{'A' 'B' 'C'},...
            [-1 -1 1],'k',0.2);
    layout = setMedia(layout,'A',1);
    layout = setMedia(layout,'B',2);
    ```

3. Add an enzyme-catalyzed reaction of the form F, catalyzed by the enzyme E, with turnover rate $20s^{-1}$ and $K_M$=0.01mmol.

   ```
   layout = addExternalReaction(layout,'enz_reaction',{'F' 'G'},...
           [-1 1],'enzyme','E','kcat',2,'km',0.25);
   layout = setMedia(layout,'F',1);
   layout = setMedia(layout,'E',0.1);
   ```
   ▲ **CRITICAL STEP** Setting these parameters correctly is critical for the outcome of the simulation.

Run the COMETS simulation

4. Set a few parameters and execute the simulation.

   ```
   layout.params.timeStep = 1/3600; %timestep = 1 second
   layout.params.writeMediaLog = 1;
   layout.params.maxCycles = 25;
   runComets(layout)
   ```
   ? TROUBLESHOOTING

Plot the data

5. Load the results and plot the data.

   ```
   media = parseMediaLog(layout.params.mediaLogName,{'A' 'B' 'C' 'E'...
           'F' 'G'});

   plotMediaTimecourse(media,{'A' 'B' 'C'});
   title('A + B -> C');

   plotMediaTimecourse(media,{'E' 'F' 'G'});
   title('Enzymatic: F -> G');
   ```

The REACTIONS block in the comets_layout.txt file generated by the runComets command contains the following:

```
reactions
    reactants
        1 15 1 2.000000e-01
        1 16 1 2.000000e-01
        2 19 2.500000e-01
    enzymes
        2 18 2
    products
        1 17 1
        2 20 1
//
```

The first column under each heading denotes the ID of the reaction each metabolite is participating in. The second column is the index of the metabolite in the WORLD_MEDIA list (the values here are not 1-6 because the media also contains exchange metabolites from the metabolic model that was included at the start of the script).

Under the REACTANTS heading, the subsequent columns differ for mass-balance and enzymatic reactions. For mass-balance reaction 1, the third value is the stoichiometry of the reactant and the fourth value is the reaction rate constant (which is the same for all reactants). For enzymatic reaction two, the third value is the turnover rate and there is no fourth value (it is assumed that the stoichiometry of the substrate in an enzymatic reaction is always 1).

The ENZYMES heading only includes a row for the enzymatic reaction. The values here are the reaction number, the index of the metabolite that acts as the enzyme, and the millimolar half-saturation concentration.

The PRODUCTS heading contains values for the reaction number, the index of the produced metabolite, and the product's stoichiometry.

The figures produced by the plotMediaTimecourse function should appear as in Fig. 5.

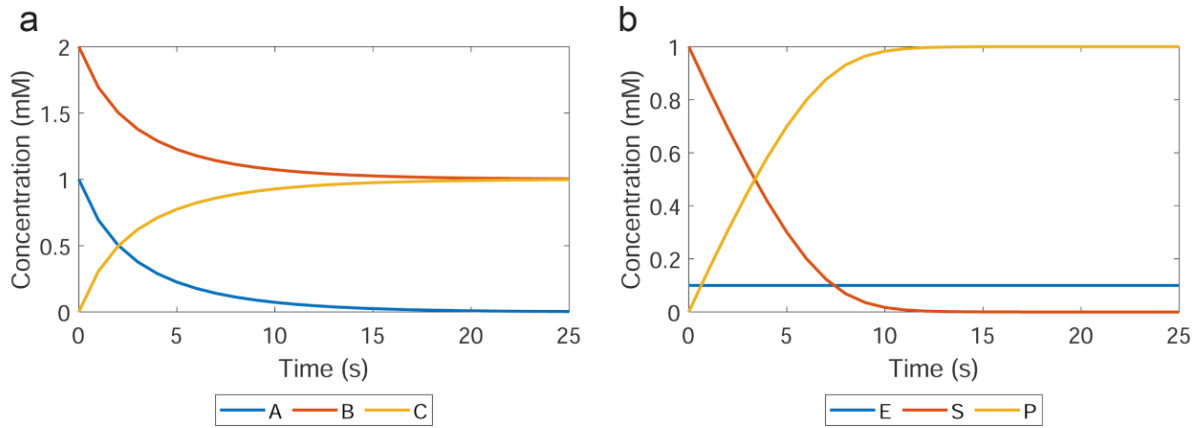

**Figure 5. Media concentrations over time during simulations demonstrating extracellular reactions.** a) A binding reaction of the form *A+B→C*, with rate *v=v_max·[A]·[B]* where $v_{max}$ = 0.2 $s^{-1}$. b) An enzyme-catalyzed reaction of the form *E+S→E+P*, with rate according to the Michaelis-Menten equation *v=v_max·[E]·[S]/(K_M+[S])*, where $v_{max}$ = 2 $s^{-1}$ and $K_M$ = 0.25 *mmol*.

## Case Study #7: Growth of multiple bacterial organisms in well-mixed conditions

[M] ● TIMING setup: 15 min, simulation: 1.5hr

This protocol demonstrates the capacity of COMETS to simulate multispecies community dynamics and metabolic exchange using the MATLAB toolbox.

The metabolic models used in this simulation modelsCommunity.mat and the growth medium mediumCommunity.mat can be downloaded from the COMETS GitHub repository. In this example, the growth phenotypes of *Bacillus subtilis*[82], *Escherichia coli*[83], *Klebsiella pneumoniae*[84], *Lactococcus lactis*[85], *Methylobacterium extorquens*[86], *Pseudomonas aeruginosa*[87], *Porphyromonas gingivalis*[88], *Rhodobacter sphaeroides*[89], *Shigella boydii*[90], *Saccharomyces cerevisiae*[81], *Salmonella enterica*[91], *Shewanella oneidensis*[92], *Synechocystis* sp. PCC6803[93], and *Zymomonas mobilis*[94] are simulated. These organisms, whose growth and metabolic exchange phenotypes were previously analyzed *in silico*[21], were selected to represent a broad cross-section of nutrient utilization capabilities. Here, the growth profile of a multispecies combination of these organisms in the presence of D-glucose and L-alanine over 12 hours will be analyzed.

Import the models and medium conditions, and initialize the COMETS layout

1. Unzip the 'modelsCommunity.mat.zip' file. If this is unzipped as a directory, it will be necessary to move the 'modelsCommunity.mat' file back into the top directory.
2. In MATLAB, import the metabolic model MATLAB structure.

```
>> load modelsCommunity.mat
```

This will result in a structure models being loaded into the workspace.

3. Initialize a COMETS layout object using the CometsLayout class and add the models using the addModel function.

```
>> layout = CometsLayout();
modelNames = fieldnames(models);
for m = 1:length(modelNames)
layout = addModel(layout,models.(modelNames{m}));
end
```

This step results in a CometsLayout object layout, with the attributes below, being added to the workspace.

layout =

CometsLayout with properties:

models: {1×14 cell}
xdim: 1
ydim: 1
mets: {725 cell}
media_amt: [725×1 double]
params: [1×1 CometsParams]
diffusion_constants: [725×2 double]
global_media_refresh: [725×1 double]
media_refresh: [725×1 double]
global_static_media: [725×2 double]
static_media: [725×1 double]
initial_media: 0
barrier: 0
initial_pop: 0
external_rxns: [0×0 table]
external_rxn_mets: [0×0 table]

Here, the object property models is an array with the individual metabolic models.

4. Load the medium file.

```
>> load mediumCommunity.mat
```

The mediumCommunity file contains four objects that will be added to the workspace:

minMed: A cell array of 32 molecules (in BIGG format) that are contained in the minimal medium to be added at nonlimiting concentrations:

```
minMed =

  32×1 cell array

    '4abz[e]'
    'btn[e]'
    'ca2[e]'
    'cbl1[e]'
    'chol[e]'
    'cl[e]'
    'cobalt2[e]'
    'cu2[e]'
    'fe2[e]'
    'fe3[e]'
    'fol[e]'
    'h2[e]'
    'h2o[e]'
    'k[e]'
    'lipoate[e]'
    'mg2[e]'
    'mn2[e]'
    'mobd[e]'
    'na1[e]'
    'ncam[e]'
    'nh4[e]'
    'ni2[e]'
    'no3[e]'
    'o2[e]'
    'pi[e]'
    'pnto-R[e]'
    'pydx[e]'
    'ribflv[e]'
    'slnt[e]'
    'so4[e]'
    'thm[e]'
    'zn2[e]'
```

Nutrients: A cell array of five carbon sources (in BIGG format) that will be added at limiting concentrations:

```
nutrients =

  1×2 cell array

    'glc-D[e]'    'ala-L[e]'
```

nutrientNames: A cell array containing human-readable names of the five carbon sources:

nutrientNames =

  1×5 cell array

   'D-glucose'    'L-alanine'

numCarbons: A numerical vector containing the number of carbon atoms contained in each of the five carbon sources:

numCarbons =

   6    3

5. Add the minimal media components at nonlimiting concentrations using the setMedia function.

```
>> for mm = 1:length(minMed)
      layout = layout.setMedia(minMed{mm},1000);
   end
```

The quantity 1000 is used to denote 1000mmol, an amount of molecule that is effectively non limiting for this simulation. If a continuous culture environment is desired, the setStaticMedia function can be used within the for loop to fix the amount of each minimal medium molecule at 1000mmol.

```
>> layout = setStaticMedia(layout,1,1,minMed{mm},1000);
```

6. Add the carbon sources at limiting amounts ($5e^{-4}$ mmol) using the **setMedia** function.

```
>> for n = 1:length(nutrients)
       layout = layout.setMedia(nutrients{n},5e-4);
   end
```

Alternatively, each nutrient can be added at equal carbon ratios within the for loop:

```
>> layout = layout.setMedia(nutrients{n},5e-4/numCarbons(n)/length(nutrients));
```

If a continuous culture environment is desired, the global_media_refresh layout parameter can be used within the for loop in addition to the setMedia function to refresh the amount of each carbon source.

```
>> layout.global_media_refresh(find(ismember(layout.mets,nutrients(n)))) = 50/(12/0.1)
```

The global_media_refresh parameter defines the amount of additional nutrient to be added at each time step, analogous to a constant influx of fresh nutrients in a continuous culture device.

7. Define the initial organism populations.

```
>> layout.initial_pop = ones(length(modelNames),1).*1.e-7;
```

Here, each organism is added to the environment at an abundance of $1.0 \cdot 10^{-7}$ grams dry weight. By default, the layout is initialized with dimensions of 1 x 1 cells to simulate a well-mixed environment.

Set the COMETS parameters and run the simulation

8. Define the COMETS working directory, log file names, and simulation parameters.

```
>> cometsDirectory = 'CometsRunDir';

layout.params.writeBiomassLog = true;
layout.params.biomassLogRate = 1;
layout.params.biomassLogName = 'biomassLog';
layout.params.biomassLogFormat = 'MATLAB';
layout.params.writeMediaLog = true;
layout.params.mediaLogRate = 1;
layout.params.mediaLogName = 'mediaLog';
layout.params.mediaLogFormat = 'MATLAB';
layout.params.writeFluxLog = true;
layout.params.fluxLogRate = 1;
layout.params.fluxLogName = 'fluxLog.m';
layout.params.fluxLogFormat = 'MATLAB';

layout.params.maxSpaceBiomass = 1e3;
layout.params.timeStep = 0.01;
layout.params.maxCycles = 1200;
layout.params.deathRate = 0.1;
```

Here, the simulation records the biomass of each organism, the amounts of medium components, and the fluxes of all organisms at each time step. The time step is set to 0.01 1/h and the maximum number of cycles is set to 2400 for a total simulation time of 24 hours. The death rate is set to a fixed value of 0.1, under which 10% of the population of each organism will be eliminated from the simulation at each time step. This death rate can also be set to approximate an organism dilution rate in a continuous culture environment.
▲ **CRITICAL STEP** Setting the correct parameter values is critical for the outcome of the simulation.

9. Prepare the metabolic models for the COMETS simulation. Here, the lower bounds of certain reactions within the models are being altered in order to allow uptake of medium components. Specifically, exchange reactions associated with uptake of the minimal medium components are being unconstrained (lower bound set to -1000), while the exchange reactions associated with uptake of the limiting carbon sources are being opened (lower bound set to -10). The reactions to be unconstrained are located on a model-by-model basis by first identifying the index of a specific external metabolite and matching the appropriate exchange reaction to it via the S matrix.

```
for m = 1:length(modelNames)
   modelCurr = models.(modelNames{m});
   minMedMets = find(ismember(modelCurr.mets,minMed));
   for i = 1:length(minMedMets)
            modelCurr.lb(intersect(find(findExcRxns(modelCurr)),find(modelCurr.S(minMedMets(i),:))
      )) = -1000; % Allow unlimited uptake of nonlimiting nutrients
   end
   limitingMets = find(ismember(modelCurr.mets,nutrients));
   for i = 1:length(limitingMets)
            modelCurr.lb(intersect(find(findExcRxns(modelCurr)),find(modelCurr.S(limitingMets(i),:))))
      = -10; % Allow limited uptake of limiting nutrients
   end
   models.(modelNames{m}) = modelCurr;
end
```

10. Run the COMETS simulation.

```
>> runComets(layout,cometsDirectory)
```
▲ **CRITICAL STEP** If the environment variable COMETS_HOME is not defined, COMETS will not run. Refer to troubleshooting guide on how to correctly define environment variables.
? TROUBLESHOOTING

Parse the COMETS output logs and visualize data

11. Parse the media log using the parseBiomassLog function and format the log into a matrix.

```
>> biomassLogRaw = parseBiomassLog([cometsDirectory '/' layout.params.biomassLogName]);
biomassLog = zeros(size(biomassLogRaw,1)/length(modelNames),length(modelNames));
for i = 1:length(modelNames)
   biomassLog(:,i) = biomassLogRaw.biomass(i:length(modelNames):end);
end
```

12. Format the names of the organisms for plotting and plot the biomass over time.

```
>> modelNamesFormatted = cell(length(modelNames),1);
for m = 1:length(modelNames)
   s = split(modelNames{m},'_');
   modelNamesFormatted{m} = [s{1} '. ' s{2}];
end

close all
figure
plotColors = parula(length(modelNames));

for m = 1:length(modelNames)
plot([1:layout.params.maxCycles+1]*layout.params.timeStep,biomassLog(:,m),'LineWidth',4,'Color',plotColors(m,:))
end

set(gca,'FontSize',16)
ylabel('Nutrient Amount (mmol)')
xlabel('Time (h)')
legend(modelNamesFormatted)
```

The results of this action are shown in Figure 6a.

13. Parse the media log using the parseMediaLog function and format the log into a matrix.

```
>> allMetsFromModels = layout.mets;
COMETSCycles = layout.params.maxCycles;
mediaLogMat = zeros(length(allMetsFromModels),COMETSCycles);

mediaLogRaw = parseMediaLog([cometsDirectory '/' layout.params.mediaLogName]);
mediaLogMetOrder = zeros(length(allMetsFromModels),1);

% Re-order the medium components to match the list in layout.mets
for i = 1:length(allMetsFromModels)
     mediaLogMetOrder(i) = find(ismember(mediaLogRaw.metname(1:length(allMetsFromModels)),allMetsFromModels(i)));
end

for i = 1:COMETSCycles
currentMedia = mediaLogRaw.amt(find(mediaLogRaw.t == i));
mediaLogMat(:,i) = currentMedia(mediaLogMetOrder);
end
```

The resulting mediaLogMat matrix has dimensions M x N, where M is the number of metabolites in the COMETS simulation and N is the number of time steps.

14. Plot the nutrient abundances.

```
>> % Make a list of indices from allMetsFromModels ordered according to the elements in nutrients
nutrientsToPlot = zeros(length(nutrients),1);
for i = 1:length(nutrients)
    nutrientsToPlot(i) = find(ismember(allMetsFromModels,nutrients{i}));
end

figure
plot([1:layout.params.maxCycles]*layout.params.timeStep,mediaLogMat(nutrientsToPlot,:)', 'LineWidth',4)
set(gca,'FontSize',16)
ylabel('Nutrient Amount (mmol)')
xlabel('Time (h)')
legend(nutrientNames)
```

> This action allows visualization of the nutrient abundances, shown in Figure 6b.
>
> These results, along with directly examining the nutrient abundances in mediaLogMat, allow us to infer that *M. extorquens* and *S. oneidensis* initially grew to high abundances on D-glucose and L-alanine, rapidly outcompeting the remaining organisms. We are also able to observe low growth of multiple other organisms, such as *S. boydii* and *Z. mobilis*, which peak upon exhausting D-glucose at hour 6. When both primary resources are exhausted between hours 6 and 7, the abundance of all organisms begins to decay.

15. Obtain the metabolites secreted, absorbed, and exchanged by each organism using the getSecAbsExcMets function.

```
>> [secMets,absMets,excTable] = getSecAbsExcMets([cometsDirectory '/'
layout.params.fluxLogName],models,layout);
```

> This function outputs two matrices, secMets and absMets, each having dimensions N x M where N is the number of metabolites present in the layout and M is the number of metabolic models. For any model-metabolite pair, the quantity present in secMets or absMets denotes the highest flux that the metabolite was secreted or absorbed by the corresponding organism in the entire simulation.
>
> The function also outputs a matrix excTable, which contains information about each metabolite that was exchanged in the simulation. The first column of excTable contains the index (corresponding to the order in models) of the organism that secreted a metabolite, the second column contains the index of the organism that absorbed that metabolite, and the third column contains the index (corresponding to allMetsFromModels) of the metabolite that was exchanged. Here, a truncated version of excTable is as follows:

excTable =

```
     2     5    17
     2    12    17
     3     5    17
     3    12    17
     6     5    17
     6    12    17
     9     5    17
     9    12    17
    11     5    17
    11    12    17
    14     5    17
    14    12    17
     6    12    36
     1     5    65
     2     5    65
     3     5    65
     5     5    65
```

The first line, for example, denotes that acetate (metabolite 17) is secreted by *E. coli* (organism 2) and absorbed by *M. extorquens* (organism 5).

16. Use the secMets matrix to identify molecule secretion.

Here, the source organism of each metabolite secreted during the simulation will be identified. First, a list of metabolites with nonzero secretion flux is generated.

>> nonzeroSecMetIndices = find(sum(secMets,2));

This action yields a list of indices that correspond to allMetsFromModels, which can be read using the command:

>> allMetsFromModels(nonzeroSecMetIndices);

This yields a list of metabolites, a truncated version of which is below:

ans =

  18×1 cell array

    '5mtr[e]'
    'ac[e]'
    'ala-L[e]'
    'co2[e]'
    'fe2[e]'
    'for[e]'
    'glyclt[e]'

Molecules of particular interest can be selected and analyzed further. Here, secretion of acetate (ac[e]) and formate (for[e]) will be analyzed.

```
>> selectSecMets = {'ac[e]','for[e]'};
selectSecMetIndices = intersect(find(ismember(allMetsFromModels,selectSecMets)),nonzeroSecMetIndices);
```

The abundances of each of these metabolites over time can be plotted:

```
>> figure
plotColors2 = winter(2);
for s = 1:length(selectSecMets)
plot([1:layout.params.maxCycles]*layout.params.timeStep,smoothdata(mediaLogMat(selectSecMetIndices(s),:)'),'LineWidth',4,'Color',plotColors2(s,:))
   hold on
end
set(gca,'FontSize',16)
ylabel('Metabolite Amount (mmol)')
xlabel('Time (h)')
legend({'Acetate','Formate'})
```

This action yields Figure 6c, showing rapid accumulation and subsequent consumption of acetate and formate.

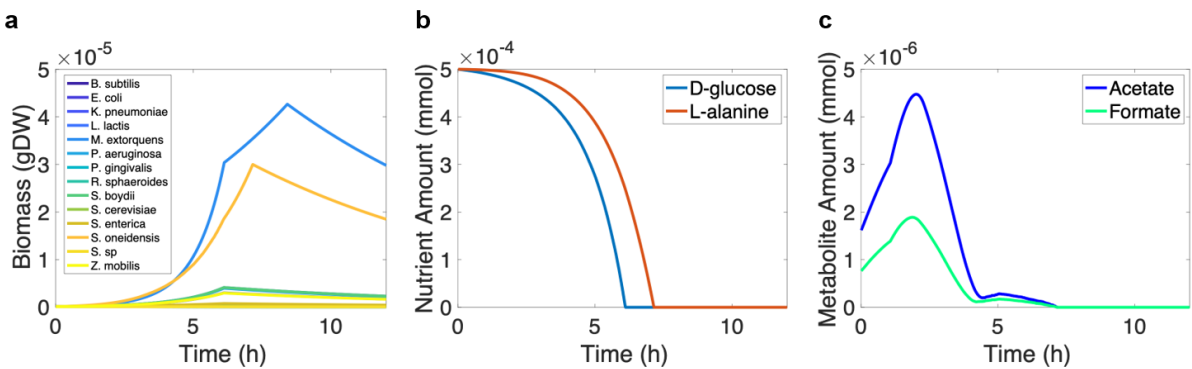

**Figure 6. Growth and metabolic exchange of 14-species microbial community.**
**a**) Biomass production of all 14 organisms over time. **b**) Consumption of limiting carbon sources over time. **c**) Secretion and consumption of metabolic byproducts over time.

*Case Study #8: Modelling the diurnal cycle* [M]  ● TIMING setup: 15 min; simulation: 1-5 min

COMETS 2 can simulate periodically changing environments, where the periodic function can be either a step function, a sine function or a half sine function (shown below). The most obvious use case for this functionality is to study how the metabolism of photosynthetic organisms change during the day / night cycle with varying sunlight (photons) and how this affects the microbes. Here we simulate one such experiment with a genome-scale model of *Prochlorococcus*[95], the most abundant marine photoautotroph.

Import the models and medium conditions, defining light and light absorption parameters and initialize the COMETS layout

1. Assuming that COBRA Toolbox is initialized the first step is to load the *Prochlorococcus* model iSO595.

```
% Load Prochlorococcus Genome-scale model
model_fn = 'iSO595v6.mat';
model = readCbModel(model_fn);
```

2. Define parameters related to light absorption and use these to calculate a model-specific absorption coefficient. We here assume a monochromatic light source at 680 nm, but it is possible to extend these calculations to a light source with a spectral distribution. All parameters are acquired from the literature [96–101].

```
% The ratio of chlorophyll is extracted from the model biomass-function
ci_dvchla = 0.0163 % g / gDW (Partensky 1993 / Casey 2016)
ci_dvchlb = 0.0013 % g / gDW (Partensky 1993 / Casey 2016)
absorption_dvchla_680 = 0.0184; % m^2 mg^-1 (Bricaud et al., 2004)
absorption_dvchlb_680 = 0.0018; % m^2 mg^-1 (Bricaud et al., 2004)
absorption_water_680 = 0.465; % m^-1 (Pope and Fry, 1997)
wavelength = 680;% nm
```
▲ **CRITICAL STEP** Setting the correct parameters is critical for the outcome of the simulation.

3. Calculate the packaging effect. This is taking into account that the light-absorbing pigments are not dissolved in the media, but contained within discrete cells [102]. The packaging effect approaches 0 asymptotically for large cells. Depending on the accuracy needed in the calculation the packaging effect can be assumed to be close to 1 for very small cells, such as *Prochlorococcus* [102].

```
diameter = 0.6; % um (Morel et al., 1993)
n_dash = 13.77*1e-3; % imaginary part of refractive index at 675 nm (Stramski et al. 2001)
size_parameter_alpha = diameter*1e3*pi/wavelength; % A variable describing the size ratio between the cell size and wavelength
rho_dash = 4*size_parameter_alpha*n_dash;
Q_a = 1+(2*exp(-rho_dash)/rho_dash)+2*(exp(-rho_dash)-1)/rho_dash^2;
packaging_effect = 1.5*Q_a/rho_dash;

% Calculate the Prochlorococcus specific biomass absorption coefficient in units m2/ g DW biomass
absorption_biomass = packaging_effect*(ci_dvchla*1e3*absorption_dvchla_680+ci_dvchlb*1e3*absorption_dvchlb_680);
```
▲ **CRITICAL STEP** Setting the correct parameters is critical for the outcome of the simulation.

4. Set the calculated absorption rate as a model parameter for the exchange reaction of light. LightEX is the exchange reaction for photons in the model.

```
absorption_matrix = zeros(1,2);
absorption_matrix(1) = absorption_water_680;
absorption_matrix(2) = absorption_biomass;
model = setLight(model, {'LightEX'}, absorption_matrix);
```

5. Create the layout using the function *CometsLayout()* from the COMETS toolbox and define parameters such as filenames for media and biomass log files, number of iterations, time-step etc.

```
% Make layout with the COMETS toolbox
layout = CometsLayout();
layout = layout.addModel(model);

% We use a single cell as the model layout, and set initial amount of cells to 1e-7 g
layout = setInitialPop(layout, '1x1', 1e-7);

% Set simulation parameters
layout.params.writeMediaLog = true;
layout.params.mediaLogName = [pwd '/mediaLog.m'];
layout.params.writeBiomassLog = true;
layout.params.biomassLogName = [pwd '/biomassLog.m'];

layout.params.maxCycles = 480;
layout.params.timeStep = 0.1;
layout.params.defaultDiffConst = 0;
layout.params.objectiveStyle = 'MAX_OBJECTIVE_MIN_TOTAL';
```

6. Define the concentration of each metabolite in the growth medium. We here set the concentration of all essential metabolites except photons to 1000 mmol because we want the growth to be only limited by the light conditions

```
% Define medium
layout = layout.setMedia('Ammonia[e]', 1000); % 1 mmol
layout = layout.setMedia('HCO3[e]', 1000);
layout = layout.setMedia('CO2[e]', 1000);
layout = layout.setMedia('H[e]', 1000);
layout = layout.setMedia('Orthophosphate[e]', 1000);
layout = layout.setMedia('H2O[e]', 1000);
layout = layout.setMedia('Cadmium[e]', 1000);
layout = layout.setMedia('Calcium_cation[e]', 1000);
layout = layout.setMedia('Chloride_ion[e]', 1000);
layout = layout.setMedia('Cobalt_ion[e]', 1000);
layout = layout.setMedia('Copper[e]', 1000);
layout = layout.setMedia('Fe2[e]', 1000);
layout = layout.setMedia('Magnesium_cation[e]', 1000);
layout = layout.setMedia('Molybdenum[e]', 1000);
layout = layout.setMedia('K[e]', 1000);
layout = layout.setMedia('Selenate[e]', 1000);
layout = layout.setMedia('Sodium_cation[e]', 1000);
layout = layout.setMedia('Strontium_cation[e]', 1000);
layout = layout.setMedia('Sulfate[e]', 1000);
layout = layout.setMedia('Zn2[e]', 1000);
layout = layout.setMedia('Hydrogen_sulfide[e]',1000);
```

7. Define light conditions. Light is modelled as individual photons the value is the number of photons (in mmol) absorbed by the organism at each timepoint. While it is treated like any other metabolite, it is more reasonable to consider it light intensity than a concentration which the other compounds in the media are. To model natural light conditions, we use the periodic function called half sin which is equal to

$$max(f(t,0))$$

where

$$f(t) = A\sin(\omega t + C)$$

Here A is the amplitude, ω the angular frequency

$$\omega = 2\pi/T$$

T the period, ϕ the phase and C the offset. Other periodic functions are available: step function, sine and cosine. In this example we define global light conditions, but when running a simulation with one or more spatial dimension it is possible to set different light conditions in each grid cell by using the function setDetailedPeriodicMedia. We here define the period to 24 hours with an amplitude of 0.04 mmol photons per meter squared per second.

```
% Set light conditions by defining parameters
amplitude = 0.04; % mmol photons / m^2 / s
function_name = 'half_sin';
period = 24; % In hours
phase = 0;
offset = 0;
photon_metabolite_id = 'Photon[e]';

% Set globally changing light conditions
layout = layout.setGlobalPeriodicMedia(photon_metabolite_id, function_name, amplitude, period,phase,offset);
```
▲ **CRITICAL STEP** Setting the correct parameters is critical for the outcome of the simulation.

Run the COMETS simulation

8. Run Comets by using the function *runComets()* in the Comets toolbox.

    If you want to run this layout in the Comets GUI export the layout and model by using the function *createCometsFiles(layout, pwd)* and import the layout in the GUI.

```
% Runs comets and produce the output files mediaLog.m and biomassLog.m
runComets(layout)
```
? TROUBLESHOOTING

Load and plot the results

9. Load the results, i.e. the biomass and media log files.

```
media = parseMediaLog('mediaLog.m');
biomass = parseBiomassLog('biomassLog.m');
```

10. Plot the output, e.g. the biomass and the light intensity. Other media components can be plotted by using the function plotMediaTimecourse()

```
% Plot the light intensity
figure('Name', 'Biomass and Light');
p = plotMediaTimecourse(media, 'Photon[e]', false);
set(p,'linewidth',2, 'color', 'b');
legend('Location','northeast')
ylabel('Light flux [mmol photons m^{-2} s^{-1}]');
ylim([0 0.045]);

% Plot the biomass in the same figure
yyaxis right
p1 = plot(biomass.biomass, 'DisplayName', 'Biomass')
set(p1,'linewidth',2, 'color', 'r');
xlabel('Timestep');
ylabel('Biomass [g DW]');
ax = gca;
ax.YAxis(2).Color = 'k';
ylim([1e-7 1.16e-7]);
```

The result is shown in Fig. 7.

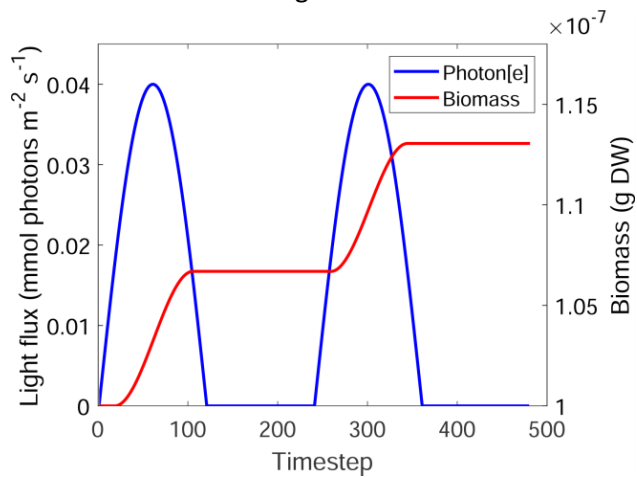

**Figure 7. Simulations of the diurnal cycle of the marine photoautotrophic bacteria Prochlorococcus.** The organism is in an environment with a time-dependent light environment replicating the day - night cycle. The growth of the biomass is evident only during daytime.

## The basics of COMETS using the Python Toolbox [P] ● TIMING 10 min

We will first walk through the basic functionalities of COMETS using the Python Toolbox, and more specific examples of usage will be provided in the next sections. Once cometspy has been installed, a user can implement this and all other python protocols in one of two ways: 1) The user can copy the code into a python script and run it. 2) We have also provided the protocols in Jupyter notebook format. To use these included files, start the Jupyter notebook by typing the following command line:

jupyter notebook

In Windows this is best done by going to the start menu, and running "Anaconda Powershell Prompt". The above command can be run from the Anaconda Powershell.

This will launch a browser tab. On this tab, load the corresponding protocol through file-open and finding the [name of protocol].ipynb file (for example, chemostat.ipynb). To run each Jupyter notebook click on the kernel tab and then click Restart & Run All. Warning: for some protocols, this may take from five minutes to several hours.

Create of the COMETS input files

1. Import the python comets toolbox, the cobra toolbox and the cobra.test tools.

```
import cobra
import cobra.test
import cometspy as c
```

2. Load an existing model using COBRAPy. Here, we use the custom function cobra.test.create_test_model() from the COBRAPy toolbox to load the *E. coli* model.

```
# Load a textbook example model using the COBRAPy toolbox
test_model = cobra.test.create_test_model('textbook')
```

3. Use the loaded COBRA model to build a COMETS model class, which allows us to change COMETS-specific model parameters, such as initial population sizes.

```
# Use the above model to create a COMETS model and open exchanges
test_model = c.model(test_model)

test_model.open_exchanges()
```

```
# Change comets specific parameters, e.g. the initial biomass of the model
test_model.initial_pop = [0, 0, 1e-7]
```

4. Use the params class to generate a list (i.e. a python dict object) containing the default parameter values.

```
# Create a parameters object with default values
my_params = c.params()
```

5. Change the parameter values as desired

```
# Change the parameter "maxCycles" corresponding to the number of iterations in our simulation
my_params.set_param('maxCycles', 100)
```

6. Check which other parameters are available and their current value.

```
# See available parameters and their values
my_params.show_params()
```

7. Use the layout class to generate a layout with the previously prepared model as input. Then, add minimal media components.

```
my_layout = c.layout(test_model)
my_layout.set_specific_metabolite('glc__D_e', 0.011)
my_layout.set_specific_metabolite('o2_e',1000);
my_layout.set_specific_metabolite('nh4_e',1000);
my_layout.set_specific_metabolite('pi_e',1000);
my_layout.set_specific_metabolite('h2o_e',1000);
my_layout.set_specific_metabolite('h_e',1000);
```

8. Visualize the media composition and relevant COMETS parameters (diffusion constants, "static" and "refresh" values), which is stored as a pandas dataframe:

My_layout.media # this shows a pandas data.frame

Run the COMETS simulation

9. Define the comets object by passing the previously created layout and parameters.

my_simulation = c.comets(my_layout, my_params)

10. Run the simulation. (Note that in this example, there will be no growth as we did not define a media allowing for it, e.g. the carbon source):

my_simulation.run()

11. Access the output of the COMETS simulation run

print(my_simulation.run_output) # this shows initialization and biomasses at each time step. This also shows a Java stack trace if COMETS had an internal error

12. Access the errors of the COMETS simulation run

print(my_simulation.run_errors) # should be empty if everything worked

The results of the successful simulation are stored in several fields in the comets object, depending on whether the parameters writeTotalBiomasslog, writeBiomassLog, writeFluxLog and writeMediaLog were set to true.

- The field total_biomass stores the total biomass (summed up over all coordinates) for each timepoint and species.
- The field biomass stores detailed biomass values for each timepoint, coordinate and species.
- The field media stores the composition of the media at each timepoint.
- The field fluxes stores the metabolic fluxes for each species, coordinate and timepoint.

Additionally, specific comets models will have additional output fields; for instance, specificMedia will contain the concentration of specific media components if set up. Similarly, if we run a simulation with evolution, the field genotypes will store information about each species such as its ancestor and which mutation it suffered.

All of the output files are stored as pandas dataframes which can be further analyzed or plotted using standard Python tools.

## *Case Study #9: Bacterial growth in a test tube* [P] ● **TIMING** setup: 30 min - 1hr, simulation: 1-5 min

In this example, we will simulate anaerobic fermentation in minimal media with glucose as the only carbon source.

Load COMETS and dependencies

1. Import required libraries.

```python
import cometspy as c
import cobra.test
import matplotlib.pyplot as plt
```

Create a "test tube"

2. Create an empty layout with default parameters, i.e. no models, an empty, well mixed space (called "cell") with volume 1cm$^3$.

```python
# Create empty 1x1 layout
test_tube = c.layout()
```

3. Modify this by setting the media composition.

```python
# Add 11mM glucose and remove o2
test_tube.set_specific_metabolite('glc__D_e', 0.011)
test_tube.set_specific_metabolite('o2_e', 0)

# Add the rest of nutrients unlimited (ammonia, phosphate, water and protons)
test_tube.set_specific_metabolite('nh4_e', 1000);
test_tube.set_specific_metabolite('pi_e', 1000);
test_tube.set_specific_metabolite('h2o_e', 1000);
test_tube.set_specific_metabolite('h_e', 1000);
```
▲ **CRITICAL STEP** It is critical to set the nutrient amounts to the correct values.

Inoculate the test tube and set experiment conditions

4. Load the model and add 1e-6 gr. biomass to the test tube.

```python
# create the model using CobraPy functionality
e_coli_cobra = cobra.test.create_test_model('textbook')

# use the loaded model to build a comets model
e_coli = c.model(e_coli_cobra)

# remove the bounds from glucose import (will be set dynamically by COMETS)
e_coli.change_bounds('EX_glc__D_e', -1000, 1000)

# set the model's initial biomass
e_coli.initial_pop = [0, 0, 5e-6]

# add it to the test_tube
test_tube.add_model(e_coli)
```
▲ **CRITICAL STEP** It is critical to add the initial biomass.

5. Set the parameters for the simulation (only those we want to be different from the default). For information on all available simulation parameters, refer to the parameter list.

```
# Create a default parameter set
sim_params = c.params()

# Change the desired parameters
sim_params.set_param('defaultVmax', 18.5)
sim_params.set_param('defaultKm', 0.000015)
sim_params.set_param('maxCycles', 1000)
sim_params.set_param('timeStep', 0.01)
sim_params.set_param('spaceWidth', 1)
sim_params.set_param('maxSpaceBiomass', 10)
sim_params.set_param('minSpaceBiomass', 1e-11)
sim_params.set_param('writeMediaLog', True)
```

Run the simulation

6. Create a COMETS simulation by instantiating the *comets* class with the layout (test_tube) and parameters as input.

```
experiment = c.comets(test_tube, sim_params)
```

7. Run the simulation.

```
experiment.run()
```
   ? TROUBLESHOOTING

Analyze the results

8. Plot the biomass growth over time.

```
ax = experiment.total_biomass.plot(x = 'cycle')
ax.set_ylabel("Biomass (gr.)")
```

9. Plot the composition of the media. In this case, we will limit the plot to those components that are not added to the layout in unlimited amounts ("static" compounds, e.g. ammonia, phosphate, water, etc in this simulation). To do this, we will plot only the compounds with a concentration lower than 900mM (the constant concentration of "static" components is set to 1M).

```
media = experiment.media.copy()
media = media[media.conc_mmol<900]

fig, ax = plt.subplots()
media.groupby('metabolite').plot(x='cycle', ax =ax, y='conc_mmol')
ax.legend(('acetate','ethanol', 'formate', 'glucose'))
ax.set_ylabel("Concentration (mmol)")
```

The results are shown in Supplementary Fig. 2.

## Case Study #10: Competition assay and competitive exclusion in serial transfers [P]
● **TIMING** setup: 30 min - 1hr, simulation: 3-10 min

Competition experiments are frequently performed in the laboratory to assay, for example, the fitness of a mutant in competition to the wild-type. Here, we simulate one such experiment involving *E. coli* and a nonessential but deleterious mutation involving the deletion of the triose phosphate isomerase reaction from glycolysis.

Prepare the models and create a mutant

1. Load the *E. coli* "core" model and create the mutant in triose phosphate isomerase by setting both upper and lower bounds to zero. We will add both models to our test_tube layout.

```
# Start by loading required packages, including the COMETS toolbox
import cometspy as c
import cobra
import cobra.test
import pandas as pd
import matplotlib.pyplot as plt

# load the models and perform the mutation
wt = c.model(cobra.test.create_test_model("ecoli"))
wt.id = 'wt'
mut = c.model(cobra.test.create_test_model("ecoli"))
mut.change_bounds('TPI', 0,0)
mut.id = 'TPI_KO'

# set its initial biomass, 5e-6 gr at coordinate [0,0]
wt.initial_pop = [0, 0, 5e-8]
mut.initial_pop = [0, 0, 5e-8]
```
▲ **CRITICAL STEP** It is critical to set the initial biomass.

Create layout, add models and set up media composition

2. Create an empty layout ("test_tube") and set the initial nutrient supply.

```
# create an empty layout
test_tube = c.layout()

# add the models to the test tube
test_tube.add_model(wt)
test_tube.add_model(mut)
```
▲ **CRITICAL STEP** The models must be added to the layout.

3. Set the media composition by adding glucose and the inorganic nutrients required for this model (ammonia, phosphate) and oxygen. These inorganic nutrients will be considered as "static" by the simulation, with a value of 1000 that never depletes. Considering metabolites as "static" is the way COMETS has to simulate an unlimited supply of metabolites.

```
# Add glucose to the media
test_tube.set_specific_metabolite('glc__D_e', 0.01)

# Add typical trace metabolites and oxygen coli as static
trace_metabolites = ['ca2_e', 'cl_e', 'cobalt2_e', 'cu2_e','fe2_e','fe3_e', 'h_e', 'k_e', 'h2o_e', 'mg2_e',
'mn2_e', 'mobd_e', 'na1_e', 'ni2_e', 'nh4_e', 'o2_e', 'pi_e', 'so4_e', 'zn2_e']

for i in trace_metabolites:
    test_tube.set_specific_metabolite(i, 1000)
    test_tube.set_specific_static(i, 1000)
```
▲ **CRITICAL STEP** If the media is not defined correctly, models will not be able to grow.

Set up simulation parameters

4. Create a parameters object and modify needed parameters - in this case only the number of cycles the simulation runs.

```
comp_params = c.params()
comp_params.set_param('maxCycles', 240)
```

Run the simulation

5. Create the comets object using the above created layout and parameters, and run the competition assay.

```
# Create comets object
comp_assay = c.comets(test_tube, comp_params)

# Run simulation
comp_assay.run()
```
? TROUBLESHOOTING

### Visualize the results

6. Plot the biomasses of these two genotypes in coculture.

```
biomass = comp_assay.total_biomass
biomass['t'] = biomass['cycle'] * comp_assay.parameters.all_params['timeStep']

myplot = biomass.drop(columns=['cycle']).plot(x = 't')
myplot.set_ylabel("Biomass (gr.)")
```

### Simulating serial transfers

Using COMETS we can also simulate a serial transfer competition between these two mutants.

7. We will just modify the parameters, increasing the total simulation time and including batch transfers of 1% every 24h, but we will use the same test_tube layout as before.

```
serial_params = c.params()
# 25 transfers of 240 cycles each, i.e. 24hr.
serial_params.set_param('maxCycles', 240*25)
serial_params.set_param('batchDilution', True)
serial_params.set_param('dilFactor', 0.01)
serial_params.set_param('dilTime', 24)
```
▲ **CRITICAL STEP** Setting these parameters to correct values determines the outcome.

### Run the COMETS simulation

8. Run the simulation

```
serial_expt = c.comets(test_tube, serial_params)
serial_expt.run()
```
? **TROUBLESHOOTING**

### Plot the biomass

9. Plot the biomasses of the two species during the experiment

```
biomass = serial_expt.total_biomass
biomass['transfer'] = biomass['cycle'] *comp_assay.parameters.all_params['timeStep']/24
```

```
myplot = biomass.drop(columns=['cycle']).plot(x = 'transfer')
myplot.set_ylabel("Biomass (gr.)")
```

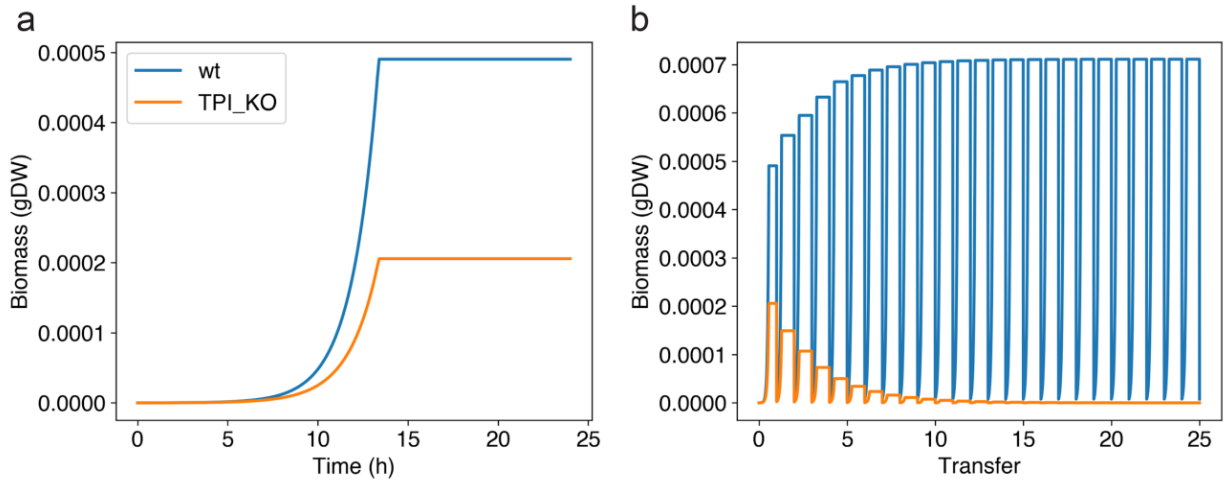

**Figure 8. Competition assay and competitive exclusion.** Results from two experiments simulated with the python toolbox. In both experiments, growth is assayed in aerobic glucose minimal media (10 mM) in a volume of 1uL. **a**) Competition assay between wild-type E. coli and a mutant in which the glycolytic enzyme triose phosphate isomerase has been knocked out. **b**) Serial transfers were performed each 24h using the same setting as in panel **a**.

## Case Study #11: Simulating cross-feeding in a chemostat [P] ● TIMING setup: 30 min - 1hr, simulation: 5-15 min

COMETS provides the functionality to run simulations in a chemostat. Here, we use the python toolbox to generate a chemostat simulation in two ways: 1) by manually assigning the initial metabolite concentrations, the metabolite inflow rate, metabolite dilution rate, and model death rate, or 2) by using a helper function that performs these steps simultaneously.

Here we are going to simulate a chemostat with lactose as the sole carbon resource and two strains of *E. coli*: one which is deficient in the ability to uptake lactose, and one which is deficient in the ability to metabolize galactose. We will use the ijo1366 model provided as part of COBRAPy.

Prepare the models

1. Load two copies of the *E. coli* ijo1366 model and knockout the relevant reactions in each of them. In the first model, we will knock out galE to prevent metabolism of galactose, which will cause galactose to be secreted during metabolism of lactose. In the second model, because lactose transport to the periplasm can be accomplished with multiple genes, we will knock out the reaction instead of all the genes individually.

```
# Start by loading the dependencies
import cobra
import cobra.test
import sys
import cometspy as c

# Load two instances of E. coli iJO1366 model
E_no_galE = cobra.test.create_test_model("ecoli")
E_no_LCTStex = E_no_galE.copy() # this model will have lactose uptake KO'd
# Perform galE KO in the first model
E_no_galE.genes.b0759.knock_out()
# Perform LCTStex reaction KO in the second model
E_no_LCTStex.reactions.LCTStex.knock_out()
```

2. Test that the knockouts perform as expected by trying to grow them in media containing lactose and galactose. We do this using COBRAPy.

```
# Create the medium for COBRAPy growth tests
medium = E_no_galE.medium
medium["EX_glc__D_e"] = 0.
medium["EX_lcts_e"] = 1.
medium["EX_gal_e"] = 1.
print(medium)
```

Expected output:

{'EX_ca2_e': 1000.0, 'EX_cbl1_e': 0.01, 'EX_cl_e': 1000.0, 'EX_co2_e': 1000.0, 'EX_cobalt2_e': 1000.0, 'EX_cu2_e': 1000.0, 'EX_fe2_e': 1000.0, 'EX_fe3_e': 1000.0, 'EX_glc__D_e': 0.0, 'EX_h_e': 1000.0, 'EX_h2o_e': 1000.0, 'EX_k_e': 1000.0, 'EX_mg2_e': 1000.0, 'EX_mn2_e': 1000.0, 'EX_mobd_e': 1000.0, 'EX_na1_e': 1000.0, 'EX_nh4_e': 1000.0, 'EX_ni2_e': 1000.0, 'EX_o2_e': 1000.0, 'EX_pi_e': 1000.0, 'EX_sel_e': 1000.0, 'EX_slnt_e': 1000.0, 'EX_so4_e': 1000.0, 'EX_tungs_e': 1000.0, 'EX_zn2_e': 1000.0, 'EX_lcts_e': 1.0, 'EX_gal_e': 1.0}

# Apply the created medium to both models
E_no_galE.medium = medium
E_no_LCTStex.medium = medium
# examine growth and uptake in the galE knockout shows galactose is excreted

E_no_galE.summary()# prints a cobrapy table, showing a growth rate of 0.086
E_no_LCTStex.summary() # prints a cobrapy table with a growth rate of 0.085

> Now that we are satisfied we have made our models correctly, we can set up a COMETS simulation. Let's intend that the medium above is the reservoir medium (except that we will remove galactose first), and that the input rate and output rate are 10% per hour.

3. Generate the COMETS models and set their initial population size

```python
# chemostat parameters
initial_pop = 1.e-3 # gDW
```

```python
# Right now both models have the same ID, which will confuse COMETS, so we must give them unique IDs.
E_no_galE.id = "galE_KO"
E_no_LCTStex.id = "LCTStex_KO"

# Create galE model
galE_comets = c.model(E_no_galE)
galE_comets.initial_pop = [0,0,initial_pop] # x, y, gDW

# Create no_LCTStex model
lcts_comets = c.model(E_no_LCTStex)
lcts_comets.initial_pop = [0,0,initial_pop] # x, y, gDW
```

```python
# Set the exchange lower bounds to -1000 so that COMETS can alter these based upon media concentrations.
galE_comets.reactions.loc[galE_comets.reactions.EXCH, "LB"] = -1000
lcts_comets.reactions.loc[lcts_comets.reactions.EXCH, "LB"] = -1000
```

Manually set up a chemostat

We are first going to use the manual method for making a chemostat.

4. Create the layout for the chemostat by providing models

```python
layout = c.layout([galE_comets, lcts_comets])
```

5. Create the media composition for the chemostat. Recall that while cobrapy media are set using exchange reaction IDs, COMETS media are set using metabolite IDs. We can easily take care of this difference with a dictionary comprehension. Here we do that, then generate a layout, and add the media components to that layout.

```python
# re-write media (while removing galactose) and add it to layout
comets_media = {key[3:]: value for key, value in medium.items()
        if key != "EX_gal_e"}

for key, value in comets_media.items():
    layout.set_specific_metabolite(key, value)
```

6. Set the media input into the chemostat from the reservoir. The input of fresh media is done using media_refresh. Metabolites with a media_refresh value are replenished at the specified amount per-hour. Since we are diluting at 0.1 per hour, we multiply the reservoir concentration by this rate.

```
# Set the dilution rate
dilution_rate = 0.1  # / hr

# Apply the dilution rate to all metabolites
for key, value in comets_media.items():
    layout.set_specific_refresh(key, value * dilution_rate)
```
▲ **CRITICAL STEP** It is critical to set up correctly the dilution rate of the chemostat.

7. Set up the chemostat outflow.

This is done using the parameters metaboliteDilutionRate and deathRate. These should be set equal to the desired dilution rate. Here we generate a parameters object and set these values.

```
params = c.params()
params.set_param("deathRate", dilution_rate)
params.set_param("metaboliteDilutionRate", dilution_rate)
```
▲ **CRITICAL STEP** It is critical to set up correctly the outflow from the chemostat.

Set up additional simulation parameters

8. Adjust time step, maximum biomass allowed per simulation cell, and duration of the simulation.

```
params.set_param("timeStep", 0.1)        # hours
params.set_param("maxSpaceBiomass", 10.)   # gDW
params.set_param("maxCycles", 2000) # duration of simulation in time steps
```

9. Adjust parameters to keep track of specific metabolites in the simulation: lactose and galactose. We do this using the specificMedia log, and choosing the metabolites with a comma-separated string with no spaces.

```
params.set_param("writeSpecificMediaLog", True)
params.set_param("specificMediaLogRate", 1)     # time steps
params.set_param("specificMedia", "lcts_e,gal_e") # metabolites to track
```

Run the simulation

10. Create a COMETS simulation and run it

```
sim = c.comets(layout, params)
sim.run()
```
? TROUBLESHOOTING

*Visualize the results*

11. Plot the results. Note how we specify the axes, otherwise "cycle", "x", and "y" will be assumed to be state variables (Fig. 10).

```
sim.total_biomass.plot(x = "cycle", logy = True)
sim.specific_media.plot(x = "cycle",y = ["lcts_e","gal_e"])
```

*Automatic setting of a chemostat*

The above code required setting chemostat parameters in multiple places. We offer this functionality so that researchers can create complex setups that may, for example, have different initial concentrations than reservoir concentrations, and different inflow rates than outflow rates. However, we expect most chemostat simulations will function like above, where a single dilution parameter dictates the behavior of the system. For this typical use-case, a helper function is available to generate a layout and parameters objects with the correct setup.

12. Set up a chemostat layout and parameters using the chemostat function imported from the utilities class.

```
from cometspy.utils import chemostat
layout, params = chemostat([galE_comets, lcts_comets], comets_media, dilution_rate)
```

13. We can still adjust the parameters as desired.

```
params.set_param("timeStep", 0.1) # hours
params.set_param("maxSpaceBiomass", 10.) # gDW
params.set_param("maxCycles", 500) # duration of simulation in time steps
params.set_param("writeSpecificMediaLog", True)
params.set_param("specificMediaLogRate", 1) # time steps
params.set_param("specificMedia", "lcts_e,gal_e") # metabolites to track
```

14. Run the simulation

```
sim = c.comets(layout, params)
sim.run()
sim.total_biomass.plot(x = "cycle")
```
? TROUBLESHOOTING

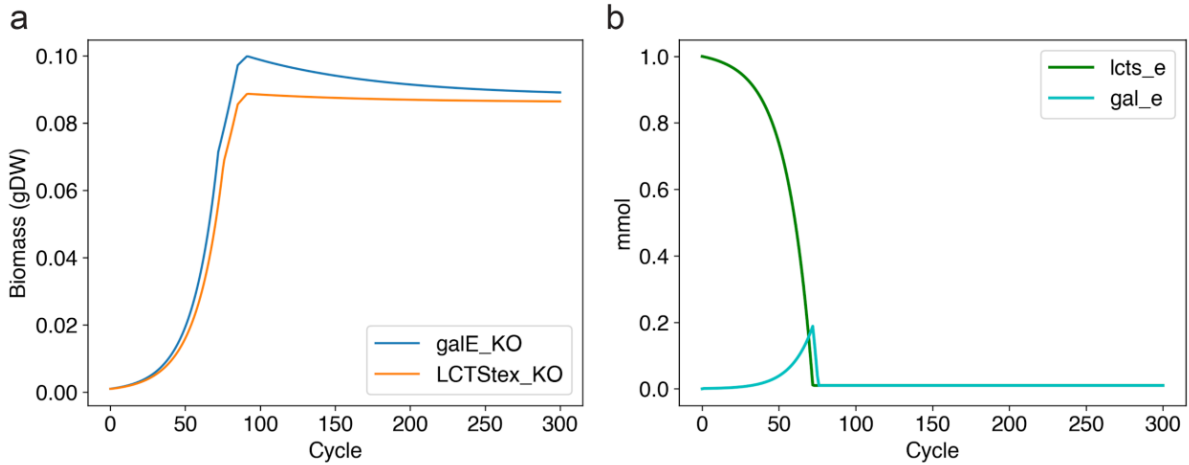

**Figure 9. Chemostat simulation.** Results from a chemostat simulation, prepared with the python toolbox, in which one strain unable to uptake lactose (LCTStex_KO) crossfeeds galactose from a different strain unable to metabolize galactose (galE_KO). The media environment was a constant supply of lactose (lcts_e), ammonia, and trace nutrients. Galactose (gal_e) was not supplied but entered the media as galE_KO grew. **a**) Biomass of the strains over time. **b**) Amounts of the two key metabolites over time. Note that it is typical for limiting nutrients to have near-zero concentrations in a chemostat.

*Case Study #12: Simulating evolutionary processes in microbial populations* [P] ●
**TIMING** setup: 30 min - 1hr, simulation: 5min-indefinite

COMETS is able to perform simulations that include the appearance of mutants containing reaction deletions and additions. In this small example, we will perform a serial transfer experiment starting with a clonal *Escherichia coli* population, and simulate the random appearance of reaction deletion mutants. (The addition of reactions is also available through the parameter addRate, bud models need to be prepared accordingly, please see documentation online).

Load the model

1. Import the necessary libraries and load the *E. coli* model.

```python
import cometspy as c
import cobra.test
import os
import pandas as pd
import matplotlib.pyplot as plt

# load model
wt = cobra.test.create_test_model("ecoli")
```

2. Remove the bounds for all exchange reactions in the model to allow them to be controlled dynamically by COMETS.

```python
# Remove bounds from exchange reactions
for i in wt.reactions:
    if 'EX_' in i.id:
        i.lower_bound =-1000.0
```

Set up the layout

3. Create a well-mixed environment with a glucose minimal media. Here, we use the add_typical_trace_metabolites method to add trace metabolites (ions, metals etc) in unlimited amounts (static flag).

```python
# generate layout
test_tube = c.layout()
test_tube.set_specific_metabolite('glc__D_e', 0.0001)
test_tube.add_typical_trace_metabolites(amount=1000)

# add model
wt = c.model(wt)
wt.initial_pop = [0, 0, 1e-7]
test_tube.add_model(wt)
```

Set up simulation parameters

4. Create a params object, and modify the needed parameters. The simulation in this example simulation consists of 10 days of experiment, with a 1:2 transfer every 3h. The mutation rate will be $10^{-7}$ deletion events per reaction and generation. The cellSize parameter sets the amount of biomass that appears when a mutant occurs (i.e., one mutant cell appears).

```python
# Load parameters and layout from file
evo_params = c.params()

# Set relevant parameters
evo_params.set_param('timeStep', 0.1)          # hours
evo_params.set_param('maxCycles', 2400)
evo_params.set_param('batchDilution', True)
evo_params.set_param('dilFactor', 0.5)
evo_params.set_param('dilTime', 3)             # hours
evo_params.set_param('evolution', True)
evo_params.set_param('mutRate', 1e-8)          # /generation /reaction
evo_params.set_param('cellSize', 1e-10)
evo_params.set_param('minSpaceBiomass', 1e-11)
evo_params.set_param('BiomassLogRate', 1)
```

▲ **CRITICAL STEP** Simulations including evolution are very sensitive to parameters such as mutRate (and addRate if additions are modeled), cellSize (the dry weight of one cell in grams).

Run the simulation

5. Create the COMETS object using the above layout and parameters, and run the simulation.

```python
# create comets object from the loaded parameters and layout
evo_simulation = c.comets(test_tube, evo_params)

# run comets simulation
evo_simulation.run()
```

? TROUBLESHOOTING

Visualize the results

6. Plot the population dynamics of all species over time (color coded) using standard Python plotting tools.

```python
fig, ax = plt.subplots(figsize=(15, 5))
for key, grp in evo_simulation.biomass.groupby(['species']):
    ax = grp.plot(ax=ax, kind='line', x='cycle', y='biomass')
ax.get_legend().remove()
plt.yscale('log')
plt.ylabel("Biomass (gr.)")
```

7. In order to analyze the results, it is also helpful to visualize the genotypes data frame, which contains all the mutants that ever appeared during the simulation.

evo_simulation.genotypes

Expected results:

A data frame containing three columns: The ancestor, the mutation (reaction number in the model), and the name of the resulting genotype, which is assigned as a random hash when the mutant is born. Using this data together with the population dynamics (plotted above), one can reconstruct the entire phylogeny of a simulation, and know exactly what mutations each individual contains.

| | Ancestor | Mutation | Species |
|---|---|---|---|
| 0 | NO_ANCESTOR | NO_MUT | iJO1366.cmd |
| 1 | iJO1366.cmd | del_1618 | ff37d064-b32a-4a4f-9990-adee15665d20 |
| 3 | iJO1366.cmd | del_524 | 41cfb561-75b1-4941-84bc-992de5b45a3c |
| 4 | iJO1366.cmd | del_594 | 8a41bfd2-1c58-4d8c-b08d-064f031706aa |
| ... | ... | ... | ... |
| 92 | iJO1366.cmd | del_1338 | d519772d-ebe5-4697-a4a6-5ecde288bd11 |
| 93 | iJO1366.cmd | del_1026 | 25bb819f-5b79-459c-9fdf-921b9a018b4 |
| 94 | iJO1366.cmd | del_2556 | c9435324-61c8-416b-a1cb-c29dc1c76c21 |
| 95 | iJO1366.cmd | del_1757 | 73450a19-50ca-4b53-ad2c-ed32573c9270 |
| 96 | iJO1366.cmd | del_1772 | 691548f9-5765-4259-bd89-1f6b4e94b69 |

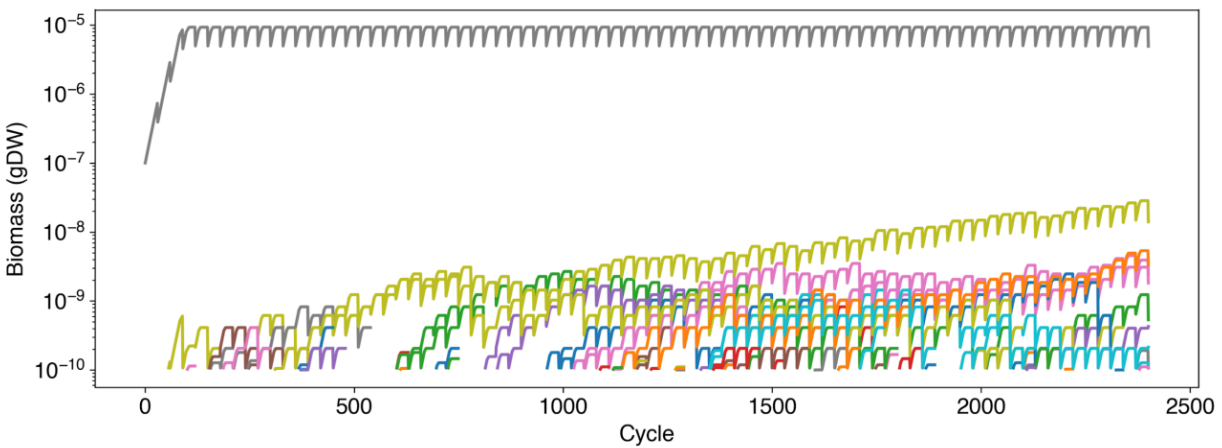

**Figure 10. Simulation of evolutionary processes.** An *Escherichia coli* model was seeded in 1uL of glucose minimal media (0.1mM) and transferred each 3hr. To fresh media using a dilution factor of 1:2 during 10 days. Mutations (reaction knock-outs) were allowed to happen in this population at a rate of $10^{-8}$ knock-outs appearing per gene and generation. The cyan line represents the ancestor, which remains at high density, and other colors are used to represent different mutations that appear, persist during variable periods and extinguish stochastically.

## *Case Study #13: Simulating the sequence of mutations involved in an evolutionary innovation* [P] ● **TIMING** setup: 30 min - 1hr, simulation: 20-40 min

Here we will use COMETS to simulate the sequence of mutations involved in an evolutionary process. Specifically, we will use as a case study one of the most well-known evolution experiments, the *E. coli*

Long Term evolution experiment. After about 33,000 generations, a large population expansion was observed in one of the replicates (Ara-3) of the *E. coli* long term evolution experiment[103]. This population expansion is associated with two key mutations that together enable the evolution of strong aerobic citrate use (Cit++ phenotype). The first mutation (occurring roughly 31,000 generations) caused the aerobic expression of the citT transporter, producing a weak citrate growth phenotype (Cit+)[103]. A subsequent mutation (occurring roughly 33,000 generations in) caused high-level, constitutive expression of dctA, a proton-driven dicarboxylic acid transporter[103,104]. Because these two mutations introduce known reactions into the *E. coli* metabolic network, we can simulate them using COMETS.

Set up the layout

1. Load the necessary libraries

```
import cometspy as c
import matplotlib as plt
import cobra.test
import cobra
import pandas as pd
import numpy as np
```

2. Set up a layout, which is in this case a flask containing DM25 medium. We model the flask as a single well mixed compartment.

```
# Create empty test tube
flask = c.layout()

# Set up DM25 media
flask.add_typical_trace_metabolites()
flask.set_specific_metabolite('glc__D_e', 0.000139)
flask.set_specific_metabolite('cit_e', 0.0017)
```

▲ **CRITICAL STEP** It is critical to set the nutrient amounts to the correct values.

Construct the necessary genotypes

3. Load the E. coli iJO1366 model

```python
# Load the E. coli iJO1366 model
model  = cobra.test.create_test_model('ecoli')

# Set exchange reaction lower bounds to -1000 to allow them being controlled by COMETS
for i in model.reactions:
    if 'EX_' in i.id:
        i.lower_bound =-1000.0
```

4. Generate the genotypes using COBRAPy functionality.

   Unlike the LTEE ancestral strain REL606 (and *E. coli* in general), which possess the necessary genes for citrate utilization but do not express them in aerobic conditions, iJO1366 is able to use both citrate and succinate as these reactions are unbounded by default. Thus, the ancestral phenotype can be recreated by knocking out three reactions CITt7pp (citT), SUCCt2_2pp (dctA) and SUCCt2_3pp (dcuA or dcuB).

```python
# SUCCt2_3pp reaction is inactive in all genotypes; change its bounds to 0
model.reactions.SUCCt2_3pp.upper_bound=0.0

# copy the model to create the genotype with both citT and dctA available
CitTdctA = model.copy()
CitTdctA.id = 'Cit++'

# now make dctA unavailable to create the mutant only expressing citT
model.reactions.SUCCt2_2pp.upper_bound =0.0
CitT = model.copy()
CitT.id = 'Cit+'

# finally, make citT unavailable to create the wild-type genotype
model.reactions.CITt7pp.upper_bound =0.0
WT = model.copy()
WT.id= 'Ancestor'
```

5. Using the above COBRAPy models, generate COMETS models and set initial population sizes. We will start with a population of 100 cells of the WT genotype

```python
# Generate comets model for the WT
wt = c.model(WT)
wt.initial_pop = [0, 0, 3.9e-11]

citT = c.model(CitT)
citT.initial_pop = [0, 0, 0] # not present at start

citTdctA = c.model(CitTdctA)
citTdctA.initial_pop = [0, 0, 0] # not present at start
```

6. Set the Vmax for oxygen, nitrogen and protons so that model growth rate is carbon limited

```
# set vmax for WT
wt.change_vmax('EX_nh4_e',1000)
wt.change_vmax('EX_o2_e',1000)
wt.change_vmax('EX_h_e',1000)

# set vmax for citT
citT.change_vmax('EX_nh4_e',1000)
citT.change_vmax('EX_o2_e',1000)
citT.change_vmax('EX_h_e',1000)

citTdctA.change_vmax('EX_nh4_e',1000)
citTdctA.change_vmax('EX_o2_e',1000)
citTdctA.change_vmax('EX_h_e',1000)
```

7. Add the models to the simulation

```
flask.add_model(wt)
flask.add_model(citT)
flask.add_model(citTdctA)
```

Set the experiment conditions (simulation parameters)

8. Create a parameter list with default values and change the relevant ones. We use 1 hr as the COMETS timestep to speed up the simulation. Shortening this to the more commonly used 0.1 hr does not substantially affect the final result in this case, but it does significantly increase the simulation timing.

```
# Setting paramaters for the simulation
b_params = c.params()
b_params.set_param('timeStep', 1.0)
b_params.set_param('deathRate', 0.01)
b_params.set_param('batchDilution', True)
b_params.set_param('dilTime', 24)
b_params.set_param('dilFactor', 100)

one_cell = 3.9e-13 # gr. dry weight of an E. coli cell
b_params.set_param('cellSize', one_cell)  b_params.set_param('minSpaceBiomass', 3.8e-13)  # <1cell
```

Run the simulation

We will divide our simulation in three actual COMETS runs. We will start the simulation at generation 25,000 and run for around 6000 generations. At roughly generation 31,000, we introduce the CitT genotype and run for around 2000 Generations. Finally, roughly at generation 33,000 we introduce the CitTdctA Genotype and run for a final 6000 generations. For each run, we will input the final biomass composition of the previous run. Each sub-simulation stores the biomass data in a separate dataframe that we will then join together for analysis.

9. Run the first portion of the simulation, growth of wt in DM25 with daily transfers, for 6000 generations.

```
# number of simulation cycles per day
cycles_per_day = 24.0/b_params.all_params['timeStep']

sim = c.comets(flask, b_params)
sim.parameters.set_param('maxCycles', int(900*cycles_per_day))
sim.run()
# Create dataframe with population dynamics from phase 1
phase_1 = pd.DataFrame({'Ancestor' : sim.total_biomass.Ancestor/one_cell,
            'CitT' : sim.total_biomass['Cit+']/one_cell,
            'CitTdctA' : sim.total_biomass['Cit++']/one_cell,
            'Generations' : 6.67*(sim.total_biomass.cycle+1)/cycles_per_day + 25000})
```
? TROUBLESHOOTING

10. At this point, introduce the CitT genotype and run for another 2000 generations

```
# The initial population size for the wild type will be its final population size from the previous phase.
sim.layout.models[0].initial_pop = [0, 0, float(sim.total_biomass.Ancestor.tail(1))]

# Introduce new genotypes 100 cells at a time to avoid the risk of them drifting to extinction
sim.layout.models[1].initial_pop = [0, 0, one_cell*100]

# Running build_initial_pop is essential when we change the initial population sizes of models that are already loaded into a layout
sim.layout.build_initial_pop()

# Change the cycles
sim.parameters.set_param('maxCycles', int(300*cycles_per_day))
sim.run()
phase_2 = pd.DataFrame({'Ancestor' : sim.total_biomass.Ancestor/one_cell,
            'CitT' : sim.total_biomass['Cit+']/one_cell,
            'CitTdctA' : sim.total_biomass['Cit++']/one_cell,
            'Generations' : 6.67*(sim.total_biomass.cycle)/cycles_per_day + max(phase_1.Generations)})
```

11. At roughly Generation 33,000 we introduce the CitTdctA Genotype and run for a final 6000 generations.

```
sim.layout.models[0].initial_pop = [0, 0, float(sim.total_biomass.Ancestor.tail(1))]
sim.layout.models[1].initial_pop = [0, 0, float(sim.total_biomass['Cit+'].tail(1))]
sim.layout.models[2].initial_pop = [0, 0, 3.9e-11]
sim.layout.build_initial_pop()
```

```
sim.parameters.set_param('maxCycles', int(900*cycles_per_day))
sim.run()
phase_3 = pd.DataFrame({'Ancestor' : sim.total_biomass.Ancestor/(3.9e-13),
            'CitT' : sim.total_biomass['Cit+']/(3.9e-13),
            'CitTdctA' : sim.total_biomass['Cit++']/(3.9e-13),
            'Generations' : 6.67*(sim.total_biomass.cycle)/cycles_per_day +
max(phase_2.Generations) })
```

Visualize the results

12. Group all the results from the three runs together and plot the stationary phase population size through time.

```
# Remove the final timepoint from each phase and merge them together to plot
phase_1.drop(phase_1.tail(1).index, inplace=True)
phase_2.drop(phase_2.tail(1).index, inplace=True)
phase_3.drop(phase_3.tail(1).index, inplace=True)
final_df = pd.concat([phase_1,phase_2,phase_3])
final_df.reindex()

# Subset to only plot final timepoint within each transfer and convet generation into thousands
final_df = final_df[np.round((final_df.Generations - 25000) % 6.67,3) == 6.67]
final_df.Generations = final_df.Generations/1000

# Plot
fig = final_df.plot(x='Generations')
fig.set_xlabel('Generation (in thousands)')
fig.set_ylabel('Population Size / ml')
fig.set_yticks([0,1e+08,2e+08,3e+08])
fig.set_yticklabels(['0','1 x $10^8$','2 x $10^8$','3 x $10^8$
```

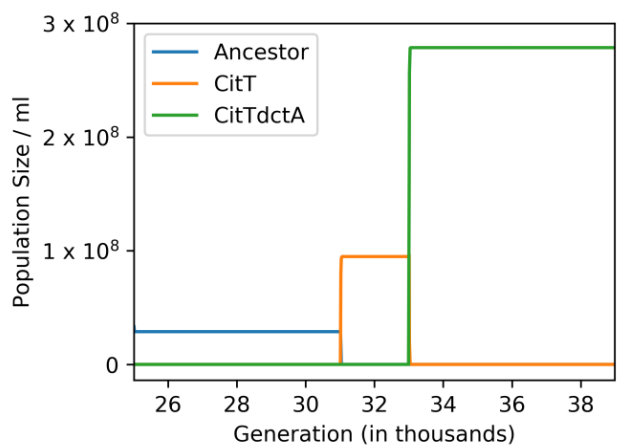

**Figure 11. Evolution of citrate utilization.** Mimicking the classic Lenski long-term evolution experiment, an *Escherichia coli* model (Ancestor) was simulated for many thousands of generations. At generation ~31000, a mutant was added to the simulation which was capable of growing on citrate (CitT), and which outcompeted the ancestor. At generation ~33000, a double mutant was introduced (CitTdctA), which outcompeted citT.

## Case Study #14: Soil-air interface simulation [P] ● TIMING setup: 30 min - 1hr, simulation: 3-6 hr

Using the functionality of COMETS, one can design simulations which go beyond in silico corollaries of lab experiments to make predictions for environments mimicking natural ecosystems, which is a necessary step for understanding natural ecology from first principles. In this example, we consider a spatially structured simulation of a soil environment. We use source-and-sink functions to model how a root provides organic acids to the environment while removing ammonia (Huang et al 2014 Botany). While we restrict the root functionality to a source/sink, one could use functionality demonstrated above, for example extracellular enzymes, to generate feedback loops between microbe-produced metabolites and root exudation. We use fixed metabolite concentrations to mimic the largely unchanging air interface above a root, which generates an oxygen gradient. Additionally, since soil is characterized by strong spatial structure with many impenetrable barriers that localized interactions[105], we place "rock" barriers throughout the simulation area (Fig 12). A more complex simulation could use varying diffusion constants for metabolites, as described in a previous protocol (Virtual Petri Dish). Similarly, we use standard diffusion of biomass, but this could be changed to use pushing force or non-linear diffusion for a potentially increased realism.

Prepare models

1. Import the necessary libraries

```
import cobra
import cobra.test # for the ijo1366 model
import sys
import copy
import numpy as np
from matplotlib import pyplot as plt
import cometspy as c
```

2. We will use three well-curated soil bacteria, Pseudomonas putida (model iJN747), Bacillus subtilis (model iYO844), and Methanosarcina barkeri (model iAF629), which are available in the BIGG database [35] and also provided in a subdirectory called "models" from where this ipynb file is located. Upon loading, the biomass functions of these models had positive lower bounds, which we set to zero.

```
iJN = cobra.io.read_sbml_model('./models/iJN746.xml')
iJN.reactions.get_by_id('BIOMASS_KT_TEMP').lower_bound = 0
iJN.reactions.get_by_id('BIOMASS_KT_TEMP').upper_bound = 1000

iAF = cobra.io.read_sbml_model('./models/iAF692.xml')
iAF.reactions.get_by_id('BIOMASS_Mb_30').lower_bound = 0
iAF.reactions.get_by_id('BIOMASS_Mb_30').upper_bound = 1000

iYO = cobra.io.read_sbml_model('./models/iYO844.xml')
iYO.reactions.get_by_id('BIOMASS_BS_10').lower_bound = 0
iYO.reactions.get_by_id('BIOMASS_BS_10').upper_bound = 1000
```

3. Convert these into COMETS models. We also use some helper functions to a) ensure the COMETS model does not think "sink" reactions are exchange reactions (as they are unbalanced in Cobra models, and therefore appear similar to exchanges), and b) open all exchange reaction bounds, to make sure that COMETS is in control of the media composition.

```
iJN_comets = c.model(iJN)
iJN_comets.ensure_sinks_are_not_exchanges()
iJN_comets.open_exchanges()

iAF_comets = c.model(iAF)
iAF_comets.ensure_sinks_are_not_exchanges()
iAF_comets.open_exchanges()

iYO_comets = c.model(iYO)
iYO_comets.ensure_sinks_are_not_exchanges()
iYO_comets.open_exchanges()
```

▲ CRITICAL STEP Often Cobra models are saved with media definitions. If we do not open exchanges, these media definitions will overrule the media we specify in the layout.

Set up the layout

Our simulated world will be a 100x100 box lattice. The left-hand side, where x = 0, will be the root. The top, where y = 0, will be the air. The biomass and the rocks will be distributed elsewhere.

4. To ensure that we don't attempt to place biomass where rocks are placed, we first determine the rock locations. Specifically, we will create 70 rocks which are impervious to any biomass or metabolite. These rocks will have an average size of 15 boxes. To pick these locations, we use the helper function grow_rocks:

```python
from cometspy.utils import grow_rocks, pick_random_locations

grid_size = 30 # 100
n_rocks = 50 # 70
rock_locs = grow_rocks(n = n_rocks, xrange = [1,grid_size-1],yrange = [1,grid_size-1],mean_size = 5)
```

5. Each species will have biomass seeded at 60 different locations, with no overlap. We will use the helper function pick_random_locations for this, which is useful as it can take in a previously-generated list of tuples of x-y locations as "forbidden" locations, such as the rock locations.

```python
# First, make a copy of the rock_locs so we don't accidentally alter it, and call this copy forbidden_locs.
forbidden_locs = copy.deepcopy(rock_locs)

# Pick the random locations for each species, adding these locations to the forbidden locs as we go so as
to prevent overlap.
founders_per_species = 60
iJN_locs = pick_random_locations(n = founders_per_species,
                    xrange = [1,grid_size], yrange = [1,grid_size],
                    forbidden_locs = forbidden_locs)
forbidden_locs.extend(iJN_locs)
iYO_locs = pick_random_locations(founders_per_species, [1,grid_size],[1,grid_size], forbidden_locs)
forbidden_locs.extend(iYO_locs)
iAF_locs = pick_random_locations(founders_per_species, [1,grid_size],[1,grid_size], forbidden_locs)
forbidden_locs.extend(iAF_locs)
```
▲ **CRITICAL STEP** Using the forbidden_locs argument is critical to ensuring biomass doesn't overlap initially, or overlap with rocks.

6. Optional: visually inspect what locations were chosen by making an image with matplotlib:

```python
initial_image = np.zeros((grid_size,grid_size,3))
for rock in rock_locs:
    initial_image[rock[1]-1,rock[0]-1,0:3] = 0.5
for loc in iJN_locs:
    initial_image[loc[1]-1,loc[0]-1,0] = 1
for loc in iYO_locs:
    initial_image[loc[1]-1,loc[0]-1,1] = 1
for loc in iAF_locs:
    initial_image[loc[1]-1,loc[0]-1,2] = 1
# plt.imshow(initial_image)
```

The plot should look like grey regions (rocks) on a black background with dots of color designating founder locations. Right now, the plotting function is commented out. See Fig 13 ("Initial COMETS state").

7. Create the layout and set the dimensions. Then, we add the rock barriers to the layout.

```
layout = c.layout()
layout.grid = [grid_size,grid_size]
layout.add_barriers(rock_locs)
```

8. We set the initial population for each species by using a python list comprehension. Note that locations are properties of the model.

```
iJN_comets.initial_pop = [[loc[0],loc[1],1e-8] for loc in iJN_locs]
iYO_comets.initial_pop = [[loc[0],loc[1],1e-8] for loc in iYO_locs]
iAF_comets.initial_pop = [[loc[0],loc[1],1e-8] for loc in iAF_locs]
```

9. After setting initial populations, the models are finished, so we add them to the layout.

```
layout.add_model(iJN_comets)
layout.add_model(iYO_comets)
layout.add_model(iAF_comets)
```

10. We want some metabolites available initially. These include all the typical trace nutrients needed, so we start with the helper function add_typical_trace_metabolites. After that, however, we want oxygen to mostly diffuse from the air, so we set that value lower. We also add a few other trace metabolites homogeneously throughout the environment that were not added with the helper function.

```
layout.add_typical_trace_metabolites()
layout.set_specific_metabolite('o2_e',0.00001)
layout.set_specific_metabolite('hco3_e',1000)
layout.set_specific_metabolite('co2_e',1000)
layout.set_specific_metabolite('h2_e',1000)
layout.set_specific_metabolite('so3_e',1000)
layout.set_specific_metabolite('nh4_e',0.000001)

layout.set_specific_metabolite('glc__D_e',0.0000001)
layout.set_specific_metabolite('meoh_e',0.00000001)
layout.set_specific_metabolite('cys__L_e',0.0000001)
layout.set_specific_metabolite('4abz_e',0.0000001)
layout.set_specific_metabolite('nac_e',0.00000001)
```

11. To make the air layer, we set its concentration to "static" at the top, which will keep that metabolite at a fixed value and therefore act as a source. We also set a static level of zero oxygen at the "bottom," to mimic continuous downwards diffusion. In a similar fashion, we set a static sink of ammonium where the "root" is.

```python
# set static media of O2 and CO2 at the top and bottom-- the "air" and continuous loss of O2 downwards
for x in range(grid_size):
    layout.set_specific_static_at_location('o2_e', (x,0), .0001) # top
    layout.set_specific_static_at_location('o2_e', (x,grid_size-1), 0.) # bottom
    layout.set_specific_static_at_location('co2_e', (x,0), .0001)

for x in range(grid_size):
    layout.set_specific_static_at_location('nh4_e', (0,x), 0.0)
```

12. In contrast to ammonia, which we assume is always entirely consumed by the root, we assume the root drips organic acids and methanol into the environment at fixed rate, so we use a refresh function.

```python
for x in range(grid_size):
    layout.set_specific_refresh_at_location('cit_e', (0,x), .000001)
    layout.set_specific_refresh_at_location('meoh_e', (0,x), .000001)
    layout.set_specific_refresh_at_location('succ_e', (0,x), .000001)
```

Set up the simulation parameters

13. Set up the experimental conditions. note the positive death rate.

```python
params = c.params()
params.set_param('timeStep', 0.1
params.set_param('maxCycles', 5000
params.set_param('maxSpaceBiomass', 10
params.set_param('deathRate', 0.0001 # die at rate of 1/10000 per hour
params.set_param('writeBiomassLog', True
params.set_param('BiomassLogRate', 500
params.set_param('writeMediaLog', True
params.set_param('MediaLogRate', 500)
params.set_param("writeFluxLog", True)
params.set_param("FluxLogRate", 500)
params.set_param('numRunThreads', 3)
params.set_param('defaultKm', 0.000001)
```

Run the simulation

14. Load the layout and parameters into a COMETS object and run the simulation

```python
sim = c.comets(layout, params)
sim.run(False)
```
      ? TROUBLESHOOTING

Visualize the results

15. We can visualize the results using the helper get_biomass_image function to get each species' biomass at the final cycle, and combining these into a single RGB image which we view with matplotlib (Fig. 13).

```
im = sim.get_biomass_image('iJN746', params.all_params['maxCycles'])
im2 = sim.get_biomass_image('iYO844',params.all_params['maxCycles'])
im3 = sim.get_biomass_image('iAF692',params.all_params['maxCycles'])

final = np.zeros((grid_size,grid_size,3))
final[:,:,0] = im / np.max(im)
final[:,:,1] = im2 / np.max(im2)
final[:,:,2] = im3 / np.max(im3)
for rock in rock_locs:
    final[rock[1]-1,rock[0]-1,0:3] = 0.5
from matplotlib import pyplot as plt
import matplotlib.colors, matplotlib.cm
my_cmap = matplotlib.cm.get_cmap("copper")
my_cmap.set_bad((0,0,0))

plt.imshow(final)
```

16. We can also see metabolite images using a similar helper function (Fig 13).

```
plt.imshow(sim.get_metabolite_image("succ_e",params.all_params['maxCycles']+1))
```

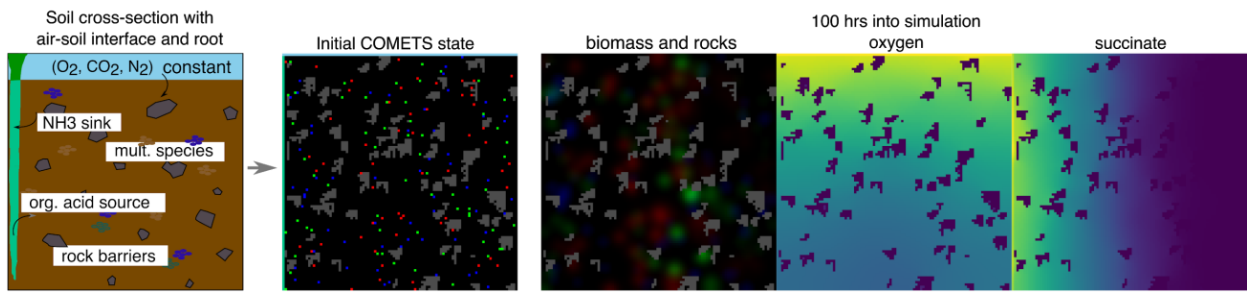

**Figure 12. Soil-air interface simulation.** Results from a spatial simulation, prepared with the python toolbox, containing multiple strains, rock-like barriers, and metabolite sources and sinks which mimic a root and the air. (Far-left) Schematic detailing common features of a soil microhabitat, which are set in COMETS using simple commands to specify metabolite concentrations, and different ways of maintaining or supplementing those concentrations, in specific spatial locations. The right three images show the biomass expanding in space, and the spatial gradients arising from the O2 source at the top and the succinate source at the left.

## Timing

A typical time to prepare and run a COMETS simulation is from 30 minutes to a few hours. This time however may vary depending on the complexity and size of the simulation layout grid, the number of models present in the simulation, the size of the S matrix in each model, and certainly the speed and number of available CPUs. The preparation of the input files (in MATLAB and Python) takes not more than 30 minutes. The simulation of a small model such as one consisting only of the core *E. coli* metabolism, for about 1000 time steps, in a layout consisting of a single grid point will take a few minutes. A single FBA calculation on the *E. coli* core model takes less than 0.02 seconds on a typical laptop PC. A simulation of a 400 by 400 grid layout, with several models consisting of a few thousands of reactions and metabolites, parallelized over 10 CPUs, can take several days. A simulation including evolution, for example, can run indefinitely, as long as the experimenter wants.

## Troubleshooting

| Step | Problem | Possible reasons | Possible solution |
|---|---|---|---|
| Attempting to create a model file from a cobra file in the Matlab toolbox. Many steps: Example step: CS#1, step 5 | initCobraToolbox() or any Cobra command not found. | The Cobra toolbox was not installed, or the path to it was not added to MATLAB. | Install Cobra and add the path. See the installation section. |
| Attempting to start a COMETS simulation or the COMETS gui. Many steps. Examples step: CS#1, step 13 | COMETS fails to launch. | Java is not installed. | Install Java from https://www.oracle.com/technetwork/java/javase/downloads/index.html. |
| | | 32-bit version of Java is also installed. | Check and make sure 64-bit Java has priority in the Path environment variable, putting it before the 32-bit version. |
| | | The COMETS_HOME environment variable is | Windows: Set the environment variable |

| | | not set. | COMETS_HOME.  Unix: Add the line `export COMETS_HOME=[path to COMETS]` to either .bashrc (Linux) or .bash_profile (Mac). |
|---|---|---|---|
| Attempting to start a COMETS simulation from the command line or using a toolbox.  Many steps. Example steps: CS#2, step 4 CS#3, step 5 | COMETS fails to launch. "Comets usage:..." or "COMETS requires at least one argument…" message is printed. | Wrong usage syntax. Wrong list of arguments. | Follow the instructions in the "Comets usage:" message. |
| | COMETS launches but halts upon loading a model, a parameters file, or a layout file, with a "classnotfound" error | Java class libraries are not in the path | Make sure installed libraries are in the expected path downstream from COMETS_HOME. If you are running a cloned version of the github version, make sure you download and install all dependencies. |
| Attempting to start the COMETS GUI.  CS#1, step 13 | GUI launched on a remote system does not show up. | X11 forwarding is not enabled. | When connecting to the remote Linux system via ssh, use the -Y option: ssh -Y *remote_address.* |
| Attempting to start COMETS simulation from the command line or load layout or model file in GUI.  Many steps: Example steps: CS#1, step 14 CS#3, step 5 | Layout or model file fails to load. | Wrong syntax or inconsistent input. | Check the error message on the COMETS console, or in the standard error file, for syntax errors. Check the consistency of the numbers of reactions and metabolites in the model file. |
| | | Gurobi class not found. | Install Gurobi, obtain a license. Check if the GUROBI_HOME environment variable is set. |

| Attempting to run a COMETS simulation with the Python toolbox. Many steps. Example step: The basics of COMETS using the Python Toolbox Step 10. | Initial test of a model after load fails. | The optimizer is not installed or the optimizer license was not installed. | Install Gurobi from gurobi.com. Check the OPTIMIZER block in the model file. Install license. |
|---|---|---|---|
| | "Error occurred while loading parameter file" message is printed. | Parameter file names do not match. | Check the parameter file name(s). |
| | "Parsing error in parameter file" message. | Parameter file keyword or value is wrong. | Check the parameter file(s). |
| | The propagation of biomass or media is numerically unstable. | Poor choice of time and spatial steps parameters values. | Change the time step to a lower value, or set a coarser spatial grid. |
| | Error in starting script mode. "Error running script file" message printed. | Error in the script file syntax. | Check the script file. Check the names of the parameters and layout files. |
| CS#14, step 2 | Model does not grow/uptake the metabolites provided in the media. | COMETS can only set model bounds dynamically at a given exchange reaction if they are within the bounds specified in the loded model. | Make sure that the model loaded into COMETS has the relevant exchange reactions with widely open bounds (e.g. -1000, 1000). In the python toolbox, this can be accomplished with the model.open_exchanges() function. |
| Running a COMETS simulation. Many steps. Example steps: CS#3, step 5 | COMETS simulation stops after 1 timestep with zero biomass. | Initial biomass has not been set and therefore defaults to 0 initial biomass. | Change initial biomass. In the python toolbox, model.initial_pop = [x, y, biomass] In the MATLAB toolbox, setInitialPop(world, '1x1', 5e-6); |
| Visualizing | In the python toolbox, | The models have different | Make unique model |

| biomass results data. Many steps in case studies using the Python toolbox. | models grow with identical rates when they are expected not to. | id variables. If models have the same id, the last COMETS model in the layout will be used for all models. | names. E.g. Ecoli1.id = 'e_coli_core_1' Ecoli2.id = 'e_coli_core_2' |
|---|---|---|---|
| Attempting to create a video. CS#2, step 5 CS#4, step 6 CS#5, step 4 | The application convert not found. | The ImageMagic application is not installed. | Install ImageMagic from https://imagemagick.org. |

Table 3. Troubleshooting

## Anticipated results

In this protocol we presented several case studies with the goal to illustrate the capabilities of COMETS, and provide starting points for future users. Here we will summarize anticipated results from the simulations described in the protocols.

### Growth of bacteria in well mixed conditions

In the first protocol we reproduced a classic FBA result, simulation of the dynamics of the core *E. coli* metabolism[43,77]. In this simplest case, we created a batch culture of *E. coli*, supplied with minimal media in homogeneous, or well mixed, conditions. This simulation is an example of a dFBA run, in its original form. The results are shown in Fig. 2. As mentioned above, Fig. 2a shows the growth of the *E. coli* biomass, while Fig. 2b shows the concentrations of glucose and the three products of glucose fermentation. The growth period of the biomass, as well as the secretion of the metabolic products, coincide with the depletion period of glucose.

### Modeling bacterial colonies: growth and propagation on flat surfaces

In this simulation we used a metabolically trivial model, one that uptakes a single nutrient which results in growth of the biomass. The goal of this protocol was to illustrate the complexity of formation of bacterial colonies using only simple models, both for the metabolic activity and the propagation of the biomass. In this case, the biomass is free to spread in two dimensions. The model for biomass propagation simulates the mobility of the bacteria due to simple pushing, as described in the methods/development of the protocol section. Although very simple, this model is capable of reproducing the transition of the colony morphology from a spherical one, to a more complex branching colony. The results obtained by performing two simulations, with two different values of the dense packing parameter are shown on Fig. 3a and 3b. For a lower value of this parameter, 0.5e4, we obtain a circular, non-branching colony. Setting the same parameter to 1.5e4 we obtain a branching colony, one where the front propagation is not stable[70]. The dynamics of the growth for the two cases is further illustrated in Supplementary videos 1a and 1b.

## Virtual petri dish

This protocol illustrates the capabilities of COMETS, as a virtual lab. The spatial layout is defined as a simulated Petri dish. The dish is split in two halves with different physical properties. The nutrient diffusivity in one half of the dish is ten times smaller than the other half. This simulation layout corresponds to an experimental layout with, for example, agar or natural substrate with two different densities. This layout results with colonies of different size in the two spatial parts.

The anticipated result is shown in Fig. 4. The top layer shows the biomass distribution after simulated 40 hours. The layout is split in the two regions with different friction constants, and consequently difference in the size of the colonies. Below the growth rate visualization shows that the colonies are growing predominantly at the leading edges. The two lower slices show the depletion of glucose and buildup of acetate in the spatial layout.

## Demographic noise and cooperative biomass propagation

Single species: In this simulation we included two biological properties modeled in COMETS: demographic noise and cooperative biomass propagation. These two biological features produce colonies that simulate the ones found in nature in a more realistic way than the ones obtained in Case Study #2. The morphology of the growth front of a colony of *E. coli* in this case is dendritic rather than smooth, due to the presence of growth instabilities in the model of cooperative biomass propagation. This simulation also provides an example of the visualization of the biomass, biomass growth rate and the glucose spatial profiles shown in Fig. 3c, 3e and 3f. It is clear from the figures that while the font where growth is observable is relatively thin and rough as shown in Fig. 3e , the nutrient is spread in a very different way with no sharp boundary at the depleted region, as shown in Fig. 3f. This illustrates the difference between the ordinary diffusion model implemented for the nutrient, and the model of cooperative diffusion implemented for the biomass.

Five species: In this protocol we employ five identical strains of *E. coli*. The model of biomass propagation is the same one as in the previous case study, cooperative biomass propagation. A change of the parameter that regulates the portion of the biomass that is actively propagating, results in a more dendritic morphology, with clearly separated branches. Here too we include demographic noise. The presence of demographic noise, in this case, leads to formation of sectors populated by a lower number of strains. As time progresses, the strains segregate completely, each one occupying its own spatial branch in the dendritic morphology. The decay of the heterozygosity is complete, and it has been observed in experiments and studied theoretically[79,80]. The result is shown in Supplementary video 3. A sample image is shown in Fig. 3d.

## Simulations including extracellular reactions

In this protocol we introduced a method for modeling the secretion and catalytic functions of extracellular enzymes. This new capability of COMETS simulates the costly production and secretion of enzymes by the cell. The secreted enzymes are free to diffuse into the environment. Figure 5 illustrates two cases. In the first case, shown in Fig. 5a, a simple reaction is defined with the form *A+B→C*. In the second case, shown in Fig. 5b, an enzyme, *E*, catalyzes the conversion of *F→G*. Here we used the Scerevisiae_iMM904.mat[81] stoichiometric model. The two panels show the difference of the outcome when an extracellular enzyme is involved in the reaction, and when it is not.

## Growth of multiple bacterial organisms in well-mixed conditions

Here, we illustrate the capability of COMETS to simulate the growth and metabolic exchange patterns of complex multispecies communities of microbes. We use a 14-species *in silico* community which, although still much smaller than natural microbiota, nonetheless showcases some of the temporally-dependent nutrient utilization patterns exhibited by multispecies consortia. For this simulation, we selected genome-scale models of a diverse set of organisms that span a number of taxonomic categories and metabolic utilization patterns[81–94]. We provide these *in silico* organisms with a simple two-carbon source medium to illustrate in detail how specific nutrients are consumed, though this framework can easily be extended to examine how different environmental compositions can affect the properties of this diverse multispecies community.

Beginning with our 14 organisms at equal starting abundances, we track their growth over a 12-hour period. Under these conditions, we observe that *M. extorquens* and *S. oneidensis* quickly overtake the community by consuming the two provided limiting nutrients, shown in Fig. 6a and 6b. A number of other organisms grow to lower abundances, and the remainder experience no growth. By using functions in the MATLAB toolbox, we show how to parse the COMETS media log file to infer patterns of metabolite secretion and exchange. In this example, we focus on acetate and formate, which we observed to be initially secreted at high rates as the primary nutrients were being consumed, though these byproducts were quickly utilized by the community members. This result is shown in Figure 6c.

## Modelling the diurnal cycle

In this protocol we implemented the possibility to simulate periodic changes in the environment. Here we implemented half sine function as a model for the day/night changes in the cycle of a photosynthetic organism. More generally, the periodic function can be either a step function, a sine function or a half sine function. We simulate one such experiment with a genome-scale model of *Prochlorococcus*[100,102], one of the most abundant marine photoautotrophs. The result is shown in Fig. 7. The biomass growth follows this light cycle with periods of growth and resting in a simulation of the day/night periodicity.

## Competition assay and competitive exclusion in serial transfers

This protocol illustrates the ability to use COMETS 2 to perform in-silico competitive fitness assays or test for competitive exclusion/coexistence in batch culture. In Fig. 8 we compete WT *E. coli* (iJO1366) with a deleterious mutant (a Triose-Phosphate Isomerase knock-out) in a well mixed (1x1 cell) aerobic environment containing minimal glucose. Both WT and mutant are introduced at equal abundance. In Fig. 8a we show growth over a single batch culture (as would be done when performing a competitive fitness assay). As expected the WT grew faster and reached a higher abundance than the Mutant. In Fig. 8b we grew the two strains for 25 transfers and show that the WT strain excludes the mutant.

## Simulating cross-feeding in a chemostat

This protocol shows how a chemostat simulation may be achieved with COMETS, as well as how cross-feeding can emerge in a co-culture with compatible genetic knockouts. The results of the simulation show that both strains survive at very similar densities (Fig. 9). This is to be expected, because lactose molecules consist of a monomer of glucose bound to a monomer of galactose, each of which can be converted to biomass with similar efficiency. Since the galE_KO strain can only metabolize the glucose portion, and the LCTStex_KO therefore has access only to the galactose portion, both strains reach similar density. Additionally, these growth-limiting metabolites, as is typical in a chemostat, are in very low concentrations once equilibrium is reached.

## Simulating evolutionary processes in microbial populations

COMETS 2 includes the ability to simulate evolutionary processes through the generation of metabolic mutants that have gained or lost novel reactions. In Fig. 10 we illustrate a small example of such a simulation. We start with a clonal population of E. coli and grow them in batch culture in minimal glucose. Over the course of the growth cycle new mutants arise (in this example through loss of function mutations (reaction deletions). Most of these mutants are lost due to drift (during passaging) but some are able to increase in frequency due to a modest reduction in genome-size (and associated genome-size cost).

## Simulating the sequence of mutations involved in an evolutionary innovation

Here we illustrate how COMETS can be used to simulate the ecological dynamics of evolutionary innovations. When the mutations and molecular mechanisms underlying a novel metabolic innovation are known, we can use this information to construct modified genome-scale metabolic models and then simulate the ecological dynamics of these models using COMETS 2. We illustrate this in the context of the Lenski long-term evolution experiment using the evolution of citrate utilisation as an example[103,104]. In this experiment 12 (almost) replicate populations of E.coli have been evolved in DM25 minimal glucose for more than 70,000 generations with 100 fold daily dilution. In one of the populations after ~33,000 generations a ~10 fold population-expansion was observed. This population expansion was associated with a novel metabolic innovation, the ability to consume citrate in aerobic conditions.

This innovation required two sequential gain of function mutations, the first leading to expression of the citT antiporter, and the second leading to expression of the dctA symporter[103,104]. The large population expansion is associated with the second of these mutations. To simulate this sequence of events we first simulate the growth of an E.coli model (lacking the CitT and dctA transporters) in batch culture with DM25 minimal media. A citT was mutant introduced at low abundance and grew to fixation. A CitdctA mutant was then introduced and also fixed, leading to a larger population expansion. The result is shown in Fig. 11. This example simulation uses the same models, parameters and reaction knockouts outlined in Bajic et al. 2018[57].

## Soil-air interface simulation

In this simulation, we set up source/sink metabolite pools along opposite axes: the "root," at the left-side, produced nutritious organic acids and removed ammonia, while the "air," at the top-side, provided a constant concentration of oxygen and nitrogen. Therefore, growth tends to be concentrated in the quadrant closest to the root and the air, as the soil bacterial models require both organic acids and oxygen to grow (Fig. 12). Furthermore, as time progresses, strong gradients become established due to the interaction between the nutrient source/sink pools and bacterial metabolism.

## Data availability

The COMETS Protocols GitHub repository (https://github.com/segrelab/COMETS_Protocols) contains all input files and jupyter notebooks from which one can reproduce the results presented in this protocol. The data is distributed under the Creative Commons CC0 1.0 Universal license.

## Code availability

COMETS (https://www.runcomets.org) is an open source code and it is available at https://github.com/segrelab/comets. The code is distributed under the GNU General Public License Version 3. The documentation is available at https://segrelab.github.io/comets-manual/ and https://comets-manual.readthedocs.io, which is structured as a tutorial and contains all the examples shown in this protocol. The Matlab toolbox is available at https://github.com/segrelab/comets-toolbox, distributed under the GNU General Public License Version 3. The COMETS Python toolbox is available at https://github.com/segrelab/cometspy, distributed under the GNU General Public License Version 3.

## Figures, videos and tables legends

**Figure 1. Overview of the COMETS platform.** Virtual experiments in COMETS combine a variety of environments and biochemical inputs. These combinations can be quickly generated using one of the provided interfaces, which feed into the COMETS core engine. The engine simulates the spatio-temporal dynamics of the ecosystem and outputs microbial biomass information, metabolic fluxes, and media concentration over time. Downstream analysis, either within the toolboxes or with the user's software of choice, can then be applied to further visualize and characterize the results.

**Figure 2. Growth of *E. coli* (core model) batch culture in minimal medium, with glucose as the only carbon source. a**) Plot of biomass vs. time. **b**) Plot of the key metabolites vs. time. The biomass growth stops when the glucose is completely depleted. The production of the typical products of fermentation also coincides with the growth of the biomass.

**Figure 3. A variety of morphologies simulated by COMETS.** Types of bacterial colony morphologies, simulated using the Convection 2D biomass propagation model: **a**) Circular colony with stable front

propagation simulated with the Convection 2D biomass propagation model. The value of the packedDensity parameter was set to a value below the critical for emergence of unstable growth front. **b**) Dendritic colony with unstable front propagation. In this case the value of packedDensity was greater than the critical. Panels **c**) and **d**) show two morphologies obtained by running the ConvNonLinDiff 2D biomass propagation model with demographic noise: **c**) Single strain colony, **d**) Five-strains colony. In the five strain colony case we see the segregation of the strains due to the presence of the demographic noise.

**Figure 4. Virtual Petri dish.** The four layers show the spatial distribution of the biomass, biomass growth rate, and concentrations of acetate and glucose, respectively from top to bottom. The blue color in the bottom two layers represents a depleted, and the red color represents an enriched metabolite region.

**Figure 5. Media concentrations over time during simulations demonstrating extracellular reactions.** a) A binding reaction of the form *A+B→C*, with rate $v=v_{max} \cdot [A] \cdot [B]$ where $v_{max} = 0.2\ s^{-1}$. b) An enzyme-catalyzed reaction of the form *E+S→E+P*, with rate according to the Michaelis-Menten equation $v=v_{max} \cdot [E] \cdot [S]/(K_M+[S])$, where $v_{max} = 2\ s^{-1}$ and $K_M = 0.25\ mmol$.

**Figure 6. Growth and metabolic exchange of 14-species microbial community.**
**a**) Biomass production of all 14 organisms over time. **b**) Consumption of limiting carbon sources over time. **c**) Secretion and consumption of metabolic byproducts over time.

**Figure 7. Simulations of the diurnal cycle of the marine photoautotrophic bacteria Prochlorococcus.** The organism is in an environment with a time-dependent light environment replicating the day - night cycle. The growth of the biomass is evident only during daytime.

**Figure 8. Competition assay and competitive exclusion.** Results from two experiments simulated with the python toolbox. In both experiments, growth is assayed in aerobic glucose minimal media (10 mM) in a volume of 1uL. **a**) Competition assay between wild-type E. coli and a mutant in which the glycolytic enzyme triose phosphate isomerase has been knocked out. **b**) Serial transfers were performed each 24h using the same setting as in panel **a**.

**Figure 9. Chemostat simulation.** Results from a chemostat simulation, prepared with the python toolbox, in which one strain unable to uptake lactose (LCTStex_KO) crossfeeds galactose from a different strain unable to metabolize galactose (galE_KO). The media environment was a constant supply of lactose (lcts_e), ammonia, and trace nutrients. Galactose (gal_e) was not supplied but entered the media as galE_KO grew. **a**) Biomass of the strains over time. **b**) Amounts of the two key metabolites over time. Note that it is typical for limiting nutrients to have near-zero concentrations in a chemostat.

**Figure 10. Simulation of evolutionary processes.** An *Escherichia coli* model was seeded in 1uL of glucose minimal media (0.1mM) and transferred each 3hr. To fresh media using a dilution factor of 1:2 during 10 days. Mutations (reaction knock-outs) were allowed to happen in this population at a rate of $10^{-8}$ knock-outs appearing per gene and generation. The cyan line represents the ancestor, which remains at high

density, and other colors are used to represent different mutations that appear, persist during variable periods and extinguish stochastically.

**Figure 11. Evolution of citrate utilization.** Mimicking the classic Lenski long-term evolution experiment, an *Escherichia coli* model (Ancestor) was simulated for many thousands of generations. At generation ~31000, a mutant was added to the simulation which was capable of growing on citrate (CitT), and which outcompeted the ancestor. At generation ~33000, a double mutant was introduced (CitTdctA), which outcompeted citT.

**Figure 12. Soil-air interface simulation.** Results from a spatial simulation, prepared with the python toolbox, containing multiple strains, rock-like barriers, and metabolite sources and sinks which mimic a root and the air. (Far-left) Schematic detailing common features of a soil microhabitat, which are set in COMETS using simple commands to specify metabolite concentrations, and different ways of maintaining or supplementing those concentrations, in specific spatial locations. The right three images show the biomass expanding in space, and the spatial gradients arising from the O2 source at the top and the succinate source at the left.

**Table 1.** List of COMETS capabilities in terms of biophysical settings one can apply and an example reason one may choose to use them.

**Table2.** Comparison of COMETS capabilities with previous version and other FBA based software packages. The black checkmark labels a fully functional capability. The blue checkmark labels a limited capability, one that requires additional programming/script writing.

**Appendix 2 Table.** COMETS parameters, with units, default value and a short definition.

**Supplementary Figure 1. The Graphical User Interface of COMETS.** COMETS simulations can be started from the GUI by loading a previously prepared layout, models and parameters files. It is meant mostly as a training tool with limited functionality. Future development of COMETS will focus on the development of a comprehensive GUI.

**Supplementary Figure 2. Growth of *E. coli* (core model) simulated using the Python toolbox.** The result in this simulation is identical to the one shown in Fig. 2. However, while Fig. 2 is a result of using the MATLAB toolbox, this figure is a result of a simulation using the Python toolbox.

**Supplementary video 1. Circular and branching colonies simulated utilizing a simple model of bacterial metabolism. a) Circular colony b) Branching colony.**

**Supplementary video 2. Simulations of a branching colony of *E. coli*.**

**Supplementary video 3. Simulations of branching and formation of sectors in a population of five identical strains of *E. coli*.**

# Appendix 1: Detailed structure of Output Files

## Console output

The standard console output of COMETS is either displayed on the GUI console or saved in an output file if run on a queueing system. The console output format is typically:

```
-script
running script file: comets_script
Loading layout file 'layout.txt'...
Found 2 model files!
Loading 'e_coli_core1.txt' ...
Loading 'e_coli_core1.txt' ...
Done!
 Testing default parameters...
Done!
Optimizer status code = 5 (looks ok!)
objective solution = [D@99e8be2
Loading 'e_coli_core2.txt' ...
Loading 'e_coli_core2.txt' ...
Done!
 Testing default parameters...
Done!
Optimizer status code = 5 (looks ok!)
objective solution = [D@7f1a75d
Constructing world...
Done!
medialist    ac[e]   acald[e]   akg[e]   co2[e]   etoh[e]   for[e]   fru[e]   fum[e]   glc__D[e]   gln__L[e]   glu__L[e]   h2o[e]   h[e]   lac__D[e]   mal__L[e]   nh4[e]   o2[e]   pi[e]   pyr[e]   succ[e]
WRITING MEDIA LOG
Cycle 1
Total biomass:
Model 0: 2.5117253201512343E-6
Model 1: 2.51172552547532E-6
Cycle complete in 0.695s
Cycle 2
Total biomass:
Model 0: 2.52350543670516E-6
Model 1: 2.523505417783118E-6
Cycle complete in 0.254s
…
Cycle 10000
Total biomass:
Model 0: 1.8129744692017875E-5
Model 1: 1.8466722853528566E-5
WRITING MEDIA LOG
Cycle complete in 0.315s
Cycle 10001
End of simulation
Total time = 1343.506s
```

In addition to the console output, if errors are detected, they are written in the standard error output file. The possible error messages are documented in the Troubleshooting section.

## Output files

The generation of the output file is optional and is controlled with the corresponding parameter in the global parameters input file. The rate of the output recording is also controlled by an input parameter.

### Total biomass file.

This is a space-delimited text format file with the first column containing the simulation step, and the integrated total biomass for each model in separate columns:

```
0   5E-6
1   5.0234319275E-6
2   5.0469736628E-6
3   5.0706257205E-6
4   5.0943886175E-6
```

The biomass unit is grams.

### Biomass file.

This is a MATLAB .m format file with the record of the spatial layout of the biomass. The variable is of the following format:

biomass_<step>_<model> (<xcoordinate>,<ycoordinate>)= <amount>;

Example of a biomass file of a 100x100 points layout, containing two models with initial population at the center of the layout. The biomass recording rate is each 100 simulation steps.

```
biomass_0_0 = sparse(100, 100);
biomass_0_0(51, 51) = 2.5E-6;
biomass_0_1 = sparse(100, 100);
biomass_0_1(51, 51) = 2.5E-6;
biomass_100_0 = sparse(100, 100);
biomass_100_0(51, 51) = 3.5426406048E-6;
biomass_100_1 = sparse(100, 100);
biomass_100_1(51, 51) = 3.4688274362E-6;
```

The biomass unit is grams.

### Media file

The media file is a MATLAB .m file format with the record of all external metabolite amounts in mmol units. The first line is an array of all metabolite names. The format is:

media_<time>{<metabolite index>}(<xcoordinate>, <ycoordinate>) = <amount>;

media_names = { 'ac[e]', 'acald[e]', 'akg[e]', 'co2[e]', 'etoh[e]', 'for[e]', 'fru[e]', 'fum[e]', 'glc__D[e]', 'gln__L[e]', 'glu__L[e]', 'h2o[e]', 'h[e]', 'lac__D[e]', 'mal__L[e]', 'nh4[e]', 'o2[e]', 'pi[e]', 'pyr[e]', 'succ[e]'};
media_0{1} = sparse(zeros(100, 100));
media_0{2} = sparse(zeros(100, 100));
media_0{3} = sparse(zeros(100, 100));
…
media_10000{18}(100, 97) = 1E0;
media_10000{18}(100, 98) = 1E0;
media_10000{18}(100, 99) = 1E0;
media_10000{18}(100, 100) = 1E0;
media_10000{19} = sparse(zeros(100, 100));
media_10000{20} = sparse(zeros(100, 100));

Fluxes file

This MATLAB .m format file contains the record of all fluxes of all models in each spatial point for a recorded time. The format is:

fluxes{<time>}{<x coordinate>}{<y coordinate>}{<model index>} = [ <flux values>];

fluxes{10}{1}{1}{1} = [-1.5084028022E1 0E0 -1.4742953314E1 5.0561349588E-1 5.0561349588E-1 -1.4742953314E1 0E0 0E0 0E0 -1.5084028022E1 8.39E0 -1.0840773838E1 4.6863796078E-1 0E0 5.0561349588E-1 0E0 0E0 3.5002712702E1 -1.5084028022E1 1.4742953314E1 -0E0 -0E0 0E0 1.5084028022E1 3.1251640737E1 -0E0 -0E0 -1.8474787691E1 -0E0 -0E0 5.5395471544E1 -1.167028166E1 -0E0 -0E0 -2.5553890725E0 -0E0 -1.7239784663E0 -0E0 -0E0];

Complete record file

This is a file in the MATLAB .mat format that contains all the input and output information for a given COMETS run. This file can be of a very large size and is meant to be used only for archiving purposes.

```
Matfile_example = 

  struct with fields:

            allowCellOverlap: 'true'
                   deathRate: 0
                 defaultHill: 1
                   defaultKm: 0.0100
                 defaultVmax: 10
                exchangestyle: 'Monod Style'
                 flowDiffRate: 3.0000e-10
                        flux: [5-D double]
               growthDiffRate: 0
              maxSpaceBiomass: 8.8000e-06
              minSpaceBiomass: 1.0000e-10
                numDiffPerStep: 10
                numRunThreads: 10
                showCycleCount: 'true'
                 showCycleTime: 'true'
                    spaceWidth: 0.0200
                      timeStep: 0.0100
                 timeStepsSaved: [11×1 double]
                  toroidalWorld: 'false'
                   total_biomass: [11×2 double]
```

# Appendix 2: Table of input parameters

Simulation parameters

| Parameter | Unit | Default value | Definition or notes |
| --- | --- | --- | --- |
| timeStep | hour | 1.0 | The amount of time between two consecutive simulation updates. |
| spaceWidth | cm | 0.1 | Width of one side of |
| | | | the 3d box in the 2D or 3D grid. Therefore, volume of a box = spaceWidth$^3$. Warning: this value matters for molarity calculations. |
| maxCycles | steps | Unlimited | Number of DFBA iterations (steps) for the simulation. The total simulation time will be timeStep * maxCycles. |
| deathRate | fraction/timepoint | 0.1 | The rate of biomass removal per time step. |
| maxSpaceBiomass | gr | 10 | Maximum biomass allowed in one grid box. |
| minSpaceBiomass | gr | 1e-10 | Minimum biomass in one grid box not considered zero. |
| cellSize | gr | 4.3e-13 | Grams in one cell. Relevant in simulations with serial dilutions or mutations. |
| exchangeStyle | Standard FBA, Monod Style, Pseudo-Monod Style | Standard FBA | The uptake function for the exchange reactions. |
| defaultVmax | mmol (gCDW)$^{-1}$ (hour)$^{-1}$ | 10 | Default maximum uptake rate of a metabolite for the Monod Style exchange. This overrides exchange reaction boundaries with greater magnitude, when using Monod updating. |
| defaultKm | mmol (cm$^3$)$^{-1}$ | 5 | Default concentration of a metabolite in which uptake is half-maximal. This value is |

| Parameter | Units/Values allowed | Default | Description |
|---|---|---|---|
| | | | compared with the metabolite concentration / spaceWidth$^3$ when computing Monod uptake. |
| defaultHill | | 2 | Hill coefficient. Alters the shape of the Monod uptake curve. |
| defaultAlpha | 1/(mmol (cm$^3$)$^{-1}$) | 1 | The default Alpha coefficient (slope) for the Pseudo-Monod style exchange. |
| defaultW | mmol (gCDW)$^{-1}$ (hour)$^{-1}$ | 10 | The default W coefficient (plateau) for the Pseudo-Monod style exchange. |
| minConcentration | mmol (cm$^3$)$^{-1}$ | 1e-26 | Minimal concentration of metabolites in the media. |
| numRunThreads | | 1 | If >1, allow multithreaded computation. The number of threads to run in parallel. |
| numDiffPerStep | | 10 | Number of substeps of media diffusion per biomass update step. |
| allowCellOverlap | | FALSE | If true, allows different species to occupy the same space. |

Parameters related to spatial propagation of either biomass or metabolites

| Parameter | Units/Values allowed | Default | Description |
|---|---|---|---|
| biomassMotionStyle | Diffusion 2D(Crank-Nicolson), Diffusion 2D(Eight Point), Diffusion 3D, Convection 2D, | Diffusion 2D(Crank-Nicolson) | Sets the method used for propagation of biomass. Only one of the indicated strings is an allowed value. |

|  | Convection 3D, ConvNonlin Diffusion 2D |  |  |
|---|---|---|---|
| growthDiffRate | cm2/s | 1.00E-07 | The default diffusion constant for the actively growing biomass in the Diffusion 2D (CN and EP) model. |
| flowDiffRate | cm2/s | 1.00E-07 | The default diffusion constant for the non-growing biomass in the Diffusion 2D (CN and FP) model. |
| defaultDiffConst | cm2/s | 1.00E-05 | The default diffusion constant for extracellular metabolites. |

Parameters related to log file writing

| Parameter | Default | Description |
|---|---|---|
| useLogNameTimeStamp | TRUE | If TRUE, appends a time stamp to every log file name. |
| writeFluxLog | FALSE | If true, writes fluxes out to a log file. |
| fluxLogName | flux_log.txt | The name of the flux log file. |
| fluxLogRate | 1 | How often to write to the flux file (number of simulation steps). A value of 1 will cause writing after every step. |
| writeMediaLog | FALSE | If true, writes media information to a log file. |
| mediaLogName | media_log.txt | The name of the media log file. |
| mediaLogRate | 1 | How often to write to the media file. |
| writeSpecificMediaLog | FALSE | If true, writes the media log only for the metabolites specified by specificMedia parameter. |
| specificMediaLogName | specific_media_log.txt | The name of the specific media log file. |
| specificMedia | | Names of metabolites for which we want to store media. |
| writeBiomassLog | FALSE | If true, writes biomass information to a log file. |
| biomassLogName | biomass_log.txt | The name of the biomass log file. |
| biomassLogRate | 1 | How often to write to the biomass file. |
| writeTotalBiomassLog | FALSE | If true, writes a summation of all biomass information to a log file. |
| totalBiomassLogName | total_biomass_log.txt | The name of the total biomass log file. |
| totalBiomassLogRate | 1 | How often to write to the total biomass log file. |

Parameters related to graphical user interface and image caption

| Parameter | Default | Description |
|---|---|---|
| showGraphics | TRUE | If true, the image will be displayed. |
| colorRelative | TRUE | If true, colors each space relative to the space with the highest value. |
| showCycleTime | TRUE | If true, shows the time it took to finish the FBA cycle in the output. |
| showCycleCount | TRUE | If true, shows the current cycle number in the output. |
| pauseOnStep | TRUE (false if running a script) | If true, pauses the simulation after completing a step. |
| displayLayer | 0 | Sets the current medium component (or biomass) to be displayed. The user must determine the number of the medium or biomass from the layout. |
| pixelScale | 4 | The number of pixels to render for each space. |

| saveSlideshow | FALSE | If true, saves a graphics slideshow to a series of files. |
|---|---|---|
| slideshowName | "/path_to_directory/slideshow" | The header of the names and path of the files with saved images. The format is "name"_number.slideshowExt |
| slideshowColorValue | 10 | Sets the color of the biomass when creating and saving an image. |
| colorRelative | TRUE | Show the colors relative to each model, i.e. on an RGB palette. |
| slideshowColorRelative | TRUE | As colorRelative above, applied to the slideshow. |
| slideshowRate | 1 | The number of steps between taking a slideshow picture. |
| slideshowLayer | 0 | Sets the current medium component (or biomass) to be displayed. The user must determine the number of the medium or biomass from the layout. |
| slideshowExt | png | The file extension(format) for slideshow pictures. Currently, "png" "bmp" and "jpg" are supported. "png" is recommended. |
| barrierColor | 0xff7D7D7D (gray) | Barrier color in hex. |
| backgroundColor | 0xff000000 (black) | Background color in hex. |

Parameter related to the extracellular reactions model

| numExRxnSubsteps | | 12 | Number of extracellular reactions substeps per biomass update step. |
|---|---|---|---|

Parameters related to lag phases

| Parameter | Unit | Default value | Definition or notes |
|---|---|---|---|

| simulateActivation | | FALSE | If true, the models are activated with the set activation rate. |
|---|---|---|---|
| activateRate | h$^{-1}$ | 0.001 | The value of activation rate. |

Parameters related to specific modes of growth, such as serial dilutions or chemostat mode

| Parameter | Unit | Default value | Definition or notes |
|---|---|---|---|
| batchDilution | | FALSE | Whether to perform serial dilutions. |
| dilFactor | Dil. factor | 1e-2 | If >1, dilution factor; if <1, 1/dilution factor. |
| dilTime | h | 12 | Periodicity of serial dilutions. |
| metaboliteDilutionRate | Fraction per hour | 0 | The rate of dilution of a metabolite. |

Parameters related to evolution (mutations)

| Parameter | Unit | Default value | Definition or notes |
|---|---|---|---|
| evolution | | FALSE | If true, the simulation will perform mutations. |
| mutRate | Per genome and cycle | 1e-9 | Mutation rate for reaction deletions. |
| addRate | Per genome and cycle | 1e-9 | Mutation rate for reaction additions. |

Parameters related to genome size cost

| Parameter | Unit | Default value | Definition or notes |
|---|---|---|---|
| costlyGenome | | FALSE | Does genome size penalize growth. |
| geneFractionalCost | | 0 | How much does genome size penalize growth. The cost grows |

|   |   |   | exponentially with genome size, with an exponent of 2. |
|---|---|---|---|

Additional, less often used general simulation parameters

| Parameter | Units | Default | Description |
|---|---|---|---|
| toroidalWorld |  | FALSE | If true, creates periodic boundary conditions. |
| showCycleCount |  | TRUE | If true, shows the current number of cycles/steps on the console. |
| showCycleTime |  | TRUE | If true, shows the time of a cycle/step on the console. |
| randomSeed |  | 0 | Seed value for the semi-random number generator. |
| defaultVelocityVector | cm/s | (0,0,0) | The default value for the velocity vector in the flow model. . |
| writeVelocityLog |  | FALSE | If true, writes velocity information to a log file. |
| velocityLogRate |  | 1 | How often to write to the velocity file (number of simulation steps). A value of 1 will cause writing after every step. |
| velocityLogName |  | velocity_log.txt | The name of the velocity log file. |
| writeMatFile |  | FALSE | If true, writes all of the simulation information to a log file. |
| matFileName |  | comets_log.mat | The name of the .mat |

| | | | | log file. |
|---|---|---|---|---|
| matFileRate | | | 1 | How often to write to the .mat file (number of simulation steps). A value of 1 will cause writing after every step. Warning: writing to this file every step may result in very large .mat file. |
| biomassLogFormat | | | MATLAB | The format in which the log file will be written. The default is .m MATLAB file. If the value is COMETS, the output is written in a space separated file. |
| mediaLogFormat | | | MATLAB | The format in which the log file will be written. The default is .m MATLAB file. If the value is COMETS, the output is written in a space separated file. |
| fluxLogFormat | | | MATLAB | The format in which the log file will be written. The default is .m MATLAB file. If the value is COMETS, the output is written in a space separated file. |
| velocityLogFormat | | | MATLAB | The format in which the log file will be written. The default is .m MATLAB file. If the value is COMETS, the output is written in a space separated file. |

Model-specific, these parameters are specified in the model file

| Parameter | Unit | Default value | Definition or notes |
|---|---|---|---|
| Optimizer | GUROBI, GLPK | GUROBI | The optimizer to be used for solving the FBA optimization. |
| OBJECTIVE_STYLE | MAX_OBJECTIVE_MIN_TOTAL, MAX_OBJECTIVE | MAX_OBJECTIVE | The type of optimization to be used. |
| VMAX_VALUES | mmol (gCDW)$^{-1}$ (hour)$^{-1}$ | Same as the global default. | Maximum flux constant for the Michaelis-Menten type exchange, for each reaction. Each reaction can be assigned separate value in the model file. |
| KM_VALUES | mmol/cm$^3$ | Same as the global default. | The Michaelis constant for the Michaelis-Menten type exchange, for each reaction. Each reaction can be assigned separate value in the model file. |
| packedDensity | g/cm$^3$ | 1.0 | The biomass density of densely packed cells in the CONVECTION 2D model. |
| frictionConst | Pa sec/cm$^2$ | 1.0 | The friction constant in the Convection 2D model. |
| elasticModulus | Pa | 1.0 | The elastic constant in the Convection 2D model. |
| convDiffConstant | cm$^2$/s | 1.0 | The diffusivity constant in the Convection 2D model. |
| convNonlinDiffZero | cm$^2$/s | 1.0 | The linear diffusivity in the ConvNonlin Diffusion 2D model. |

| convNonlinDiffN | N/A It is a function of convNonlinDiffExponent | 1.0 | The non-linear diffusivity coefficient in the ConvNonlin Diffusion 2D model. |
|---|---|---|---|
| convNonlinDiffExponent | | 1 | The exponent in the ConvNonlin Diffusion 2D model. |
| convNonlinDiffHillN | | 10 | The exponent in the Hill function model of local growth dependent diffusivity. |
| convNonlinDiffHillK | N/A It is a function of convNonlinDiffHillN. | 0.9 | The K constant in the Hill function model of local growth dependent diffusivity. |
| noiseVariance | | 0.0 | The variance of the growth noise factor. |
| neutralDrift | | FALSE | The boolean switch for the demographic noise. |
| neutralDriftSigma | $g^{1/2} s^{-1}$ | | The pre-factor constant for the demographic noise. |

## Author contributions statement

Overall management of COMETS platform: DS, ID. Conceptualization of COMETS capabilities: DS, ID, AS, WH, KK. Writing and maintenance of initial COMETS code: ID, WJR. Development of current COMETS software and capabilities: ID, DB, JC, MQ. Writing of specific modules: WJR, ID, JC, DB, MQ, SS. Preparing and implementing protocols: ID, JC, DB, MQ, ARP, SS, JV. Conceptualization and preparation of the manuscript: ID, DB, JC, MQ, JV, SS, ARP, DBB, KK, AS, WH, DS.

## ORCID

DS: 0000-0003-4859-1914
ID: 0000-0002-5160-1944
ARP: 0000-0002-1128-3232
DBB: 0000-0001-6091-4021

WRH: 0000-0001-8445-2052
JMC: 0000-0002-6129-2604
JCCV: 0000-0003-0499-0200
SS: 0000-0002-3967-1683
MQ: 0000-0001-6466-4234
DB: 0000-0002-6716-7898
AS:0000-0002-2292-5608
KSK: 0000-0003-1988-0645
WJR: 0000-0002-3405-2744

# Acknowledgments

We are grateful to members of the Segrè, Sanchez, and Harcombe labs for helpful inputs and discussions at multiple stages of the development of COMETS.  We also thank Michael Hasson for his contribution to the development of the code. The development of the first version of COMETS was supported by the U.S. Department of Energy, Office of Science, Office of Biological & Environmental Research, grant DE-SC0004962 to DS. DS also acknowledges funding from the U.S. Department of Energy, Office of Science, Office of Biological & Environmental Research through the Microbial Community Analysis and Functional Evaluation in Soils SFA Program (m-CAFEs) under contract number DE-AC02-05CH11231 to Lawrence Berkeley National Laboratory; the NIH (T32GM100842, 5R01DE024468, R01GM121950), the National Science Foundation (1457695 and NSFOCE-BSF 1635070), the Human Frontiers Science Program (RGP0020/2016), and the Boston University Interdisciplinary Biomedical Research Office. ARP is supported by a Howard Hughes Medical Institute Gilliam Fellowship and a National Academies of Sciences, Engineering, and Medicine Ford Foundation Predoctoral Fellowship. SS was funded by SINTEF, the Norwegian graduate research school in bioinformatics, biostatistics and systems biology (NORBIS) and by the INBioPharm project of the Centre for Digital Life Norway (Research Council of Norway grant no. 248885). WRH acknowledges funding from RO1GM121498. Work by AS, DB and JCCV was supported by a young investigator award from the Human Frontier Science Program (RGY0077/2016), by a Packard Fellowship from the David and Lucile Packard foundation, and by the National Institutes of Health through grant 1R35 GM133467-01 to AS. KSK was supported by Simons Foundation Grants #409704 and by the Research Corporation for Science Advancement through Cottrell Scholar Award #24010.

# Competing interests

The authors declare that they have no competing financial interests.

# Supplementary figures and videos

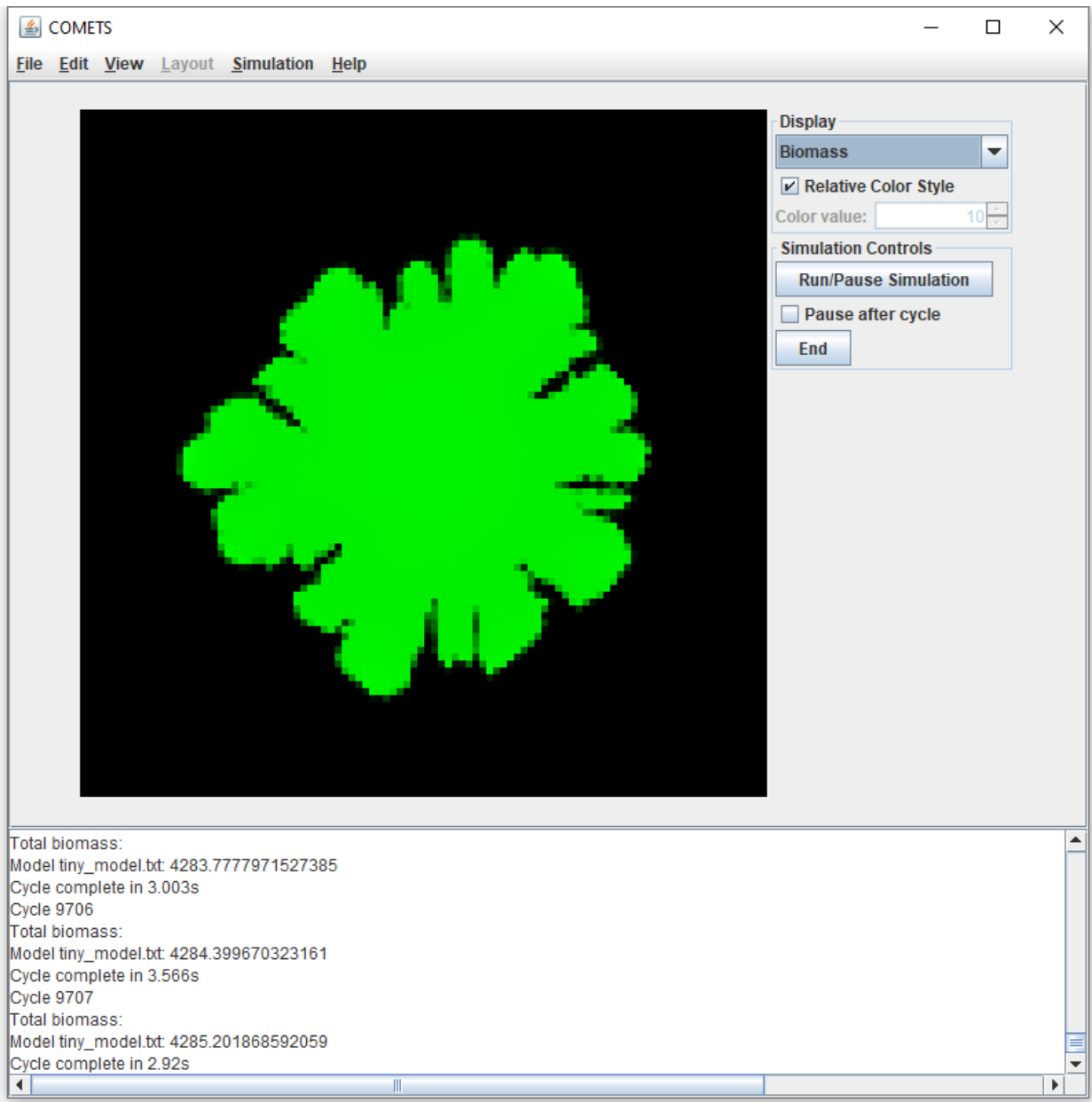

**Supplementary Figure 1. The Graphical User Interface of COMETS.** COMETS simulations can be started from the GUI by loading a previously prepared layout, models and parameters files. It is meant mostly as a training tool with limited functionality. Future development of COMETS will focus on the development of a comprehensive GUI.

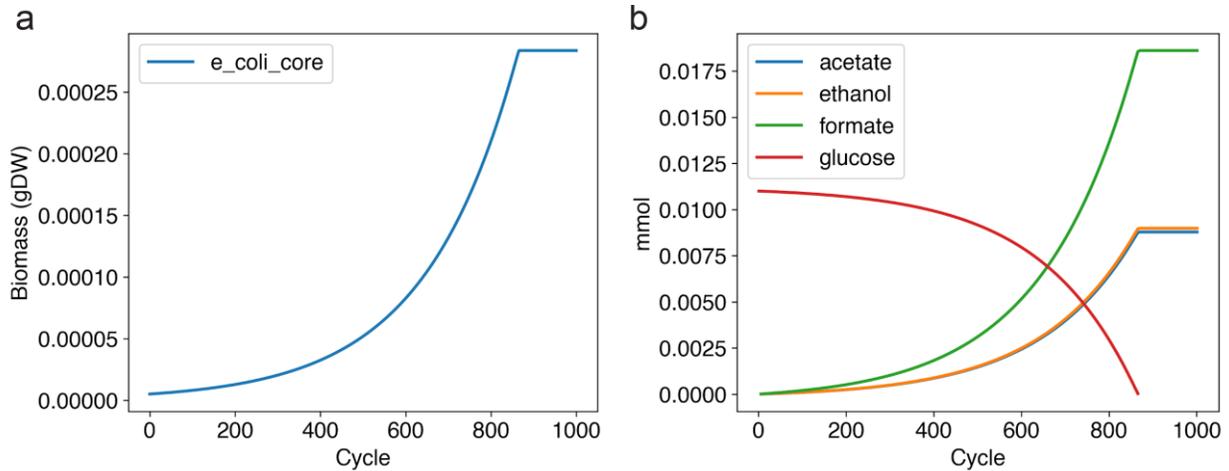

**Supplementary Figure 2. Growth of *E. coli* (core model) simulated using the Python toolbox.** The result in this simulation is identical to the one shown in Fig. 2. However, while Fig. 2 is a result of using the MATLAB toolbox, this figure is a result of a simulation using the Python toolbox.

**Supplementary video 1**. https://www.runcomets.org/video1 Circular and branching colonies simulated utilizing a simple model of bacterial metabolism: a) Circular colony, b) Branching colony.

**Supplementary video 2.** https://www.runcomets.org/video2 Simulations of a branching colony of *E. coli*.

**Supplementary video 3.** https://www.runcomets.org/video3 Simulations of branching and formation of sectors in a population of five identical strains of *E. coli*.